\pdfoutput=1
\RequirePackage{ifpdf}
\ifpdf 
\documentclass[pdftex]{sigma}
\else
\documentclass{sigma}
\fi

\usepackage{physics}
\usepackage{enumitem}
\usepackage{mathtools}
\usepackage{relsize}
\usepackage{dsfont}
\usepackage{thm-restate}
\usepackage{tikz}
\usepackage{tikz-cd}
\usetikzlibrary{braids}
\usepackage{ytableau}
\usepackage{stackengine}
\usepackage{array}
\usepackage{mathtools,xparse}
\usepackage{relsize}
\usepackage{tikz,tkz-tab}
\usepackage{authblk}
\usepackage{pgfplots}

\DeclarePairedDelimiter{\oldnormaux}{\bracevert}{\bracevert}

\NewDocumentCommand{\oldnorm}{som}{%
 \IfBooleanTF{#1}
 {\oldnormaux*{#3}}
 {\IfNoValueTF{#2}
 {\oldnormaux*{\vphantom{dq}#3}}
 {\oldnormaux[#2]{#3}}%
 }%
}

\usetikzlibrary{decorations.markings}
\tikzset{
 pt/.style={insert path={node[scale=2]{.}}},
 hdnup/.style={insert path={ [pt] .. controls +(0,-1) and +(0,1) .. +(#1,-0.75) [pt]}},
 dnup/.style={insert path={ [pt] .. controls +(0,1) and +(0,-1) .. +(#1,2) [pt]}},
 dndn/.style={insert path={ [pt] .. controls +(0,1) and +(0,1) .. +(#1,0) [pt]}},
 upup/.style={insert path={ [pt] .. controls +(0,-1) and +(0,-1) .. +(#1,0) [pt]}},
 hupup/.style={insert path={ [pt] .. controls +(0,-1.5) and +(0,-1.5) .. +(#1,0) [pt]}},
}
\tikzset{->-/.style={decoration={
 markings,
 mark=at position .5 with {\arrow{>}}},postaction={decorate}}}
\tikzset{-<-/.style={decoration={
 markings,
 mark=at position .5 with {\arrow{<}}},postaction={decorate}}}

\newcommand{\CC}{\mathbb{C}}

\newcommand{\NN}{\mathbb{N}}
\newcommand{\QQ}{\mathbb{Q}}
\newcommand{\ZZ}{\mathbb{Z}}

\newcommand{\qa}{\mathfrak{q}}
\newcommand{\qam}{\mathfrak{q}^{-1}}
\newcommand{\Uq}{U_{\mathfrak{q}}\mathfrak{sl}_2}

\newcommand{\Hs}{\mathcal{H}}
\newcommand{\Hb}{\mathcal{H}_b}
\newcommand{\Hbb}{\mathcal{H}_{2b}}
\newcommand{\Hxxz}{H_{{\rm XXZ}}}
\newcommand{\Hnd}{H_{{\rm n.d.}}}
\newcommand{\KK}{\mathsf{K}}

\newcommand{\EE}{\mathsf{E}}
\newcommand{\FF}{\mathsf{F}}
\newcommand{\HH}{\mathsf{H}}
\newcommand{\CCC}{\mathsf{C}}
\newcommand{\BB}{\mathsf{B}}
\newcommand{\YYY}{\mathsf{Y}}

\newcommand{\II}{\mathcal{I}}
\newcommand{\RRR}{\mathsf{R}}
\newcommand{\VV}{\mathcal{V}}

\newcommand{\WW}{\mathcal{W}}
\newcommand{\UU}{\mathcal{U}}
\newcommand{\XX}{\mathcal{X}}
\newcommand{\YY}{\mathcal{Y}}
\newcommand{\MM}{\mathcal{M}}

\newcommand{\TL}{\mathsf{TL}_{\delta,N}}
\newcommand{\tl}{\mathsf{TL}}
\newcommand{\Blob}{\mathsf{B}_{\delta,y,N}}

\newcommand{\twoBlob}{2\BB_{\delta,y_{l/r},Y,N}}
\newcommand{\twoBlobb}{2\BB_{\delta,\delta-y_{l/r},\delta-y_l-y_r+Y,N}}
\newcommand{\twoBlobmm}{2\BB_{\delta,\delta-y_{l/r},\delta-y_l-y_r+Y_{N-M-1},N}}
\newcommand{\twoBlobm}{2\BB_{\delta,y_{l/r},Y_M,N}}
\newcommand{\utwoBlob}{2\BB_{\delta,y_{l/r},N}^{\rm uni}}
\newcommand{\Id}{\operatorname{Id}}
\newcommand{\End}{\operatorname{End}}
\newcommand{\Hom}{\operatorname{Hom}}

\newcommand{\ddd}{{\rm d}}
\newcommand{\qtr}{\operatorname{qtr}}

\newcommand{\Ker}{\operatorname{Ker}}

\newcommand{\bbullet}{\mathlarger{\mathlarger{\mathlarger{\bullet}}}}
\newcommand{\ccirc}{\mathlarger{\mathlarger{\mathlarger{\circ}}}}

\numberwithin{equation}{section}

\newtheorem{Theorem}{Theorem}[section]
\newtheorem*{Theorem*}{Theorem}
\newtheorem{Corollary}[Theorem]{Corollary}
\newtheorem{Lemma}[Theorem]{Lemma}

 { \theoremstyle{definition}

\newtheorem{Example}[Theorem]{Example}
\newtheorem{Remark}[Theorem]{Remark} }

\begin{document}
\allowdisplaybreaks

\newcommand{\arXivNumber}{2212.09696}

\renewcommand{\PaperNumber}{046}

\FirstPageHeading

\ShortArticleName{Algebraic Bethe Ansatz for the Open XXZ Spin Chain}

\ArticleName{Algebraic Bethe Ansatz for the Open XXZ Spin Chain\\ with Non-Diagonal Boundary Terms\\ via $\boldsymbol{U_{\mathfrak{q}}\mathfrak{sl}_2}$ Symmetry}

\Author{Dmitry CHERNYAK~$^{\rm ab}$, Azat M. GAINUTDINOV~$^{\rm c}$, Jesper Lykke JACOBSEN~$^{\rm abd}$\newline and Hubert SALEUR~$^{\rm be}$}

\AuthorNameForHeading{D.~Chernyak, A.M.~Gainutdinov, J.L.~Jacobsen and H.~Saleur}

\Address{$^{\rm a)}$~Laboratoire de Physique de l'{\'E}cole Normale Sup{\'e}rieure, ENS,\\
\hphantom{$^{\rm a)}$}~Universit{\'e} PSL, CNRS, Sorbonne Universit{\'e}, Universit{\'e} de Paris, 75005 Paris, France}
\EmailD{\href{mailto:dmitry.chernyak@phys.ens.fr}{dmitry.chernyak@phys.ens.fr}, \href{mailto:jesper.jacobsen@ens.fr}{jesper.jacobsen@ens.fr} }

\Address{$^{\rm b)}$~Institut de Physique Th{\'e}orique, Paris Saclay, CEA, CNRS, 91191 Gif-sur-Yvette, France}
\EmailD{\href{mailto:hubert.saleur@ipht.fr}{hubert.saleur@ipht.fr}}

\Address{$^{\rm c)}$~Institut Denis Poisson, CNRS, Universit\'e de Tours, Parc de Grandmont, 37200 Tours, France}
\EmailD{\href{mailto:azat.gainutdinov@lmpt.univ-tours.fr}{azat.gainutdinov@lmpt.univ-tours.fr}}

\Address{$^{\rm d)}$~Sorbonne Universit{\'e}, {\'E}cole Normale Sup{\'e}rieure, CNRS, Laboratoire de Physique (LPENS),\\
\hphantom{$^{\rm d)}$}~75005 Paris, France}

\Address{$^{\rm e)}$~USC Physics and Astronomy Department, Los Angeles Ca 90089, USA}

\ArticleDates{Received January 26, 2023, in final form July 04, 2023; Published online July 16, 2023}

\Abstract{We derive by the traditional algebraic Bethe ansatz method the Bethe equations for the general open XXZ spin chain with non-diagonal boundary terms under the Nepomechie constraint [\textit{J.~Phys.~A} \textbf{37} (2004), 433--440, arXiv:hep-th/0304092]. The technical difficulties due to the breaking of $\mathsf{U}(1)$ symmetry and the absence of a reference state are overcome by an algebraic construction where the two-boundary Temperley--Lieb Hamiltonian is realised in a new $U_{\mathfrak{q}}\mathfrak{sl}_2$-invariant spin chain involving infinite-dimensional Verma modules on the edges [\textit{J.~High Energy Phys.} \textbf{2022} (2022), no.~11, 016, 64~pages, arXiv:2207.12772]. The equivalence of the two Hamiltonians is established by proving Schur--Weyl duality between $U_{\mathfrak{q}}\mathfrak{sl}_2$ and the two-boundary Temperley--Lieb algebra. In this framework, the Nepomechie condition turns out to have a simple algebraic interpretation in terms of quantum group fusion rules.}


\Classification{81R50; 81R12; 81U15; 16T25}

\section{Introduction}

Let us consider the open XXZ Hamiltonian acting on $\Hs:=(\CC^2)^{\otimes N}$ with the most general boundary fields
\begin{gather}
H_{{\rm n.d.}} := \frac{1}{2}\sum_{i=1}^{N-1} \big(\sigma^x_{i}\sigma^x_{i+1}+\sigma^y_{i}\sigma^y_{i+1}+\cosh(h)\sigma^z_{i}\sigma^z_{i+1}+\sinh(h)\big(\sigma_{i+1}^z-\sigma_i^z\big)\big) \nonumber\\
\hphantom{H_{{\rm n.d.}} := }{} +\frac{\sinh(h)}{2\sinh(h\delta_l)\cosh(h\kappa_l)}\big({\rm e}^{h\theta_l}\sigma_1^++{\rm e}^{-h\theta_l}\sigma_1^-+\sinh(h(\delta_l+\kappa_l))\sigma^z_1\big) \nonumber\\
\hphantom{H_{{\rm n.d.}} := }{} +\frac{\sinh(h)}{2\sinh(h\delta_r)\cosh(h\kappa_r)}\big({\rm e}^{h\theta_r}\sigma_N^++{\rm e}^{-h\theta_r}\sigma_N^--\sinh(h(\delta_r+\kappa_r))\sigma^z_N\big)
\label{hnd}
\end{gather}
depending on $7$ parameters $h$, $\delta_{l/r}$, $\kappa_{l/r}$ and $\theta_{l/r}$, where
\begin{equation*}
\sigma^x=\begin{pmatrix}
0 & 1\\ 1 & 0
\end{pmatrix} ,\qquad
\sigma^y=\begin{pmatrix}
0 & -{\rm i}\\ {\rm i} & \hphantom{-}0
\end{pmatrix} ,\qquad
\sigma^z=
\begin{pmatrix}
1 & \hphantom{-}0\\ 0 & -1
\end{pmatrix}
\end{equation*}
are Pauli matrices and
\begin{equation*}
\sigma^+=\frac{1}{2}\left(\sigma^x+{\rm i}\sigma^y\right)=
\begin{pmatrix}
0 & 1\\ 0 & 0
\end{pmatrix} ,\qquad
\sigma^-=\frac{1}{2}\left(\sigma^x-{\rm i}\sigma^y\right)=
\begin{pmatrix}
0 & 0\\ 1 & 0
\end{pmatrix}
\end{equation*}
are raising and lowering matrices. Performing a rotation of angle $\theta$ around the $z$-axis shifts $\theta_{l/r}$ by $\theta$ while leaving the other $5$ parameters unchanged so one can always set, for example, $\theta_r$~to~$0$. In other words, the spectrum of $H_{{\rm n.d.}}$ only depends on the difference
\begin{equation}
\label{Theta}
\Theta:=\theta_l-\theta_r
\end{equation}
and so there is actually only $6$ relevant parameters.\footnote{One can also show that the spectrum of $H_{{\rm n.d.}}$ does not depend on the sign of $\Theta$ (see end of Section~\ref{latticealg}).}

While it is known from Sklyanin's boundary Bethe ansatz formalism~\cite{sklyanin1988boundary} and the subsequent construction of boundary $K$-matrices~\cite{bk1993, bk1994} that $H_{{\rm n.d.}}$ is integrable, the rigorous implementation of this procedure for the most general choice of parameters is far from being straightforward. The main reason is that the non-diagonal boundary terms in~\eqref{hnd} containing $\sigma_1^\pm$ and $\sigma_N^\pm$ break the $\mathsf{U}(1)$ invariance of the usual XXZ model and so $\ket{\uparrow}^{\otimes N}$ is not an eigenvector of $H_{{\rm n.d.}}$ anymore and cannot be used as a reference state (also sometimes called ``pseudovacuum'') for algebraic Bethe ansatz (ABA).

In the last twenty years, various new approaches have been proposed to circumvent this problem. A first major breakthrough was made in~\cite{nepomechie2002functional, nep2003} and independently in~\cite{cao2002exact}, with the derivation of the Bethe ansatz equations (BAE) under the assumption
\begin{equation}
\label{nepcond}
\delta_l+\kappa_l+\delta_r+\kappa_r\pm\Theta=2M+1-N\in\ZZ,
\end{equation}
where $M\geq 0$ is the magnon number, a constraint informally known as the ``Nepomechie condition".\footnote{Note that the notations $\alpha_{\pm}$, $\beta_{\pm}$, $\theta_{\pm}$, $\eta$ in~\cite{nep2003} correspond here to $\delta_{l/r}$, $\kappa_{l/r}$, $\theta_{l/r}$, $h$ respectively and that $k=2M+1-N$ according to~\cite[equation~(3.31)]{nep2003}.} Later on, the BAE for any choice of parameters were derived~\cite{bookoff} and it turned out that if the constraint~\eqref{nepcond} is not satisfied they contained an additional ``inhomogeneous" term, which moreover fixed the magnon number to $M=N$ (see more details in Appendix~\ref{otherBAE}). The same equations were also obtained using various forms of coordinate Bethe ansatz~\cite{Cramp_2010, Gier_2004, Simon_2009} and the separation of variables method~\cite{Kitanine_2014}. Additionally, the closely related modified algebraic Bethe ansatz formalism was developed~\cite{Avan_2015, Belliard_2015, Belliard_2013, PBel} as well as an algebraic framework based on the so-called $\qa$-Onsager algebra~\cite{Baseilhac_2007}.

Although these methods provide a satisfactory solution to the spectral problem of $H_{{\rm n.d.}}$ it is fair to say they are rather indirect and still lack a simple representation-theoretic understanding. Indeed, in the above-mentioned works, the BAE are either derived by analytically continuing some truncated functional relations and fusion rules at roots of unity~\cite{nepomechie2002functional, nep2003}, by using an intricate dynamical gauge (or face-vertex~\cite{Filali_2011}) transformation~\cite{PBel,cao2002exact} or by writing the most general form for the eigenvalues of the transfer matrix respecting certain analyticity conditions and asymptotics~\cite{bookoff}. In all these different approaches, the Nepomechie condition~\eqref{nepcond} naturally appears at some step of the computation but its representation-theoretic meaning remains elusive. Notice that the role of Nepomechie condition~\eqref{nepcond} was also touched upon from different perspectives in several other works~\cite{Avan_2015, Bajnok_2006} (see also~\cite{Galleas_2005} for the XXX case).

In this paper, we will rigorously derive the BAE for all $H_{{\rm n.d.}}$, under the constraint~\eqref{nepcond} by standard algebraic Bethe ansatz and explain the algebraic origin of this condition.

Our first step is to reinterpret $H_{{\rm n.d.}}$ as (a representation of) an abstract element
\begin{equation*}
\mathbf{H}:=-\mu_lb_l-\mu_rb_r-\sum_{i=1}^{N-1} e_i
\end{equation*}
belonging to a certain lattice algebra, namely the two-boundary Temperley--Lieb algebra \linebreak $\twoBlob$, evaluated in a specific $2^N$-dimensional representation $(\WW_0,\rho_{\WW_0})$ called the vacuum module (see definitions in Section~\ref{latticealg}). Concretely, this means that, with some (explicit) mapping of parameters $(h,\delta_{l/r},\kappa_{l/r},\Theta)\leftrightarrow (\delta,y_{l/r},Y,\mu_{l/r})$, we have ${H_{{\rm n.d.}}=\rho_{\WW_0}(\mathbf{H})}$ (see Theorem~\ref{dgthm} which is due to~\cite{de_Gier_2009}).

The next step is to repackage all the $\Hnd$ satisfying~\eqref{nepcond} into sectors of a different spin chain whose Hilbert space
\begin{equation*}
\Hbb:=\VV_{\alpha_l}\otimes (\CC^2)^{\otimes N}\otimes\VV_{\alpha_r}
\end{equation*}
is constructed from $N$ spin-$\frac{1}{2}$ representations and two infinite-dimensional Verma modules $\VV_{\alpha_{l/r}}$ of the $\Uq$ quantum group and whose Hamiltonian $H_{2b}$ commutes with the action of~$\Uq$ on~$\Hbb$ (Sections~\ref{uqgroup} and \ref{swsec}). As a representation of $\Uq$, $\Hbb$ decomposes into an infinite direct sum of irreducible representations, with some multiplicity spaces $\Hs_M$, $M\geq 0$~\eqref{hbbdec}. It was shown in~\cite{previouspaper}\footnote{The analysis in this reference was performed for even $N$ only, however it can be extended rather straightforwardly to the odd $N$ case too.} that $\Hs_M$ carries an action of the two-boundary Temperley--Lieb algebra $\twoBlobm$ for some explicit value $Y_M$ of the $Y$ parameter~\eqref{ymlr}. Under explicit assumptions on the generic values of the bulk parameter $\qa$ and of the boundary parameters $\alpha_{l/r}$ we prove in Theorem~\ref{thm} that $\Hs_M$, as a $\twoBlobm$-module, is isomorphic to the (irreducible) vacuum module $\WW_0$ if $M\geq N$ and to an irreducible piece of $\WW_0$ if $0\leq M\leq N-1$, thereby confirming~\cite[Conjecture~1]{previouspaper}. This new Schur--Weyl duality between $\Uq$ and the two-boundary Temperley--Lieb algebra is the main algebraic result of this paper.\footnote{It is worth mentioning that a similar Schur--Weyl duality theorem but with finite-dimensional spin-$j$ representations of $\Uq$ at the boundary has been recently proven by Daugherty and Ram~\cite[Theorem 5.1]{daugherty2020calibrated} (see also~\cite{daugherty2018boundary}).}

Using this theorem, we can interpret the restriction of~$H_{2b}$ to $\Hs_M$ as a representation of~$\mathbf{H}$ and thus identify it with $H_{{\rm n.d.}}$ satisfying~\eqref{nepcond} for $M\geq N$ and with an irreducible subblock of~$H_{{\rm n.d.}}$ satisfying~\eqref{nepcond} for $0\leq M\leq N-1$ (Corollary~\ref{cor}). By a simple algebraic transformation, we are also able to reach the remaining block of $H_{{\rm n.d.}}$ for $0\leq M\leq N-1$ as well as negative values of $M$, thus realising any $H_{{\rm n.d.}}$ satisfying~\eqref{nepcond} as some subsector of $H_{2b}$ (Corollary~\ref{corr}). In other words, the spectral problem of $H_{{\rm n.d.}}$ for all values of the parameters subject to the constraint~\eqref{nepcond} is equivalent to diagonalising $H_{2b}$. But this is a simpler task, since $H_{2b}$, as a~representation of $\mathbf{H}$, is integrable~\cite{Doikou_2003}, and, as an operator acting on $\Hbb$, is $\Uq$-invariant and so has a suitable highest-weight reference eigenvector to implement ABA. This formalism also gives an algebraic interpretation of the Nepomechie condition: it is just a direct consequence of the fusion rules for the $\Uq$-modules entering the construction of $\Hbb$, in particular Verma modules, which restrict the possible values of the $Y$ parameter of $\twoBlob$ to the discrete set $\{Y_M, M\in\ZZ\}$.

It is worth mentioning that the idea to use the two-boundary Temperley--Lieb algebra to derive the BAE for $H_{{\rm n.d.}}$ and to understand the Nepomechie condition from an algebraic point of view was previously explored in~\cite{Gier_2004}. However, the lack of a suitable $\Uq$-invariant representation in that paper makes it necessary to use coordinate Bethe ansatz in a given basis and keeps the relevant underlying algebra hidden.

The paper is divided into two parts. The first (Section~\ref{algset}), purely algebraic, introduces the objects and states the theorems we need to make a precise connection between $H_{{\rm n.d.}}$ and $H_{2b}$. The second (Section~\ref{BAEsec}) is devoted to the diagonalisation of $H_{2b}$, first by implementing the ABA procedure for the simpler one-boundary Hamiltonian $H_b$ (Section~\ref{oneBA}) and then by extending it to the two-boundary Hamiltonian $H_{2b}$ (Section~\ref{twoBA}). We also discuss the completeness of the BAE in both cases. The main text is supplemented by three technical appendices, the first (Appendix~\ref{proof}) containing the proof of Theorem~\ref{thm}, the second (Appendix~\ref{genBAE}) carrying out ABA for the most general integrable $\Uq$-invariant highest-weight spin chain and the third (Appendix~\ref{otherBAE}) discussing an alternative form of BAE for $\Hnd$ also appearing in the literature.

\section*{Notations}

\begin{itemize}\itemsep=0pt
\item $N$: length of bulk of spin chains.
\item $\sigma^x=\begin{psmallmatrix} 0 & 1\\ 1 & 0\end{psmallmatrix}$, $\sigma^y=\begin{psmallmatrix} 0 & -{\rm i}\\ {\rm i} & \hphantom{-} 0\end{psmallmatrix}$, $\sigma^z=\begin{psmallmatrix} 1 & \hphantom{-} 0\\ 0 & -1\end{psmallmatrix}$: Pauli matrices.
\item $\sigma^+=\dfrac{1}{2} (\sigma^x+{\rm i}\sigma^y )=\begin{psmallmatrix} 0 & 1\\ 0 & 0\end{psmallmatrix}$, $\sigma^-=\dfrac{1}{2} (\sigma^x-{\rm i}\sigma^y )=\begin{psmallmatrix} 0 & 0\\ 1 & 0\end{psmallmatrix}$: raising and lowering matrices.
\item $\Hs:=(\CC^2)^{\otimes N}$: Hilbert space of the open XXZ spin chain of length $N$.
\item $\Hnd$: open XXZ Hamiltonian with non-diagonal boundary terms.
\item $\Hnd^{(M)}$: $\Hnd$ under the Nepomechie constraint~\eqref{nepcond} for $M\in\ZZ$.
\item $\TL$: Temperley--Lieb (TL) algebra on $N$ sites with loop weight $\delta$.
\item $e_i$, $1\leq i\leq N-1$: generators of the Temperley--Lieb algebra or their spin-chain representatives.
\item $\Blob$: blob algebra on $N$ sites with loop weight $\delta$ and blob weight $y$.
\item $b$, $\bar{b}:=1-b$: blob/anti-blob generator or its spin-chain representative.
\item $\twoBlob$: two-boundary Temperley--Lieb algebra on $N$ sites with loop weight $\delta$, left/right blob weights $y_{l/r}$ and two-blob weight $Y$.
\item $b_{l/r}$, $\bar{b}_{l/r}:=1-b_{l/r}$: left/right blob/anti-blob generators ($b_l:=b$) or their spin-chain representatives.
\item $\WW_j^{bb}$, $\WW_j^{b\bar{b}}$, $\WW_j^{\bar{b}b}$, $\WW_j^{\bar{b}\bar{b}}$, $1\leq j\leq N/2$, and $\WW_0$: standard $\twoBlob$-modules.
\item $\mathbf{H}:=-\mu_lb_l-\mu_rb_r-\sum_{i=1}^{N-1} e_i\in\twoBlob$: abstract two-boundary Temperley--Lieb Hamiltonian.
\item $\utwoBlob$: universal two-boundary Temperley--Lieb algebra on $N$ sites with loop weight $\delta$ and left/right blob weights $y_{l/r}$ and central element $\YYY$.
\item $\qa={\rm e}^h$: deformation parameter of the XXZ spin chain.
\item $[x]_\qa:=\frac{\qa^x-\qa^{-x}}{\qa-\qam}$: $\qa$-deformed numbers.
\item $\{x\}:=\qa^x-\qa^{-x}$.
\item $\Uq$: quantum group, a $\qa$-deformation of $\mathsf{SU}(2)$.
\item $\EE$, $\FF$, $\KK$, $\KK^{-1}$: generators of $\Uq$.
\item $\CCC$: Casimir element of $\Uq$.
\item $\VV_\alpha$: infinite-dimensional Verma module of $\Uq$ of highest-weight $\qa^{\alpha-1}$.
\item $H_{{\rm XXZ}}$: $\Uq$-invariant open XXZ Hamiltonian.
\item $\Hb:=\VV_\alpha\otimes(\CC^2)^{\otimes N}$: Hilbert space of the one-boundary spin chain of length $N$.
\item $H_b:=-\mu b+H_{{\rm XXZ}}$: $\Uq$-invariant one-boundary Hamiltonian with coupling $\mu$.
\item $\Hbb:=\VV_{\alpha_l}\otimes(\CC^2)^{\otimes N}\otimes\VV_{\alpha_r}$: Hilbert space of the two-boundary spin chain of length $N$.
\item $H_{2b}:=-\mu_l b_l+H_{{\rm XXZ}}-\mu_rb_r$: $\Uq$-invariant two-boundary Hamiltonian with couplings $\mu_{l}$,~$\mu_{r}$.
\item $\Hs_M$, $M\geq 0$: subspaces of highest-weight vectors of weight $\qa^{\alpha_l+\alpha_r+N-2M-2}$ of $\Hbb$.
\end{itemize}

\section{Algebraic setting}
\label{algset}

In this section, we present the necessary algebraic tools:
\begin{itemize}\itemsep=0pt
\item the relevant lattice algebras, namely the Temperley--Lieb (TL) algebra, the Blob algebra and the two-boundary Temperley--Lieb algebra and their representations (Section~\ref{latticealg}),
\item the $\Uq$ quantum group and its representations (Section~\ref{uqgroup}),
\item $\Uq$-invariant spin chains with Hilbert spaces $\Hs$, $\Hb$, $\Hbb$ and corresponding Hamiltonians~$\Hxxz$, $H_b$, $H_{2b}$ (Section~\ref{swsec}),
\item Schur--Weyl duality between the $\Uq$ and lattice algebra actions on these spin chains (Section~\ref{swsec}).
\end{itemize}
Most of the formalism and results were introduced in~\cite{previouspaper} and we will often refer to this paper for additional details. To simplify the exposition we will always assume that $N$ is strictly positive and even, but our construction can be extended to the odd $N$ case too.

\subsection{Lattice algebras}
\label{latticealg}

The TL algebra~\cite{TLalg}, denoted $\TL$, is defined by generators $(e_i)_{1\leq i\leq N-1}$ and relations
\begin{equation}
\label{TLrel}
e_i^2=\delta e_i ,\qquad e_ie_{i\pm1}e_i=e_i ,\qquad [e_i,e_j]=0\quad\forall |i-j|\geq 2 ,
\end{equation}
with $\delta\in\CC$ some parameter. If we set
\begin{equation*}
\begin{tikzpicture}[scale=0.8]
\draw (0.2,1) node {$e_i=$};
\draw (1,0) [dnup=0];
\draw (3,0) [dnup=0];
\draw (2,1) node {\dots };
\draw (4,2) [upup=1];
\draw (4,0) [dndn=1];
\draw (7,1) node {\dots };
\draw (6,0) [dnup=0];
\draw (8,0) [dnup=0];
\draw (4,-0.5) node {$i$};
\draw (5,-0.5) node {$i+1$};
\end{tikzpicture}
\end{equation*}
these relations are neatly expressed by the graphical rules{\samepage
\begin{align*}
&\begin{tikzpicture}[scale=0.5]
\draw (-1.5,2) node {$e_i^2=$};
\draw (1,0) [dnup=0];
\draw (0,1) node {\dots };
\draw (2,2) [upup=1];
\draw (2,0) [dndn=1];
\draw (5,1) node {\dots };
\draw (4,0) [dnup=0];
\draw (1,2) [dnup=0];
\draw (0,3) node {\dots };
\draw (2,4) [upup=1];
\draw (2,2) [dndn=1];
\draw (5,3) node {\dots };
\draw (4,2) [dnup=0];
\draw (6,2) node {=};
\draw (7,2) node {$\delta$};
\draw (9,1) [dnup=0];
\draw (8,2) node {\dots };
\draw (10,3) [upup=1];
\draw (10,1) [dndn=1];
\draw (13,2) node {\dots };
\draw (12,1) [dnup=0];
\end{tikzpicture}
\\[-1ex]
&
\begin{tikzpicture}[scale=0.5]
\draw (-2.5,3) node {$e_ie_{i+1}e_i=$};
\draw (0,1) node {\dots };
\draw (1,0) [dnup=0];
\draw (2,2) [upup=1];
\draw (2,0) [dndn=1];
\draw (4,0) [dnup=0];
\draw (5,0) [dnup=0];
\draw (6,1) node {\dots };
\draw (0,3) node {\dots };
\draw (1,2) [dnup=0];
\draw (2,2) [dnup=0];
\draw (3,4) [upup=1];
\draw (3,2) [dndn=1];
\draw (5,2) [dnup=0];
\draw (6,3) node {\dots };
\draw (0,5) node {\dots };
\draw (1,4) [dnup=0];
\draw (2,6) [upup=1];
\draw (2,4) [dndn=1];
\draw (4,4) [dnup=0];
\draw (5,4) [dnup=0];
\draw (6,5) node {\dots };
\draw (7,3) node {=};
\draw (8,3) node {\dots };
\draw (9,2) [dnup=0];
\draw (10,4) [upup=1];
\draw (10,2) [dndn=1];
\draw (12,2) [dnup=0];
\draw (13,2) [dnup=0];
\draw (14,3) node {\dots };
\end{tikzpicture}
\end{align*}
The parameter $\delta$ is then interpreted as the weight of a closed loop.}

The TL can be shown to be finite-dimensional and all its irreducible representations, called standard modules, have been classified. They are indexed by an integer $0\leq j\leq N/2$ interpreted as half the number of through lines propagating in a TL diagram. Concretely, the standard module $\WW_j$ has a basis of so-called link states which are half-diagrams containing exactly $2j$ through lines. For example, for $N=4$, these are given by
\begin{align*}
&\WW_0 =
\begin{tikzpicture}[scale=0.4,baseline=-2pt]
\draw (1.2,0) node {$\CC\langle$};
\draw (2,0.4) [upup=1];
\draw (4,0.4) [upup=1];
\draw (5.5,0) node {,};
\draw (6,0.4) [hupup=3];
\draw (7,0.4) [upup=1];
\draw (9.5,0) node {$\rangle$};
\end{tikzpicture}\!\!,
\\
&\WW_1 =
\begin{tikzpicture}[scale=0.4,baseline=-2pt]
\draw (1.2,0) node {$\CC\langle$};
\draw (2,0.4) [upup=1];
\draw (4,0.4) [hdnup=0];
\draw (5,0.4) [hdnup=0];
\draw (5.5,0) node {,};
\draw (6,0.4) [hdnup=0];
\draw (9,0.4) [hdnup=0];
\draw (7,0.4) [upup=1];
\draw (9.5,0) node {,};
\draw (11,0.4) [hdnup=0];
\draw (10,0.4) [hdnup=0];
\draw (12,0.4) [upup=1];
\draw (13.5,0) node {$\rangle$};
\end{tikzpicture}\!\!,
\\
&\WW_2=
\begin{tikzpicture}[scale=0.4,baseline=-2pt]
\draw (1.2,0) node {$\CC\langle$};
\draw (2,0.4) [hdnup=0];
\draw (3,0.4) [hdnup=0];
\draw (4,0.4) [hdnup=0];
\draw (5,0.4) [hdnup=0];
\draw (5.5,0) node {$\rangle$};
\end{tikzpicture}\!\!.
\end{align*}
A TL diagram then acts on these link states by propagating them with the diagrammatical rules of the TL algebra, with the additional condition that if two through lines are contracted then the diagram acts by $0$. For example,
\begin{align*}
&\begin{tikzpicture}[scale=0.5,baseline=-3.5pt]
\draw (0,0) node {$e_2$};
\draw (1,0.4) [hupup=3];
\draw (2,0.4) [upup=1];
\end{tikzpicture}
=
\begin{tikzpicture}[scale=0.5,baseline=-3.5pt]
\draw (6,0) [hupup=3];
\draw (7,0) [upup=1];
\draw (6,0) [dnup=0];
\draw (7,0) [dndn=1];
\draw (9,0) [dnup=0];
\draw (7,2) [upup=1];
\draw (10,0) node {$=~\delta$};
\draw (11,0.4) [hupup=3];
\draw (12,0.4) [upup=1];
\end{tikzpicture}\!\!,
\\
&\begin{tikzpicture}[scale=0.5,baseline=-2.0pt]
\draw (0,0) node {$e_2$};
\draw (1,0.4) [upup=1];
\draw (3,0.4) [upup=1];
\end{tikzpicture}
=
\begin{tikzpicture}[scale=0.5,baseline=-2.0pt]
\draw (6,0) [upup=1];
\draw (8,0) [upup=1];
\draw (6,0) [dnup=0];
\draw (7,0) [dndn=1];
\draw (9,0) [dnup=0];
\draw (7,2) [upup=1];
\draw (10,0) node {$=$};
\draw (11,0.4) [hupup=3];
\draw (12,0.4) [upup=1];
\end{tikzpicture}\!\!,
\\
&\begin{tikzpicture}[scale=0.5,baseline=-2.0pt]
\draw (0,0) node {$e_2$};
\draw (1,0.4) [upup=1];
\draw (3,0.4) [hdnup=0];
\draw (4,0.4) [hdnup=0];
\end{tikzpicture}
=
\begin{tikzpicture}[scale=0.5,baseline=-2.0pt]
\draw (6,0) [upup=1];
\draw (8,0) [hdnup=0];
\draw (9,0) [hdnup=0];
\draw (6,0) [dnup=0];
\draw (7,0) [dndn=1];
\draw (9,0) [dnup=0];
\draw (7,2) [upup=1];
\draw (10,0) node {$=$};
\draw (11,0.4) [hdnup=0];
\draw (14,0.4) [hdnup=0];
\draw (12,0.4) [upup=1];
\end{tikzpicture}\!\!,
\\
&\begin{tikzpicture}[scale=0.5,baseline=-3.0pt]
\draw (0,0) node {$e_3$};
\draw (1,0.4) [upup=1];
\draw (3,0.4) [hdnup=0];
\draw (4,0.4) [hdnup=0];
\end{tikzpicture}
=
\begin{tikzpicture}[scale=0.5,baseline=-3.0pt]
\draw (6,0) [upup=1];
\draw (8,0) [hdnup=0];
\draw (9,0) [hdnup=0];
\draw (6,0) [dnup=0];
\draw (8,0) [dndn=1];
\draw (7,0) [dnup=0];
\draw (8,2) [upup=1];
\draw (10,0) node {$= 0$};
\end{tikzpicture}\!\!.
\end{align*}
By some standard combinatorial arguments one can show that
\begin{equation}
\label{tlmoddim}
\dim\WW_j={N \choose \frac{N}{2}-j}-{N \choose \frac{N}{2}-j-1} .
\end{equation}

Let us now define some boundary extensions of the TL algebra. The simplest one is the blob algebra introduced in~\cite{Martin:1993jka} and denoted $\Blob$. It has an additional generator $b$ called the blob satisfying
\begin{equation}
\label{eq:blob-rel}
b^2=b ,\qquad e_1 b e_1 = y e_1 , \qquad [b,e_i]=0 \quad \text{for} \quad 2\leq i\leq N-1,
\end{equation}
with $y\in\CC$ some parameter. Graphically, $b$ is represented by
\begin{equation*}
\begin{tikzpicture}[scale=0.8]
\draw (0,1) node {$b=$};
\draw (1,0) [dnup=0];
\draw (1,1) node {$\bbullet$};
\draw (2,0) [dnup=0];
\draw (3,1) node {\dots };
\draw (4,0) [dnup=0];
\draw (5,0) [dnup=0];
\end{tikzpicture}
\end{equation*}
and the rules~\eqref{eq:blob-rel} mean that
\begin{equation}
\label{blobrulediag}\begin{split}
&\begin{tikzpicture}[scale=0.8]
\draw (0,0) [dnup=0];
\draw (0,0.75) node {$\bbullet$};
\draw (0,1.25) node {$\bbullet$};
\draw (1,1) node {$=$};
\draw (2,0) [dnup=0];
\draw (2,1) node {$\bbullet$};
\draw (5,-1) [dndn=1];
\draw (5,1) [upup=1];
\draw (5,1) [dndn=1];
\draw (5,3) [upup=1];
\draw (5,1) node {$\bbullet$};
\draw (7,1) node {$=~y$};
\draw (8,2) [upup=1];
\draw (8,0) [dndn=1];
\end{tikzpicture}
\end{split}
\end{equation}
The parameter $y$ is then interpreted as the weight of a closed loop carrying a blob. One often introduces the anti-blob $\bar{b}=1-b$, represented by
\begin{equation*}
\begin{tikzpicture}[scale=0.8]
\draw (0.2,1) node {$\bar{b}=$};
\draw (1,0) [dnup=0];
\draw (1,1) node {$\ccirc$};
\draw (2,0) [dnup=0];
\draw (3,1) node {\dots };
\draw (4,0) [dnup=0];
\draw (5,0) [dnup=0];
\end{tikzpicture}
\end{equation*}
which also satisfies relations~\eqref{eq:blob-rel} but with the blob weight $y$ replaced by $\delta-y$. Moreover, $b\bar{b}=\bar{b}b=0$ so diagrammatically
\begin{equation}
\label{antiblobrulediag}
\begin{split}
&\begin{tikzpicture}[scale=0.8]
\draw (0,0) [dnup=0];
\draw (0,0.75) node {$\bbullet$};
\draw (0,1.25) node {$\ccirc$};
\draw (1,1) node {$=$};
\draw (2,0) [dnup=0];
\draw (2,1.25) node {$\bbullet$};
\draw (2,0.75) node {$\ccirc$};
\draw (3,1) node {$=0$};
\end{tikzpicture}
\end{split}
\end{equation}
which justifies the anti-blob terminology.

The blob algebra is also finite-dimensional and the classification of its standard modules is very similar to the analogous construction in the TL algebra. They are also indexed by $0\leq j\leq N/2$ and constructed using link states with $2j$ through lines but now we also have to decorate all the cups and through lines which can touch the left boundary by blobs and anti-blobs. Since only the leftmost through line can touch it, we have two types of standard modules:~$\WW_j^b$ where the leftmost through line carries a blob and $\WW_j^{\bar{b}}$ where it carries an anti-blob. For~$j=0$, there are no through lines so we just have $\WW_0$.\footnote{We will often slightly abuse notation and denote $\WW_0$ the module with no through lines irrespectively of the lattice algebra we consider.} For example,
\begin{align*}
\begin{split}
&\begin{tikzpicture}[scale=0.4]
\draw (0,0) node {$\WW_1^b=\CC\langle$};
\draw (2,0.4) [upup=1];
\draw (2.5,-0.3) node {$\bullet$};
\draw (4,0.4) [hdnup=0];
\draw (4,0) node {$\bullet$};
\draw (5,0.4) [hdnup=0];
\draw (5.5,0) node {,};
\draw (6,0.4) [upup=1];
\draw (6.5,-0.3) node {$\circ$};
\draw (8,0.4) [hdnup=0];
\draw (8,0) node {$\bullet$};
\draw (9,0.4) [hdnup=0];
\draw (9.5,0) node {,};
\draw (10,0.4) [hdnup=0];
\draw (10,0) node {$\bullet$};
\draw (13,0.4) [hdnup=0];
\draw (11,0.4) [upup=1];
\draw (13.5,0) node {,};
\draw (14,0.4) [hdnup=0];
\draw (14,0) node {$\bullet$};
\draw (15,0.4) [hdnup=0];
\draw (16,0.4) [upup=1];
\draw (17.5,0) node {$\rangle$,};
\end{tikzpicture}
\\
&\begin{tikzpicture}[scale=0.4]
\draw (0,0) node {$\WW_1^{\bar{b}}=\CC\langle$};
\draw (2,0.4) [upup=1];
\draw (2.5,-0.3) node {$\circ$};
\draw (4,0.4) [hdnup=0];
\draw (4,0) node {$\circ$};
\draw (5,0.4) [hdnup=0];
\draw (5.5,0) node {,};
\draw (6,0.4) [upup=1];
\draw (6.5,-0.3) node {$\bullet$};
\draw (8,0.4) [hdnup=0];
\draw (8,0) node {$\circ$};
\draw (9,0.4) [hdnup=0];
\draw (9.5,0) node {,};
\draw (10,0.4) [hdnup=0];
\draw (10,0) node {$\circ$};
\draw (13,0.4) [hdnup=0];
\draw (11,0.4) [upup=1];
\draw (13.5,0) node {,};
\draw (14,0.4) [hdnup=0];
\draw (14,0) node {$\circ$};
\draw (15,0.4) [hdnup=0];
\draw (16,0.4) [upup=1];
\draw (17.5,0) node {$\rangle$.};
\end{tikzpicture}
\end{split}
\end{align*}
The action of the blob algebra on $\WW_j^b$, $\WW_j^{\bar{b}}$, with $1\leq j\leq N/2$, and $\WW_0$ is defined in the same way as for the TL algebra with the additional diagrammatical rules~\eqref{blobrulediag} and~\eqref{antiblobrulediag}. Moreover, one can show~\cite{JScombi,Martin:1993jka} that
\begin{equation}
\label{dimblob}
\dim\WW_j^b=\dim\WW_j^{\bar{b}}={N\choose N/2+j} ,\qquad\dim\WW_0={N\choose N/2} .
\end{equation}

We can further extend the blob algebra by working with two blobs, one on the left, denoted~$b_l:=b$, with weight $y_l:=y$ and one on the right, denoted $b_r$ and represented by
\begin{equation*}
\begin{tikzpicture}[scale=0.8]
\draw (0.2,1) node {$b_r=$};
\draw (1,0) [dnup=0];
\draw (5,1) node {$\blacksquare$};
\draw (2,0) [dnup=0];
\draw (3,1) node {\dots };
\draw (4,0) [dnup=0];
\draw (5,0) [dnup=0];
\end{tikzpicture}
\end{equation*}
and satisfying an analogue of~\eqref{eq:blob-rel} and \eqref{blobrulediag} but on the right
\begin{gather}
b_r^2=b_r ,\qquad e_{N-1} b_r e_{N-1} = y_r e_{N-1} , \nonumber\\
[b_l,b_r]=0 , \qquad [b_r,e_i]=0 \quad \text{for} \quad 1\leq i\leq N-2 ,\label{blobruler}
\end{gather}
with a weight $y_r$ for a loop carrying the right blob $\blacksquare$. One of course also has a right anti-blob $\bar{b}_r=1-b_r$ represented by $\square$ and satisfying~\eqref{blobruler} with weight $\delta-y_r$ together with $b_r\bar{b}_r=\bar{b}_rb_r=0$. This is not sufficient from a diagrammatical point of view however as we also need to assign some weight $Y\in\CC$ to a closed loop carrying both the left and the right blob. Formally, this non-local relation is given by
\begin{equation}
\label{2bloop}
\Bigg(\prod_{i=1}^{N/2}e_{2i-1}\Bigg)b_l\Bigg(\prod_{i=1}^{N/2-1}e_{2i}\Bigg)b_r\Bigg(\prod_{i=1}^{N/2}e_{2i-1}\Bigg)=Y\prod_{i=1}^{N/2}e_{2i-1} .
\end{equation}
The generators $(e_i)_{1\leq i\leq N-1}$ and $b_{l/r}$ with relations~\eqref{TLrel}, \eqref{eq:blob-rel}, \eqref{blobruler} and \eqref{2bloop}, then define a~fi\-nite-dimensional algebra called the two-boundary Temperley--Lieb algebra denoted $\twoBlob$.

The classification of standard modules of the two-boundary TL algebra is a natural generalisation of the blob algebra case~\cite{de_Gier_2009, dubail:tel-00555624}. Namely, for $1\leq j\leq N$ we have four types of modules, namely $\WW_j^{bb}$, $\WW_j^{\bar{b}b}$, $\WW_j^{b\bar{b}}$ and $\WW_j^{\bar{b}\bar{b}}$, with $2j$ through lines depending on whether the leftmost/rightmost line carries a blob/anti-blob, as well as a single module $\WW_0$ with no through lines, all of these being constructed from link states decorated by left/right blob/anti-blob in all allowed ways. The action of $\twoBlob$ on these representations is again given by the defining diagrammatical rules of the algebra. One can show~\cite{de_Gier_2009,dubail:tel-00555624} that
\begin{equation}
\label{dim2blob}
\dim\WW_j^{bb}=\dim\WW_j^{\bar{b}b}=\dim\WW_j^{b\bar{b}}=\dim\WW_j^{\bar{b}\bar{b}}=\sum_{k=j}^{N/2} {N\choose N/2+k} ,\qquad\dim\WW_0=2^N .
\end{equation}
$\WW_0$ is called the vacuum module. It is generically irreducible but can become reducible but indecomposable for certain values of the parameters~\cite{de_Gier_2009} which will be important for us later on (see Section~\ref{swsec} and Appendix~\ref{proof}). It is also worth mentioning that since a closed loop touching both boundaries can be formed only if there are no through lines, $\WW_0$ is the only standard module which actually depends on the value of $Y$.

The main interest of this whole formalism for the open XXZ spin chain with non-diagonal boundary terms is the following. Let us set
\begin{gather}
e_i =-\frac{1}{2}\big(\sigma^x_{i}\sigma^x_{i+1}+\sigma^y_{i}\sigma^y_{i+1}+\cosh(h)\big(\sigma^z_{i}\sigma^z_{i+1}-1\big)+\sinh(h)\big(\sigma_{i+1}^z-\sigma_i^z\big)\big) , \nonumber\\
b_{l} =\frac{1}{2\sinh(h\alpha_{l})}\big({\rm i}{\rm e}^{h\theta_l}\sigma_1^++ {\rm i}{\rm e}^{-h\theta_l}\sigma_1^-+\cosh(h\alpha_l)\sigma_1^z\big)+\frac{1}{2} , \nonumber\\
b_{r} =-\frac{1}{2\sinh(h\alpha_{r})}\big( {\rm i}{\rm e}^{h\theta_r}\sigma_N^++{\rm i}{\rm e}^{-h\theta_r}\sigma_N^-+\cosh(h\alpha_r)\sigma_N^z\big)+\frac{1}{2} , \nonumber\\
\mu_{l/r} =\frac{\sinh(h)\sinh\big(h\alpha_{l/r}\big)}{\sinh(\zeta_{l/r}-\frac{h\alpha_{l/r}}{2})\sinh(\zeta_{l/r}+\frac{h\alpha_{l/r}}{2})}\label{newpar}
\end{gather}
with some new parameters $\alpha_{l/r}$ and $\zeta_{l/r}$. Then, up to an irrelevant additive constant, we have
\begin{equation*}
H_{{\rm n.d.}}=-\mu_lb_l-\mu_rb_r-\sum_{i=1}^{N-1} e_i
\end{equation*}
with\footnote{Other choices are possible but this will not affect the end result.}
\begin{alignat}{3}
&h\delta_l=\frac{h\alpha_l}{2}-\zeta_l ,\qquad &&h\kappa_l=\frac{h\alpha_l}{2}+\zeta_l+\frac{{\rm i}\pi}{2} ,& \nonumber\\
&h\delta_r=\frac{h\alpha_r}{2}-\zeta_r ,\qquad &&h\kappa_r=\frac{h\alpha_r}{2}+\zeta_r-\frac{{\rm i}\pi}{2} ,&
\label{parameters}
\end{alignat}
and the following result holds.
\begin{Theorem}[J.\ de Gier, A.\ Nichols~\cite{de_Gier_2009}]
\label{dgthm}
The $e_i$, $1\leq i\leq N-1$, and $b_{l/r}$ from~\eqref{newpar} satisfy the relations of the two-boundary TL algebra with weights
\begin{gather}
\delta=2\cosh(h) ,\qquad y_{l/r}=\frac{\sinh\big(h(\alpha_{l/r}+1)\big)}{\sinh\big(h\alpha_{l/r}\big)} , \nonumber\\ Y=\dfrac{\sinh(h\frac{\alpha_l+\alpha_r+1\pm\Theta}{2})\sinh(h\frac{\alpha_l+\alpha_r+1\mp\Theta}{2})}{\sinh(h\alpha_{l})\sinh(h\alpha_{r})},\label{weights1}
\end{gather}
where $\Theta:=\theta_l-\theta_r$~\eqref{Theta}, and thus define a $2^N$-dimensional representation of $\twoBlob$ on $(\CC^2)^{\otimes N}$. Moreover, this representation is isomorphic to the vacuum module $\WW_0$.
\end{Theorem}

Using this theorem, we can identify $(\CC^2)^{\otimes N}$ with $\WW_0$ and interpret $H_{{\rm n.d.}}$ as an abstract element of $\twoBlob$
\begin{equation}
\label{Hbf}
\mathbf{H}:=-\mu_lb_l-\mu_rb_r-\sum_{i=1}^{N-1} e_i
\end{equation}
evaluated in the vacuum representation $\WW_0$, that is $H_{{\rm n.d.}}=\rho_{\WW_0}(\mathbf{H})$ where $\rho_{\WW_0} \colon \twoBlob\to\End_{\CC}(\WW_0)$ is the representation map of $\WW_0$. This will be an essential ingredient of our construction. In what follows we will always tacitly make the identification $(\CC^2)^{\otimes N}\cong\WW_0$. To the best of our knowledge, there is no simple way to construct this isomorphism explicitly.

Note also that the weights~\eqref{weights1} do not depend on the sign of $\Theta$, so the spectrum $H_{{\rm n.d.}}$ is invariant under the transformation $\Theta\leftrightarrow-\Theta$. In particular, the $+\Theta$ and $-\Theta$ choices in the Nepomechie condition~\eqref{nepcond} are equivalent. When convenient, we will write $\pm\Theta$ instead of $\Theta$.

\subsection[The U\_q sl\_2 quantum group]{The $\boldsymbol{\Uq}$ quantum group}
\label{uqgroup}

Let us now introduce the second main ingredient: the $\Uq$ quantum group.

The algebra $\Uq$~\cite{Drinfeld:1985rx, Jimbo:1985zk} (see also~\cite[Chapter~6.4]{qgroups} and~\cite[Chapters~VI and VII]{kassel}) is defined by gener\-a\-tors~$\EE$,~$\FF$,~$\KK$ and $\KK^{-1}$ and relations
\begin{equation}
\label{uqdefrel}
\KK\EE\KK^{-1}=\qa^2\EE ,\qquad \KK\FF\KK^{-1}=\qa^{-2}\FF ,\qquad [\EE,\FF]=\frac{\KK-\KK^{-1}}{\qa-\qam} ,\qquad \KK\KK^{-1}=\KK^{-1}\KK=\mathsf{1} .
\end{equation}
It is a $\qa$-deformation of the universal enveloping algebra of the Lie algebra $\mathfrak{sl}_2$, in the sense that we recover the commutation relations of the $\mathfrak{sl}_2$ triple $(\EE,\FF,\HH)$ in the limit $\qa\to 1$ with $\KK=\qa^\HH$. It is important for defining the action on tensor products of representations that this algebra admits the coproduct
\begin{equation}
\label{coproduct}
\Delta(\EE)=\mathsf{1}\otimes\EE+\EE\otimes\KK ,\qquad \Delta(\FF)=\KK^{-1}\otimes\FF+\FF\otimes\mathsf{1} ,\qquad \Delta\big(\KK^{\pm 1}\big)=\KK^{\pm 1}\otimes\KK^{\pm 1}\ .
\end{equation}

As $\mathfrak{sl}_2$, $\Uq$ admits ($2j+1$)-dimensional spin-$j$ representations for all $j\in\frac{1}{2}\NN$. For our purposes we will need the fundamental spin-$\frac{1}{2}$ representation $\CC^2$, where the action of the generators is given by
\begin{equation}
\label{funduq}
\EE_{\CC^2}=\sigma^+ ,\qquad\FF_{\CC^2}=\sigma^- ,\qquad \KK_{\CC^2}^{\pm 1}=\qa^{\pm\sigma^z} .
\end{equation}
Let us also introduce the Verma modules $\VV_\alpha$~\cite[Chapter~VI.3]{kassel} that we shall need to define our modified boundary conditions. For all $\alpha\in\CC$, they are given in a basis $\VV_\alpha:=\bigoplus_{0\leq n}\CC\ket{n}$ by
\begin{gather}
 \EE_{\VV_\alpha}\ket{n}=[n]_\qa[\alpha-n]_\qa\ket{n-1} , \qquad
 \FF_{\VV_\alpha}\ket{n}=\ket{n+1} , \qquad
 \KK_{\VV_\alpha}^{\pm 1}\ket{n}=\qa^{\pm(\alpha-1-2n)}\ket{n}
\label{alpharepgen}
\end{gather}
for all $n\geq 0$, with $\ket{-1}=0$, and where
\begin{equation*}
[x]_\qa:=\frac{\qa^x-\qa^{-x}}{\qa-\qam}=\frac{\{x\}}{\{1\}} ,\qquad \{x\}:=\qa^x-\qa^{-x} .
\end{equation*}
The basis vectors $\ket{n}$ diagonalise $\KK$ and their $\KK$-eigenvalue $\qa^{\alpha-1-2n}$ is called the weight. The vector $\ket{0}$ is annihilated by the raising operator $\EE$ and is thus called the highest-weight vector. Note that its weight is $\qa^{\alpha-1}$ with the $-1$ shift of $\alpha$ introduced for later convenience. When $\qa$ is not a root of unity, $\VV_\alpha$ is irreducible if and only if $\qa^\alpha\neq\pm\qa^n$ for all $n\in\NN^{*}$.\footnote{If $\qa^\alpha\in\pm\qa^{\NN^*}$, $\VV_\alpha$ contains a unique non-trivial stable subspace but is indecomposable.} If that is the case, $\VV_\alpha$ is also unique, meaning that any $\Uq$-module generated from a highest-weight vector of weight $\qa^{\alpha-1}$ is isomorphic to $\VV_\alpha$. Finally, for all $\alpha\in\CC$ such that $\qa^\alpha\neq\pm 1$ we have the fusion rule\footnote{If $\qa^\alpha=\pm 1$, the two factors $\VV_{\alpha+1}$ and $\VV_{\alpha-1}$ get ``glued" into a single indecomposable representation known as a tilting module.}
\begin{equation}
\label{fusrule}
\VV_\alpha\otimes \CC^2 \cong \VV_{\alpha+1}\oplus \VV_{\alpha-1} .
\end{equation}

The above definitions have to be slightly adapted if $\qa$ is a root of unity. This case was thoroughly treated in~\cite{previouspaper} and presents no major complications. To keep the exposition simple, from now on we will always assume $\qa$ to be generic (not a root of unity) unless otherwise stated. We refer to~\cite{previouspaper} for further details.

\subsection[U\_q sl\_2-invariant spin chains and Schur--Weyl duality]{$\boldsymbol{\Uq}$-invariant spin chains and Schur--Weyl duality}
\label{swsec}

\subsubsection{The bulk spin chain}

Applying the coproduct~\eqref{coproduct} $N-1$ times (recall that the coproduct is coassociative, and so the result does not depend on the order of its application) to the spin-$\frac{1}{2}$ representation~\eqref{funduq} we obtain a well-defined action of $\Uq$ on $\Hs:=(\CC^2)^{\otimes N}$. Now if we set
\begin{gather}
e_i =-\frac{1}{2}\bigg(\sigma^x_{i}\sigma^x_{i+1}+\sigma^y_{i}\sigma^y_{i+1}+\frac{\qa+\qa^{-1}}{2}\big(\sigma^z_{i}\sigma^z_{i+1}-1\big)\bigg)-\frac{\qa-\qa^{-1}}{4}\big(\sigma_{i+1}^z-\sigma_{i}^z\big)
 \nonumber\\
\hphantom{e_i}{} =
\begin{pmatrix}
0 & \hphantom{-}0 & \hphantom{-}0 & 0\\
0 & \hphantom{-}\qa & -1 & 0\\
0 & -1 & \hphantom{-}\qam & 0\\
0 & \hphantom{-}0 & \hphantom{-}0 & 0
\end{pmatrix}\label{eiexp}
\end{gather}
as in~\eqref{newpar}, with $\qa={\rm e}^h$, it turns out that the $e_i$ commute with this $\Uq$ action~\cite{PASQUIER1990523}. This implies that the Hamiltonian
\begin{align}
H_{{\rm XXZ}} &=\frac{1}{2}\sum_{i=1}^{N-1} \bigg(\sigma^x_{i}\sigma^x_{i+1}+\sigma^y_{i}\sigma^y_{i+1}+\frac{\qa+\qa^{-1}}{2}\sigma^z_{i}\sigma^z_{i+1}\bigg)+\frac{\qa-\qa^{-1}}{4}\big(\sigma_N^z-\sigma_1^z\big) \nonumber\\
& =\frac{\qa+\qa^{-1}}{2}(N-1)-\sum_{i=1}^{N-1}e_i
\label{XXZham}
\end{align}
(which is just the first line of $\Hnd$ from~\eqref{hnd}, i.e., without the two boundary terms) is $\Uq$-invariant. More generally, the whole $\TL$-action on $\Hs$ generated by the $e_i$ commutes with~$\Uq$. Actually, not only they commute, but they are even mutual maximal centralisers and $\Hs$ decomposes as a $(\TL,\Uq)$-bimodule~\cite{goodman, Jimbo1986AQO, Martin1992ONSD, martin1992commutants}
\begin{equation}
\label{swtl}
\Hs=\bigoplus_{j=0}^{N/2}\WW_j\otimes\CC^{2j+1} ,
\end{equation}
where $\WW_j$ are the standard $\TL$-modules introduced above and $\CC^{2j+1}$ are spin-$j$ representations of $\Uq$. This result is known as (quantum) Schur--Weyl duality. This result is essential, as it reduces the study of $\Hxxz$ on $\Hs:=(\CC^2)^{\otimes N}$ to its restriction on standard $\TL$-modules.

Unfortunately, we cannot use this method for $\Hnd$ because the boundary terms~\eqref{newpar}
\begin{gather*}
b_{l} =\frac{1}{\{\alpha_{l}\}}\left({\rm i}\qa^{\theta_l}\sigma_1^++ {\rm i}\qa^{-\theta_l}\sigma_1^-+\frac{\qa^{\alpha_l}+\qa^{-\alpha_l}}{2}\sigma_1^z\right)+\frac{1}{2}=\frac{1}{\{\alpha_{l}\}}\begin{pmatrix}
\qa^{\alpha_l} & {\rm i}\qa^{\theta_l}\\
{\rm i}\qa^{-\theta_l} & -\qa^{-\alpha_l}
\end{pmatrix} ,\\
b_{r} =-\frac{1}{\{\alpha_{r}\}}\left({\rm i}\qa^{\theta_r}\sigma_N^++ {\rm i}\qa^{-\theta_r}\sigma_N^-+\frac{\qa^{\alpha_r}+\qa^{-\alpha_r}}{2}\sigma_N^z\right)+\frac{1}{2}=\frac{1}{\{\alpha_{r}\}}\begin{pmatrix}
-\qa^{-\alpha_r} & -{\rm i}\qa^{\theta_r}\\
-{\rm i}\qa^{-\theta_r} & \hphantom{-}\qa^{\alpha_r}
\end{pmatrix}
\end{gather*}
break the $\Uq$ symmetry. This is why we are going to build a different Hamiltonian~\eqref{twob} which does preserve the quantum group symmetry and then show that it can be related back to~$\Hnd$ through the two-boundary TL algebra.

\subsubsection{The one-boundary system}

Following~\cite{previouspaper}, let us first introduce the one-boundary Hamiltonian. It is constructed by tensoring the usual spin chain $\Hs:=(\CC^2)^{\otimes N}$ with the Verma module $\VV_\alpha$ and adding a new $\Uq$-invariant boundary term acting on the two leftmost sites, $\VV_\alpha\otimes\CC^2$, of the new Hilbert space
\begin{equation*}
\Hb := \VV_\alpha\otimes(\CC^2)^{\otimes N} .
\end{equation*}
Because of the fusion rule~\eqref{fusrule} the most general such term can only be a linear combination of projectors $b_\pm$ on the direct summands $\VV_{\alpha\pm1}$, which are given by
\begin{equation}
\label{blobqgen}
b_\pm =\frac{\pm 1}{\{\alpha\}}
\begin{pmatrix}
 -\qam \KK^{-1}+\qa^{\pm \alpha} & \{1\}\FF\\
 \qa\{1\}\KK^{-1}\EE & \qa\KK^{-1}-\qa^{\mp \alpha}
\end{pmatrix} .
\end{equation}
In the expression above, $b_\pm$ are operators acting on $\VV_\alpha\otimes\CC^2$ which we have written as $2\times 2$ matrices with entries in $\End(\VV_\alpha)$. Since $b_++b_-=1$, it is sufficient, up to irrelevant additive terms in the Hamiltonian, to consider boundary couplings of the form $-\mu b$, with $b:=b_+$ and $\mu\in\CC$ a coupling constant. The one-boundary Hamiltonian on $\Hb$ is then defined as
\begin{equation}
\label{oneb}
H_b:=-\mu b-\sum_{i=1}^{N-1} e_i .
\end{equation}
One can also check that $b$ satisfies
\begin{equation}
\label{blobrule}
b^2=b ,\qquad e_1 b e_1 = y e_1 , \qquad [b,e_i]=0 \quad \text{for} \quad 2\leq i\leq N-1
\end{equation}
with
\begin{equation}
\label{blobwa}
y=\frac{[\alpha+1]_\qa}{[\alpha]_\qa} .
\end{equation}
Together with~\eqref{TLrel} this means that $\Hb$ carries a representation of the blob algebra. By construction, the actions of $\Uq$ and $\Blob$ on $\Hb$ commute with each other. One can actually show~\cite{previouspaper} that if $\qa^\alpha\notin\pm \qa^\ZZ$, we have Schur--Weyl duality, namely $\Uq$ and $\Blob$ are mutual maximal centralisers and we have the $(\Blob,\Uq)$-bimodule decomposition
\begin{equation}
\label{swb}
\Hb=\WW_0\otimes\VV_\alpha\oplus\bigoplus_{j=1}^{N/2}\big(\WW_j^b\otimes\VV_{\alpha+2j}\oplus\WW_j^{\bar{b}}\otimes\VV_{\alpha-2j}\big) .
\end{equation}

\subsubsection{The two-boundary system}

The two-boundary Hamiltonian is constructed in very much the same way, this time by tensoring the usual spin chain $\Hs=(\CC^2)^{\otimes N}$ with two Verma modules $\VV_{\alpha_l}$ and $\VV_{\alpha_r}$, on the left and on the right respectively~\cite{previouspaper}. The most general left boundary coupling is still given by $-\mu_l b_l$ with $b_l$ the projector on $\VV_{\alpha_l+1}$ from~\eqref{blobqgen}. On the other hand, the projector on the $\VV_{\alpha_r+1}$ summand of $\CC^2\otimes\VV_{\alpha_r}\cong \VV_{\alpha_r+1}\oplus\VV_{\alpha_r-1}$ is
\begin{equation}
\label{blobrqgen}
b_r =
\begin{pmatrix}
 \qa\KK - \qa^{-\alpha_r} & \qa\{1\}\KK\FF\\
 \{1\}\EE & -\qam\KK+\qa^{\alpha_r}
\end{pmatrix} .
\end{equation}
The two-boundary Hamiltonian is then defined on the Hilbert space
\begin{equation}
\label{Hbbdef}
\Hbb:=\VV_{\alpha_l}\otimes (\CC^2)^{\otimes N}\otimes\VV_{\alpha_r}
\end{equation}
as
\begin{equation}
\label{twob}
H_{2b}:=-\mu_l b_l-\mu_r b_r-\sum_{i=1}^{N-1} e_i
\end{equation}
with $\mu_r\in\CC$ a coupling constant. Similarly to $b_l$, the new generator $b_r$ satisfies
\begin{equation}
\label{blobrrule}
b_r^2=b_r ,\qquad e_{N-1} b_r e_{N-1} = y_r e_{N-1} , \qquad [b_r,e_i]=0 \quad \text{for} \quad 1\leq i\leq N-2
\end{equation}
with
\begin{equation}
\label{blobwar}
y_r=\frac{[\alpha_r+1]_\qa}{[\alpha_r]_\qa} .
\end{equation}

To obtain a representation of the two-boundary TL algebra it remains to compute the weight~$Y$ of a loop carrying both $b_r$ and $b_l$. We can indeed find such a $Y$, but it turns out that in our case it will not be a number but some non-trivial central element~\cite{previouspaper}. This is why we need to use a slightly different version of the two-boundary TL algebra, namely the \emph{universal} two-boundary TL algebra $\utwoBlob$. It is defined by the same relations as $\twoBlob$ but now $Y$ is treated as an additional generator denoted $\YYY$ and commuting with all the other generators $e_i$ and~$b_{l/r}$, i.e., it is a central extension of $\twoBlob$. If we want to recover the usual two-boundary TL at some fixed value of $Y\in\CC$, we just have to take the quotient $\utwoBlob/\langle\YYY-Y\rangle\cong\twoBlob$. In particular, this implies that any representation of $\twoBlob$ for any value of $Y\in\CC$ is automatically a representation of $\utwoBlob$ (the converse is not true in general, however). Note also that contrary to the usual two-boundary TL algebra, $\utwoBlob\supset \CC[\YYY]$ is infinite-dimensional.

Coming back to our spin chain, we showed in~\cite{previouspaper} that $\Hbb$ carries a representation of the universal two-boundary TL algebra $\utwoBlob$. Concretely, $\Uq$ admits a Casimir element
\begin{equation}
\label{casimir}
\CCC:=\{1\}^2\FF\EE+\qa\KK+\qam\KK^{-1},
\end{equation}
which commutes with $\Uq$ and moreover its action on $\Hbb$, denoted $\CCC_{\Hbb}$, commutes with the~$e_i$ and $b_{l/r}$. Then the relation~\eqref{2bloop} is satisfied for\footnote{This result, shown in~\cite{previouspaper}, has also a natural interpretation in terms of the affine Hecke algebra of type~$C$~\cite[Section 3.7]{daugherty2020calibrated}.}
\begin{equation}
\label{Ygen}
\YYY=\frac{\qa^{\alpha_l+\alpha_r+1}+\qa^{-\alpha_l-\alpha_r-1}-\CCC_{\Hbb}}{\{\alpha_l\} \{\alpha_r\}}
\end{equation}
making $\Hbb$ a representation of $\utwoBlob$. By construction, this action commutes with that of~$\Uq$.

Following~\cite{previouspaper}, we can restrict the action of $\utwoBlob$ to a $\YYY$-eigenspace to obtain a well-defined action of the usual two-boundary TL algebra $\twoBlob$ for some fixed value of $Y$. Since the Casimir $\CCC$ commutes with $\Uq$ it acts as a scalar on any irreducible representation of $\Uq$, in particular\footnote{Just evaluate it on the highest-weight vector: $\CCC\ket{0}=\{1\}^2\FF\EE\ket{0}+\qa\KK\ket{0}+\qam\KK^{-1}\ket{0}=\left(\qa^\alpha+\qa^{-\alpha}\right)\ket{0}$.}
\begin{equation}
\label{casimirva}
\CCC_{\VV_\alpha}=\qa^\alpha+\qa^{-\alpha} ,
\end{equation}
and so this amounts to computing the decomposition of $\Hbb$ into simple $\Uq$-modules. Using the fusion rule for Verma modules, valid for $\qa^{\alpha_l+\alpha_r-1}\notin\pm \qa^{\NN^*}$,\footnote{As for the fusion rule~\eqref{fusrule}, if this condition is not satisfied, some of the summands are glued into tilting modules.}
\begin{equation}
\label{vermafusrule}
\VV_{\alpha_l}\otimes\VV_{\alpha_r}\cong\bigoplus_{n\geq 0} \VV_{\alpha_l+\alpha_r-1-2n}
\end{equation}
as well as~\eqref{fusrule}, we obtain the $\Uq$-decomposition
\begin{equation}
\label{hbbdec}
\Hbb=\bigoplus_{M\geq 0}\Hs_M\otimes\VV_{\alpha_l+\alpha_r-1+N-2M} ,
\end{equation}
where $\Hs_M$ are some multiplicity spaces of dimension
\begin{equation}
\label{multqgen}
d_M:=\dim\Hs_M=
\begin{cases}
 \displaystyle\sum_{k=0}^M {N\choose k} & \text{for}\ 0\leq M\leq N,\\
 2^N & \text{for}\ M\geq N,
\end{cases}
\end{equation}
which can be identified with the subspaces of highest-weight vectors of weight $\qa^{\alpha_l+\alpha_r-2+N-2M}$. By direct computation, one then shows~\cite{previouspaper} that restricted to $\Hs_M$, $\YYY$ acts as the scalar
\begin{equation}
\label{ymlr}
Y_M=\dfrac{\left[M+1-\frac{N}{2}\right]_\qa\left[\alpha_l+\alpha_r-M+\frac{N}{2}\right]_\qa}{[\alpha_{l}]_\qa[\alpha_{r}]_\qa}
\end{equation}
and so, for all $M\geq 0$, $\Hs_M$ is a representation of the two-boundary TL algebra $\twoBlobm$.

The final question is what are these representations. Let us first recall an important result from~\cite{de_Gier_2009, dubail:tel-00555624}, valid for generic $\qa$.
\begin{Theorem}
\label{dgthm2}\quad
\begin{itemize}\itemsep=0pt
\item[$(i)$] For $0\leq M\leq N/2-1$ the vacuum module $\WW_0$ of $\twoBlobm$ is reducible but indecomposable, with a unique irreducible proper $\twoBlobm$-submodule isomorphic to $\WW_{N/2-M}^{bb}$ and an irreducible subquotient $\WW_0/\WW_{N/2-M}^{bb}$.
\item[$(ii)$] For $N/2\leq M\leq N-1$ the vacuum module $\WW_0$ of $\twoBlobm$ is reducible but indecomposable, with a unique irreducible proper $\twoBlobm$-submodule isomorphic to $\WW_{M+1-N/2}^{\bar{b}\bar{b}}$ and an irreducible subquotient $\WW_0/\WW_{M+1-N/2}^{\bar{b}\bar{b}}$.
\item[$(iii)$] For $M\geq N$ the vacuum module $\WW_0$ of $\twoBlobm$ is irreducible.
\end{itemize}
\end{Theorem}

Based on this theorem and the dimensions~\eqref{multqgen} a conjecture about the nature of the $\twoBlobm$-modules $\Hs_M$ was made in~\cite{previouspaper}. In Appendix~\ref{proof}, we prove this conjecture, so let us restate it here as a theorem:
\begin{Theorem}
\label{thm}
For $\qa\in\CC\backslash\qa^{i\pi\QQ}$, $\qa^{\alpha_l},\qa^{\alpha_r}\in\CC\backslash\{\pm\qa^\ZZ\}$ such that $\qa^{\alpha_l+\alpha_r}\notin\pm\qa^\ZZ$ and $N\in 2\mathbb{N}^*$, $\Uq$ and $\utwoBlob$ are mutual centralisers on $\Hbb$ in~\eqref{Hbbdef}, with generators~$e_i$,~$1\leq i\leq N-1$,~$b_l$,~$b_r$,~$\YYY$ acting by~\eqref{eiexp},~\eqref{blobqgen},~\eqref{blobrqgen},~\eqref{Ygen} respectively, and we have the $(\utwoBlob,\Uq)$-bimodule decomposition
\begin{equation}
\label{bimod2b}
\Hbb=\bigoplus_{M\geq 0}\Hs_M\otimes\VV_{\alpha_l+\alpha_r-1+N-2M} ,
\end{equation}
where the $\Hs_M$ are irreducible $\twoBlobm$-modules given by
\begin{equation}
\label{2bmodiso}
\begin{aligned}
\Hs_M & \cong\WW_{N/2-M}^{bb}\varsubsetneq\WW_0 \qquad\;\, \text{for}~~0\leq M\leq N/2-1 ,\\
\Hs_M & \cong\WW_0/\WW_{M+1-N/2}^{\bar{b}\bar{b}} \,\, \qquad \text{for}~~N/2\leq M\leq N-1 ,\\
\Hs_M & \cong\WW_0 \,\, \ \qquad\qquad\qquad\quad \text{for} \ N\leq M .
\end{aligned}
\end{equation}
\end{Theorem}
This theorem means, in particular, that even for $0\leq M\leq N-1$, $\Hs_M$ is always isomorphic to an irreducible piece of the vacuum module $\WW_0$ of $\twoBlobm$, either a stable subspace ($0\leq M\leq N/2-1$) or an irreducible subquotient ($N/2\leq M\leq N-1$) of $\WW_0$.

Now recall $\qa={\rm e}^h$, $\delta=2\cosh(h)=[2]_\qa$, and compare~\eqref{weights1} with~\eqref{blobwa}--\eqref{ymlr}. All the weights coincide, except for $Y$, which however matches in both cases if and only if
\begin{equation*}
\alpha_l+\alpha_r\pm\Theta=2M+1-N ,
\end{equation*}
where $\Theta:=\theta_l-\theta_r$~\eqref{Theta}. Recalling the reparametrisation~\eqref{parameters}, this is equivalent to
\begin{equation*}
\delta_l+\kappa_l+\delta_r+\kappa_r\pm\Theta=2M+1-N ,
\end{equation*}
which is exactly the Nepomechie condition~\eqref{nepcond}! Moreover, Theorems~\ref{dgthm}--\ref{thm} imply the following. Set
\begin{equation*}
\Hnd^{(M)}:=\Hnd(\delta,y_{l/r},\mu_{l/r},Y_M)
\end{equation*}
and for any $\twoBlob$-module $\MM$ denote the representation map $\rho_\MM \colon \twoBlob\to\End_\CC(\MM)$. Then
\begin{Corollary}\samepage
\label{cor}\quad
\begin{itemize}\itemsep=0pt
\item[$(i)$] For $0\leq M\leq N/2-1$,
\begin{align*}
\Hnd^{(M)}=\rho_{\WW_0}(\mathbf{H})&=
\begin{pmatrix}
\rho_{\WW_0/\WW_{N/2-M}^{bb}}(\mathbf{H}) & *\\
0 & \rho_{\WW_{N/2-M}^{bb}}(\mathbf{H})
\end{pmatrix} \\
& =
\begin{pmatrix}
\rho_{\WW_0/\WW_{N/2-M}^{bb}}(\mathbf{H}) & *\\
0 & H_{2b}|_{\Hs_M}
\end{pmatrix} .
\end{align*}
\item[$(ii)$] For $N/2\leq M\leq N-1$,
\begin{align*}
\Hnd^{(M)}=\rho_{\WW_0}(\mathbf{H})&
=
\begin{pmatrix}
\rho_{\WW_0/\WW_{M+1-N/2}^{\bar{b}\bar{b}}}(\mathbf{H}) & *\\
0 & \rho_{\WW_{M+1-N/2}^{\bar{b}\bar{b}}}(\mathbf{H})
\end{pmatrix} \\
&=
\begin{pmatrix}
H_{2b}|_{\Hs_M} & *\\
0 & \rho_{\WW_{M+1-N/2}^{\bar{b}\bar{b}}}(\mathbf{H})
\end{pmatrix} .
\end{align*}
\item[$(iii)$] For $M\geq N$,
\begin{equation*}
\Hnd^{(M)}=\rho_{\WW_0}(\mathbf{H})=H_{2b}|_{\Hs_M} .
\end{equation*}
\end{itemize}
\end{Corollary}
In other words, the spectrum of all the open non-diagonal XXZ Hamiltonians with non-diagonal boundary terms for \emph{all} the values of the parameters covered by the Nepomechie condition~\eqref{nepcond} with $M\geq N$ is contained in the irreducible sectors of a \emph{single} Hamiltonian $H_{2b}$. For $0\leq M\leq N-1$ the sectors of $H_{2b}$ only contain an irreducible block of $\Hnd$. The spectral problem of $H_{2b}$ will be solved in the next section by algebraic Bethe ansatz.

One may wonder if we can also express the remaining blocks $\rho_{\WW_0/\WW_{N\!/2-\!M}^{bb}}\!\!(\mathbf{H})$ and $\rho_{\WW_{M+1-N\!/2}^{\bar{b}\bar{b}}}\!\!(\mathbf{H})$ in terms of $H_{2b}$ to diagonalise $\Hnd$ completely, even for $0\leq M\leq N-1$. This turns out to be possible via the following trick. Let us introduce the involution
\begin{equation}
\label{inv}
\overline{\,\cdot\,}\colon \ \twoBlob\to\twoBlobb ,\qquad (e_i, b_{l/r})\mapsto (e_i, \bar{b}_{l/r}:=1-b_{l/r}) .
\end{equation}
From the definitions of the various weights one easily sees that it is an algebra isomorphism and moreover that it exchanges $\WW_{j}^{bb}$ with $\WW_{j}^{\bar{b}\bar{b}}$ \big(and also $\WW_{j}^{b\bar{b}}$ with $\WW_{j}^{\bar{b}b}$\big) for $1\leq j\leq N/2$, while leaving $\WW_0$ invariant. Let us define
\begin{equation*}
\overline{\mathbf{H}}:=-\mu_l\bar{b}_l-\mu_r\bar{b}_r-\sum_{i=1}^{N-1} e_i\in\twoBlobb .
\end{equation*}
and
\begin{equation}
\label{hamtr}
\overline{H}_{2b}:=\rho_{\overline{\Hs}_{2b}}\big(\overline{\mathbf{H}}\big)=-\mu_l-\mu_r+H_{2b}(\alpha_{l/r}\to -\alpha_{l/r}, \mu_{l/r}\to -\mu_{l/r})
\end{equation}
acting on $\overline{\Hs}_{2b}:=\VV_{-\alpha_l}\otimes (\CC^2)^{\otimes N}\otimes\VV_{-\alpha_r}$. Note that by~\eqref{blobwa}--\eqref{blobwar}
\begin{equation*}
y_{l/r}(\alpha_{l/r}\to-\alpha_{l/r})=\frac{[\alpha_{l/r}-1]_\qa}{[\alpha_{l/r}]_\qa}=\delta-y_{l/r}
\end{equation*}
and using~\eqref{ymlr} one easily checks that
\begin{equation}
\label{ymlrtr}
Y_M(\alpha_{l/r}\to-\alpha_{l/r})=\dfrac{\left[\frac{N}{2}-M-1\right]_\qa\left[\alpha_l+\alpha_r-\frac{N}{2}+M\right]_\qa}{[\alpha_{l}]_\qa[\alpha_{r}]_\qa}=\delta-y_l-y_r+Y_{N-M-1},
\end{equation}
so by Theorem~\ref{thm}, $\overline{\Hs}_{2b}$ decomposes as
\begin{equation*}
\overline{\Hs}_{2b}=\bigoplus_{M\geq 0}\overline{\Hs}_M\otimes\VV_{-\alpha_l-\alpha_r-1+N-2M},
\end{equation*}
where the $\overline{\Hs}_M$ are irreducible $\twoBlobmm$-modules given by~\eqref{2bmodiso}. Pulling back by the algebra isomorphism~\eqref{inv}, we can thus generalise Corollary~\ref{cor}.
\begin{Corollary}
\label{corr}\quad
\begin{itemize}\itemsep=0pt
\item[$(i)$] For $0\leq M\leq N/2-1$,
\begin{align*}
\Hnd^{(M)}=\rho_{\WW_0}(\mathbf{H})=\rho_{\WW_0}\big(\overline{\mathbf{H}}\big)&=
\begin{pmatrix}
\rho_{\WW_0/\WW_{N/2-M}^{\bar{b}\bar{b}}}\big(\overline{\mathbf{H}}\big) & *\\
0 & \rho_{\WW_{N/2-M}^{bb}}(\mathbf{H})
\end{pmatrix} \\
&=
\begin{pmatrix}
\overline{H}_{2b}|_{\overline{\Hs}_{N-M-1}} & *\\
0 & H_{2b}|_{\Hs_M}
\end{pmatrix} .
\end{align*}
\item[$(ii)$] For $N/2\leq M\leq N-1$,
\begin{align*}
\Hnd^{(M)}=\rho_{\WW_0}(\mathbf{H})=\rho_{\WW_0}\big(\overline{\mathbf{H}}\big)&=
\begin{pmatrix}
\rho_{\WW_0/\WW_{M+1-N/2}^{\bar{b}\bar{b}}}(\mathbf{H}) & *\\
0 & \rho_{\WW_{M+1-N/2}^{bb}}\big(\overline{\mathbf{H}}\big)
\end{pmatrix} \\
&=
\begin{pmatrix}
H_{2b}|_{\Hs_M} & *\\
0 & \overline{H}_{2b}|_{\overline{\Hs}_{N-M-1}}
\end{pmatrix} .
\end{align*}
\item[$(iii)$] For $M\geq N$,
\begin{equation*}
\Hnd^{(M)}=\rho_{\WW_0}(\mathbf{H})=H_{2b}|_{\Hs_M} .
\end{equation*}
\item[$(iv)$] For $M\leq -1$,
\begin{equation*}
\Hnd^{(M)}=\rho_{\WW_0}\big(\overline{\mathbf{H}}\big)=\overline{H}_{2b}|_{\overline{\Hs}_{N-M-1}} .
\end{equation*}
\end{itemize}
\end{Corollary}
Note that, because of~\eqref{ymlrtr}, the magnon number $M$ labelling the sectors $\Hs_M$ of $\Hbb$ is mapped to the dual magnon number $\overline{M}:=N-M-1$ in the $\overline{\Hs}_{2b}$ spin chain (and vice versa) by the involution~\eqref{inv}. It is this purely algebraic observation that enables us to reach the missing blocks of $\Hnd^{(M)}$ for $0\leq M\leq N-1$ as well as negative values of $M$, which correspond to $\overline{M}\geq N$. As we will see, this will be essential to establish a complete set of Bethe ansatz equations for the degenerate cases $0\leq M\leq N-1$.

To summarise, we have reduced the spectral problem of all the $\Hnd^{(M)}$, $M\in\ZZ$, satisfying the Nepomechie condition~\eqref{nepcond} to the spectral problem of $H_{2b}$.\footnote{Note that $\overline{H}_{2b}$ is related to $H_{2b}$ by the transformation~\eqref{hamtr} so it is sufficient to consider $H_{2b}$ only.} This may not seem like a big step, but actually it is: $H_{2b}$ is $\Uq$-invariant and as such has a natural reference state $\ket{0}\otimes\ket{\uparrow}^{\otimes N}\otimes\ket{0}$ which will enable us to compute its spectrum using standard algebraic Bethe ansatz.

\section{Bethe ansatz}
\label{BAEsec}

We now turn to the computation of the spectrum of $H_{2b}$~\eqref{twob} using the algebraic boundary Bethe ansatz formalism first developed by Sklyanin~\cite{sklyanin1988boundary}. Here, we no longer assume that $N$ is even. As a warm-up, we will first treat the one-boundary Hamiltonian $H_b$~\eqref{oneb}.

\subsection{The one-boundary system}
\label{oneBA}

In the basis $\{\ket{\uparrow\uparrow},\ket{\uparrow\downarrow},\ket{\downarrow\uparrow},\ket{\downarrow\downarrow}\}$ the $\Uq$-invariant (affine) $R$-matrix is given by
\begin{equation}
\label{afrmat}
R(u):=
\begin{pmatrix}
\sinh(u+h) & 0 & 0 & 0\\
0 & \sinh(u) & \sinh(h){\rm e}^{u} & 0\\
0 & \sinh(h){\rm e}^{-u} & \sinh(u) & 0\\
0 & 0 & 0 & \sinh(u+h)
\end{pmatrix} .
\end{equation}
It is easy to check that
\begin{equation}
\label{Rcheck}
\check{R}_{i,i+1}(u):=P_{i,i+1}R_{i,i+1}(u)=\sinh(u+h)-\sinh(u)e_i,
\end{equation}
where $P_{i,i+1}$ is the operator permuting the $i$ and $(i+1)$-th sites of the spin chain. For all $u,v\in\CC$, the $R$-matrix $R(u)$ satisfies the Yang--Baxter equation (YBE)
\begin{equation}
\label{YBE}
R_{1,2}(u-v)R_{1,3}(u)R_{2,3}(v)=R_{2,3}(v)R_{1,3}(u)R_{1,2}(u-v),
\end{equation}
where $R_{i,j}(u)$ denotes $R(u)$ acting the $i$-th and $j$-th tensor factors of $(\CC^2)^{\otimes 3}$ or, equivalently,
\begin{equation*}
\check{R}_{1,2}(u-v)\check{R}_{2,3}(u)\check{R}_{1,2}(v)=\check{R}_{2,3}(v)\check{R}_{1,2}(u)\check{R}_{2,3}(u-v) .
\end{equation*}
Note also that
\begin{equation}
\label{invR}
R_{1,2}(u)R_{2,1}(-u)=\sinh(h+u)\sinh(h-u)\Id_{\CC^2\otimes\CC^2},
\end{equation}
so $R(u)$ is invertible if $u\neq\pm h$.

For boundary Bethe ansatz, one also needs an additional ingredient: a so-called $K$-matrix. It is a $2\times 2$ matrix $K(u)$ with entries in some (possibly non-commutative) algebra satisfying the boundary Yang--Baxter equation (bYBE)
\begin{equation}
\label{BYBE}
R_{1,2}(u-v)K_1(u)R_{2,1}(u+v)K_2(v)=K_2(v)R_{1,2}(u+v)K_1(u)R_{2,1}(u-v)
\end{equation}
or, written differently,
\begin{equation}
\label{BYBE2}
\check{R}_{1,2}(u-v)K_1(u)\check{R}_{1,2}(u+v)K_1(v)=K_1(v)\check{R}_{1,2}(u+v)K_1(u)\check{R}_{1,2}(u-v) .
\end{equation}
For example, because
\begin{equation}
\label{idsol}
R_{1,2}(u-v)R_{2,1}(u+v)=R_{1,2}(u+v)R_{2,1}(u-v) ,
\end{equation}
$K(u)=\Id_{\CC^2}$ is a solution of~\eqref{BYBE}. This can also be seen directly from~\eqref{Rcheck}--\eqref{BYBE2}.

Although many other solutions to the bYBE~\eqref{BYBE} are known~\cite{bk1993, bk1994, Doikou_2003, sklyanin1988boundary}, it is not always possible to find a $K$-matrix yielding precisely the boundary conditions we want to impose. In our case however, it is possible to use the symmetry of the one-boundary spin chain to find a~suitable solution of the bYBE~\eqref{BYBE}. Actually, there are even two independent constructions: one based on the $\Blob$-module structure of the spin chain and the other on the $\Uq$ symmetry.

The first was described in detail~\cite{Doikou_2003}. Namely, for any $b$ satisfying the blob algebra relations~\eqref{blobrule} there exists a solution of the bYBE~\eqref{BYBE} for $H_b$. With our choice of $b$ this $K$-matrix reads
\begin{align}
K(u) & =1-\frac{\mu}{\sinh(h)\sinh(h\alpha)}\left(\sinh(u-h\alpha)\sinh(u)+\sinh(h\alpha)\sinh(2u)b\right)\nonumber\\
& =\Id_{\VV_\alpha\otimes\CC^2}-\frac{\mu\sinh(u)\cosh(u)}{\sinh(h)\sinh(h\alpha)}\!
\begin{pmatrix}
-{\rm e}^{-h}\KK^{-1}+\frac{{\rm e}^u\cosh(h\alpha)}{\cosh(u)} & 2\sinh(h)\FF\\
2{\rm e}^h\sinh(h)\KK^{-1}\EE & {\rm e}^{h}\KK^{-1}-\frac{{\rm e}^{-u}\cosh(h\alpha)}{\cosh(u)}
\end{pmatrix} ,\!\!\!\label{kmatrix}
\end{align}
where again we have written it as a $2\times 2$ matrix with entries in $\End(\VV_\alpha)$. The second construction based on the $\Uq$ symmetry of the spin chain will be used for the two-boundary case where it is most convenient. For the time being, let us work with the $K$-matrix~\eqref{kmatrix}.

Let us define the transfer matrix\footnote{The $\frac{h}{2}$ shift is introduced to make the final result neater.}
\begin{equation}
\label{trmat}
t_b(u):=\qtr_0{T(u-h/2)K(u-h/2)\hat{T}(u-h/2)},
\end{equation}
where $K$ is treated as a $2\times 2$ matrix with entries in $\End(\VV_\alpha)$ acting on the auxiliary $\CC^2$ space (with index $0$),
\begin{gather}
T(u) :=R_{0,N}(u)\cdots R_{0,1}(u) , \qquad \hat{T}(u) :=R_{1,0}(u)\cdots R_{N,0}(u)
\label{trmatp}
\end{gather}
and $\qtr_0(-):=\tr_0\bigl(\qa^{\sigma^z_0}-\bigr)$ denotes the partial quantum trace over the auxiliary space. This is the natural trace for $\Uq$-invariant objects and it ensures that $t_b(u)$ commutes with $\Uq$ as it should (see~\cite{Doikou_2003} for more details\footnote{The formalism in this paper is a bit more general with $J$ corresponding to our $\qa^{\sigma^z_0}$.}).

Now by~\eqref{YBE} and~\eqref{BYBE}
\begin{equation*}
[t_b(u), t_b(v)]=0
\end{equation*}
and, moreover,~\cite{sklyanin1988boundary}
\begin{align*}
\frac{\ddd}{\ddd u}\bigg\rvert_{u=\frac{h}{2}} t_b(u) ={}& 2\sinh^{2N-1}(h)\tr_0\big(\qa^{\sigma^z}\big)\sum_{i=1}^{N-1}\check{R}'_{i,i+1}(0)\\
&{} +2\sinh^{2N-1}(h)\qtr_0\big(\check{R}'_{N,0}(0)\big)
 +\sinh^{2N}(h)\tr_0\big(\qa^{\sigma^z}\big)P_{0,1}K'(0)P_{0,1}\\
={}& 4\sinh^{2N}(h)\coth(h)\sum_{i=1}^{N-1}\cosh(h)-e_i\\
&{} + 4\sinh^{2N}(h)\coth(h)\left(\cosh(h)-\frac{1}{2\cosh(h)}\right)\\
&{} +4\sinh^{2N}(h)\coth(h)\left(-\mu b+\frac{\mu}{2}\right),
\end{align*}
so
\begin{equation}
\label{Hbtm}
H_b=-\mu b-\sum_{i=1}^{N-1}e_i=\frac{\tanh(h)}{4\sinh^{2N}(h)}\frac{\ddd}{\ddd u}\bigg\rvert_{u=\frac{h}{2}} t_b(u)-N\cosh(h)+\frac{1}{2\cosh(h)}-\frac{\mu}{2} .
\end{equation}
Therefore, we are reduced to computing the spectrum of $t_b(u)$.

Let us now define the monodromy
\begin{equation*}
\mathcal{T}(u):=T(u-h/2)K(u-h/2)\hat{T}(u-h/2)=
\begin{pmatrix}
\mathcal{A}(u) & \mathcal{B}(u) \\
\mathcal{C}(u) & \mathcal{D}(u)
\end{pmatrix},
\end{equation*}
where the coefficients of this auxiliary space $2\times 2$ matrix are in $\End(\Hb)$. Repeatedly applying the YBE~\eqref{YBE}, one finds that $T(u)$ satisfies the RTT relation
\begin{equation}
\label{RTT}
R_{0,\bar{0}}(u-v)T_0(u)T_{\bar{0}}(v) =T_{\bar{0}}(v)T_0(u)R_{0,\bar{0}}(u-v)
\end{equation}
involving two different auxiliary $\CC^2$ spaces with index $0$ and $\bar{0}$. Note also~\eqref{invR} implies that $\hat{T}(u)\propto T(-u)^{-1}$. By a general result~\cite[Proposition 2]{sklyanin1988boundary}, for any $T(u)$ satisfying~\eqref{RTT} and any solution $K(u)$ of the bYBE~\eqref{BYBE}, the product $T(u)K(u)T(-u)^{-1}$ is also a solution of~\eqref{BYBE}. Therefore, $\mathcal{T}(u+h/2)$ satisfies~\eqref{BYBE}\footnote{This statement can also be proved by induction on $N$ using only the YBE~\eqref{YBE} and~\eqref{idsol}.}, that is,
\begin{gather}
R_{1,2}(u-v)\begin{pmatrix}
\mathcal{A}(u) & \mathcal{B}(u) \\
\mathcal{C}(u) & \mathcal{D}(u)
\end{pmatrix}_1 R_{2,1}(u+v-h)\begin{pmatrix}
\mathcal{A}(v) & \mathcal{B}(v) \\
\mathcal{C}(v) & \mathcal{D}(v)
\end{pmatrix}_2 \nonumber\\
\qquad{}=\begin{pmatrix}
\mathcal{A}(v) & \mathcal{B}(v) \\
\mathcal{C}(v) & \mathcal{D}(v)
\end{pmatrix}_2 R_{1,2}(u+v-h)\begin{pmatrix}
\mathcal{A}(u) & \mathcal{B}(u) \\
\mathcal{C}(u) & \mathcal{D}(u)
\end{pmatrix}_1 R_{2,1}(u-v) .\label{BYBEexp}
\end{gather}
From the explicit expression of $R(u)$~\eqref{afrmat}, we can then derive the relevant commutation relations between $\mathcal{A}$, $\mathcal{B}$, $\mathcal{C}$ and $\mathcal{D}$ at different values of the spectral parameter. Doing so, we obtain
\begin{equation}
\label{bcomrel}
[\mathcal{B}(u),\mathcal{B}(v)]=0
\end{equation}
for all $u,v\in\CC$ and, moreover,
\begin{align*}
\mathcal{A}(u)\mathcal{B}(v) = & \ \frac{\sinh(u-v-h)\sinh(u+v-h)}{\sinh(u-v)\sinh(u+v)}\mathcal{B}(v)\mathcal{A}(u)\\
 & +\frac{{\rm e}^{u-v}\sinh(h)\sinh(u+v-h)}{\sinh(u-v)\sinh(u+v)}\mathcal{B}(u)\mathcal{A}(v)
 -\frac{{\rm e}^{u+v-h}\sinh(h)}{\sinh(u+v)}\mathcal{B}(u)\mathcal{D}(v) , \\
\mathcal{D}(u)\mathcal{B}(v) = & \ \frac{\sinh(u-v+h)\sinh(u+v+h)}{\sinh(u-v)\sinh(u+v)}\mathcal{B}(v)\mathcal{D}(u)\\
 & -\frac{{\rm e}^{v-u}\sinh(h)\sinh(u+v+h)}{\sinh(u-v)\sinh(u+v)}\mathcal{B}(u)\mathcal{D}(v)\\
& +\frac{{\rm e}^{-u-v+h}\sinh(h)\sinh(u-v+2h)}{\sinh(u-v)\sinh(u+v)}\mathcal{B}(u)\mathcal{A}(v)\\
& -\frac{2{\rm e}^{-2u+h}\sinh^2(h)\cosh(h)}{\sinh(u-v)\sinh(u+v)}\mathcal{B}(v)\mathcal{A}(u) .
\end{align*}
Introducing
\begin{equation}
\label{dbar}
\bar{\mathcal{D}}(u)=\frac{{\rm e}^{-h}\sinh(2u)}{\sinh(2u-h)}\mathcal{D}(u)-\frac{{\rm e}^{-2u}\sinh(h)}{\sinh(2u-h)}\mathcal{A}(u),
\end{equation}
these equations become
\begin{gather}
\mathcal{A}(u)\mathcal{B}(v) = \frac{\sinh(u-v-h)\sinh(u+v-h)}{\sinh(u-v)\sinh(u+v)}\mathcal{B}(v)\mathcal{A}(u)\nonumber\\
\hphantom{\mathcal{A}(u)\mathcal{B}(v) =}{} +f_1(u,v)\mathcal{B}(u)\mathcal{A}(v)+f_2(u,v)\mathcal{B}(u)\bar{\mathcal{D}}(v) ,\label{comrelA}
\\
\bar{\mathcal{D}}(u)\mathcal{B}(v) = \frac{\sinh(u-v+h)\sinh(u+v+h)}{\sinh(u-v)\sinh(u+v)}\mathcal{B}(v)\bar{\mathcal{D}}(u)\nonumber\\
\hphantom{\bar{\mathcal{D}}(u)\mathcal{B}(v) =}{} +g_1(u,v)\mathcal{B}(u)\bar{\mathcal{D}}(v)+g_2(u,v)\mathcal{B}(u)\mathcal{A}(v) ,\label{comrelD}
\end{gather}
where
\begin{equation}
\label{expfg}
f_1(u,v) =\frac{{\rm e}^{u-v}\sinh(h)\sinh(2v-h)}{\sinh(u-v)\sinh(2v)} ,\qquad f_2(u,v) =-\frac{{\rm e}^{u+v}\sinh(h)\sinh(2v-h)}{\sinh(u+v)\sinh(2v)}
\end{equation}
and
\begin{equation}
\label{lprop}
g_1(u,v)=\frac{\sinh(2u+h)}{\sinh(2u-h)}f_1(u,v) ,\qquad g_2(u,v)=\frac{\sinh(2u+h)}{\sinh(2u-h)}f_2(u,v) .
\end{equation}
We will actually never need the explicit expressions of $f_1$, $f_2$, $g_1$ and $g_2$~\eqref{expfg} but only the relations~\eqref{lprop}. The transfer matrix reads
\begin{equation}
\label{tmatu}
t_b(u)=\qtr_0\mathcal{T}(u)={\rm e}^h\mathcal{A}(u)+{\rm e}^{-h}\mathcal{D}(u)=\frac{\sinh(2u+h)}{\sinh(2u)}\mathcal{A}(u)+\frac{\sinh(2u-h)}{\sinh(2u)}\bar{\mathcal{D}}(u) .
\end{equation}

Let us now look for eigenvectors of $t_b(u)$ of the form
\begin{equation}
\label{bansatz}
\ket{\{v_m\}}=\mathcal{B}(v_1)\cdots\mathcal{B}(v_M)\ket{\Uparrow},
\end{equation}
where
\begin{equation*}
\ket{\Uparrow}:=\ket{0}\otimes\ket{\uparrow}^{\otimes N}
\end{equation*}
is our reference state (recall that $\ket{0}$ is the highest-weight vector of the Verma module $\VV_\alpha$~\eqref{alpharepgen}), $M\geq 0$ is the magnon number and $\{v_m\}_{1\leq m\leq M}$ are some complex numbers that we want to determine. Note that because of~\eqref{bcomrel}, the order of the $v_m$ is irrelevant. Also $\mathcal{B}(u)$ decreases the $\Uq$-weight by $\qa^{-2}$ so
\begin{equation}
\label{Kweight}
\KK_{\Hb}\ket{\{v_m\}}=\qa^{-2M}\mathcal{B}(v_1)\cdots\mathcal{B}(v_M)\KK_{\Hb}\ket{\Uparrow}=\qa^{\alpha-1+N-2M}\ket{\{v_m\}}
\end{equation}
meaning that $\ket{\{v_m\}}$ has weight $\qa^{\alpha-1+N-2M}$.

The first step is to compute the eigenvalues of $\mathcal{A}(u)$ and $\bar{\mathcal{D}}(u)$ when acting on $\ket{\Uparrow}$. Re\-write~\eqref{kmatrix}--\eqref{trmatp}
\begin{gather}
T(u-h/2)=
\begin{pmatrix}
A(u) & \tilde{B}(u) \\
C(u) & D(u)
\end{pmatrix} ,\qquad
\hat{T}(u-h/2)=
\begin{pmatrix}
A(u) & B(u) \\
\tilde{C}(u) & D(u)
\end{pmatrix} ,\nonumber\\
K(u-h/2)=
\begin{pmatrix}
a(u) & b(u) \\
c(u) & d(u)
\end{pmatrix}
\label{matform}
\end{gather}
as $2\times 2$ matrices acting on the auxiliary space with coefficients in $\End(\Hb)$ such that
\begin{equation}
\label{Tdec}
\mathcal{T}(u)=
\begin{pmatrix}
\mathcal{A}(u) & \mathcal{B}(u) \\
\mathcal{C}(u) & \mathcal{D}(u)
\end{pmatrix}=
\begin{pmatrix}
A(u) & \tilde{B}(u) \\
C(u) & D(u)
\end{pmatrix}
\begin{pmatrix}
a(u) & b(u) \\
c(u) & d(u)
\end{pmatrix}
\begin{pmatrix}
A(u) & B(u) \\
\tilde{C}(u) & D(u)
\end{pmatrix}.
\end{equation}
Knowing that $C(u)\ket{\Uparrow}=\tilde{C}(u)\ket{\Uparrow}=0$, we have
\begin{gather}
\mathcal{A}(u)\ket{\Uparrow} = a(u)A(u)^2\ket{\Uparrow} , \nonumber\\
\mathcal{D}(u)\ket{\Uparrow} = (a(u)C(u)B(u)+d(u)D(u)^2)\ket{\Uparrow} .\label{hwaction}
\end{gather}
Introducing a basis $\{\ket{\uparrow_0},\ket{\downarrow_0}\}$ of the auxiliary space and the matrix entries $r_{ij}:=R(u-h/2)_{ij}$, $1\leq i,j\leq 4$, we obtain
\begin{gather*}
A(u)\ket{\Uparrow} = \bra{\uparrow_0}R_{0,N}(u-h/2)\cdots R_{0,1}(u-h/2)\ket{\uparrow_0}\otimes\ket{\Uparrow} = r_{11}^N\ket{\Uparrow} =\sinh(u+h/2)^{N}\ket{\Uparrow} ,\\
D(u)\ket{\Uparrow} = \bra{\downarrow_0}R_{0,N}(u-h/2)\cdots R_{0,1}(u-h/2)\ket{\downarrow_0}\otimes\ket{\Uparrow} = r_{33}^N\ket{\Uparrow} =\sinh(u-h/2)^{N}\ket{\Uparrow}
\end{gather*}
and
\begin{equation*}
\begin{aligned}
B(u)\ket{\Uparrow} & =\bra{\uparrow_0}R_{1,0}(u-h/2)\cdots R_{N,0}(u-h/2)\ket{\downarrow_0}\otimes\ket{\Uparrow}\\
& =\bra{\uparrow_0}R_{1,0}(u-h/2)\cdots R_{N-1,0}(u-h/2)\left(r_{22}\ket{\downarrow_0}\otimes\ket{\Uparrow}+r_{32}\ket{\uparrow_0}\otimes\sigma_N^-\ket{\Uparrow}\right)\\
& = r_{22}\bra{\uparrow_0}R_{1,0}(u-h/2)\cdots R_{N-1,0}(u-h/2)\ket{\downarrow_0}\otimes\ket{\Uparrow}+r_{11}^{N-1}r_{32}\ket{\uparrow_0}\otimes\sigma_N^-\ket{\Uparrow}\\
& =\sum_{k=0}^{N-1}r_{11}^{N-1-k}r_{32}r_{22}^k\sigma_{N-k}^-\ket{\Uparrow} ,
\end{aligned}
\end{equation*}
so
\begin{equation*}
\begin{aligned}
C(u)B(u)\ket{\Uparrow} &= \sum_{k=0}^{N-1}r_{11}^{N-1-k}r_{32}r_{22}^k\bra{\downarrow_0}R_{0,N}(u-h/2)\cdots R_{0,1}(u-h/2)\ket{\uparrow_0}\otimes\sigma_{N-k}^-\ket{\Uparrow}\\
& = \sum_{k=0}^{N-1}r_{11}^{N-1-k}r_{32}r_{33}^{k}r_{22}^{k}r_{32}r_{11}^{N-1-k}\ket{\Uparrow}\\
& =\sinh^2(h){\rm e}^{-2u+h}\sum_{k=0}^{N-1} \sinh(u+h/2)^{2N-2k-2}\sinh(u-h/2)^{2k}\ket{\Uparrow}\\
& =\frac{\sinh^{2N}(u+h/2)-\sinh^{2N}(u-h/2)}{\sinh^{2}(u+h/2)-\sinh^{2}(u-h/2)}\sinh^2(h){\rm e}^{-2u+h}\ket{\Uparrow} .
\end{aligned}
\end{equation*}
Using the explicit expression of $K(u)$~\eqref{kmatrix} and $\bar{\mathcal{D}}(u)$~\eqref{dbar} as well as~\eqref{hwaction}, we obtain
\begin{gather}
\mathcal{A}(u)\ket{\Uparrow} = \sinh^{2N}(u+h/2)\Delta(u)\ket{\Uparrow} , \label{AonUp} \\
\bar{\mathcal{D}}(u)\ket{\Uparrow} = \sinh^{2N}(u-h/2)\Delta(-u)\ket{\Uparrow} , \label{DonUp}
\end{gather}
where
\begin{equation}
\label{delta}
\Delta(u)=1-\mu\frac{\sinh(u-h/2)\sinh(u+h\alpha-h/2)}{\sinh(h)\sinh(h\alpha)} .
\end{equation}
Finally, following the standard algebraic Bethe ansatz procedure~\cite{sklyanin1988boundary}, we use the commutation relations~\eqref{comrelA} and~\eqref{comrelD} to compute $t_b(u)\ket{\{v_m\}}$. We have from (\ref{comrelA}) and (\ref{AonUp})
\begin{align}
\mathcal{A}(u)\ket{\{v_m\}} ={}& \mathcal{A}(u)\mathcal{B}(v_1)\cdots\mathcal{B}(v_M)\ket{\Uparrow}\nonumber\\
 ={}& \left(\prod_{m=1}^M\frac{\sinh(u-v_m-h)\sinh(u+v_m-h)}{\sinh(u-v_m)\sinh(u+v_m)}\mathcal{B}(v_k)\right)\mathcal{A}(u)\ket{\Uparrow}\nonumber\\
 &{}+\sum_{k=1}^M f_1(u,v_k)\mathcal{B}(u)\nonumber\\
 &\quad{}\times \left(\prod_{m\neq k}\frac{\sinh(v_k-v_m-h)\sinh(v_k+v_m-h)}{\sinh(v_k-v_m)\sinh(v_k+v_m)}\mathcal{B}(v_k)\right)\mathcal{A}(v_k)\ket{\Uparrow}\nonumber\\
 &{}+\sum_{k=1}^M f_2(u,v_k)\mathcal{B}(u)\nonumber\\
 &\quad{}\times\left(\prod_{m\neq k}\frac{\sinh(v_k-v_m+h)\sinh(v_k+v_m+h)}{\sinh(v_k-v_m)\sinh(v_k+v_m)}\mathcal{B}(v_k)\right)\bar{\mathcal{D}}(v_k)\ket{\Uparrow}\nonumber\\
 ={}& \sinh^{2N}(u+h/2)\Delta(u)\prod_{m=1}^M\frac{\sinh(u-v_m-h)\sinh(u+v_m-h)}{\sinh(u-v_m)\sinh(u+v_m)}\ket{\{v_m\}}\nonumber\\
 &{}+\sum_{k=1}^M f_1(u,v_k)\sinh^{2N}(v_k+h/2)\Delta(v_k)\nonumber\\
 &\quad{}\times\prod_{m\neq k}\frac{\sinh(v_k-v_m-h)\sinh(v_k+v_m-h)}{\sinh(v_k-v_m)\sinh(v_k+v_m)}\ket{\psi_k}\nonumber\\
 &{}+\sum_{k=1}^M f_2(u,v_k)\sinh^{2N}(v_k-h/2)\Delta(-v_k)\nonumber\\
 &\quad{}\times\prod_{m\neq k}\frac{\sinh(v_k-v_m+h)\sinh(v_k+v_m+h)}{\sinh(v_k-v_m)\sinh(v_k+v_m)}\ket{\psi_k}
\label{auxA}
\end{align}
and similarly from (\ref{comrelD}) and (\ref{DonUp})
\begin{align}
\bar{\mathcal{D}}(u)\ket{\{v_m\}} ={}& \bar{\mathcal{D}}(u)\mathcal{B}(v_1)\cdots\mathcal{B}(v_M)\ket{\Uparrow} \nonumber\\
 ={}& \sinh^{2N}(u-h/2)\Delta(-u)\prod_{m=1}^M\frac{\sinh(u-v_m+h)\sinh(u+v_m+h)}{\sinh(u-v_m)\sinh(u+v_m)}\ket{\{v_m\}}\nonumber\\
 &{} +\sum_{k=1}^M g_1(u,v_k)\sinh^{2N}(v_k-h/2)\Delta(-v_k)\nonumber\\
 &{}\quad\times\prod_{m\neq k}\frac{\sinh(v_k-v_m+h)\sinh(v_k+v_m+h)}{\sinh(v_k-v_m)\sinh(v_k+v_m)}\ket{\psi_k}\nonumber\\
 &{} +\sum_{k=1}^M g_2(u,v_k)\sinh^{2N}(v_k+h/2)\Delta(v_k)\nonumber\\
 &{}\quad\times\prod_{m\neq k}\frac{\sinh(v_k-v_m-h)\sinh(v_k+v_m-h)}{\sinh(v_k-v_m)\sinh(v_k+v_m)}\ket{\psi_k},
\label{auxD}
\end{align}
where
\begin{equation*}
\ket{\psi_k}:=\mathcal{B}(u)\prod_{m\neq k}\mathcal{B}(v_k)\ket{\Uparrow} .
\end{equation*}
Therefore, using (\ref{tmatu}),
\begin{equation*}
t_b(u)\ket{\{v_m\}}=\left(\frac{\sinh(2u+h)}{\sinh(2u)}\mathcal{A}(u)+\frac{\sinh(2u-h)}{\sinh(2u)}\bar{\mathcal{D}}(u)\right)\ket{\{v_m\}}=\Lambda_b(u)\ket{\{v_m\}}+\ket{\psi},
\end{equation*}
where
\begin{align}
\Lambda_b(\{v_m\};u)={}& \sinh^{2N}(u+h/2)\Delta(u)\frac{\sinh(2u+h)}{\sinh(2u)}\prod_{m=1}^M\frac{\sinh(u-v_m-h)\sinh(u+v_m-h)}{\sinh(u-v_m)\sinh(u+v_m)} \nonumber\\
&{}+ \sinh^{2N}(u-h/2)\Delta(-u)\frac{\sinh(2u-h)}{\sinh(2u)}\nonumber\\
&\quad{}\times\prod_{m=1}^M\frac{\sinh(u-v_m+h)\sinh(u+v_m+h)}{\sinh(u-v_m)\sinh(u+v_m)}
\label{eigenvalueBA}
\end{align}
and $\ket{\psi}$ is a linear combination of the vectors $\{\ket{\psi_k}\}_{1\leq k\leq M}$. Moreover, \eqref{auxA} and \eqref{auxD} together with~\eqref{lprop} imply that $\ket{\psi}$ vanishes if and only if $\{v_m\}_{1\leq m\leq M}$ satisfy the Bethe ansatz equations (BAE)
\begin{equation}
\label{BAEu}
\frac{\Delta(v_m)}{\Delta(-v_m)}\left(\frac{\sinh(v_m+h/2)}{\sinh(v_m-h/2)}\right)^{2N}=\prod_{\underset{k\neq m}{k=1}}^M\frac{\sinh(v_m-v_k+h)\sinh(v_m+v_k+h)}{\sinh(v_m-v_k-h)\sinh(v_m+v_k-h)}
\end{equation}
for all $1\leq m\leq M$. Thus $\ket{\{v_m\}}$ is an eigenvector of $t_b(u)$ with eigenvalue $\Lambda_b(\{v_m\};u)$ for any solution of~\eqref{BAEu}. Note the additional factor of $\Delta(v_m)/\Delta(-v_m)$ compared to the BAE of the usual open $\Uq$-invariant XXZ spin chain. It contains all the contribution of the new boundary coupling.

From~\eqref{Hbtm} and~\eqref{eigenvalueBA}, the energy corresponding to a solution $\{v_m\}_{1\leq i\leq M}$ is then given by
\begin{equation}
\label{energyu}
E_b(\{v_m\})=-\mu+\sum_{m=1}^M\frac{\sinh^2(h)}{\sinh(v_m-h/2)\sinh(v_m+h/2)} .
\end{equation}

It is also possible to introduce the variables $x:=\frac{\sinh(u+h/2)}{\sinh(u-h/2)}$ and $\lambda:=x+x^{-1}$ to rewrite the BAE~\eqref{BAEu} as (recall that $\delta:=[2]_\qa$ and $y:=\frac{[\alpha+1]_\qa}{[\alpha]_\qa}$)
\begin{align}
\frac{\lambda_m-\delta+\mu\big((\delta-y)x_m^{-1}-1\big)}{\lambda_m-\delta+\mu\big((\delta-y)x_m-1\big)}x_m^{2N} & =\prod_{\underset{k\neq m}{k=1}}^M\frac{\lambda_m+\lambda_k-\delta x_m(\lambda_k-\delta)-2\delta}{\lambda_m+\lambda_k-\delta x_m^{-1}(\lambda_k-\delta)-2\delta} \nonumber\\
& =\prod_{\underset{k\neq m}{k=1}}^M\frac{(1-\delta x_m+x_mx_k)\big(1-\delta x_k+x_kx_m^{-1}\big)}{\big(1-\delta x_m^{-1}+x_m^{-1}x_k\big)(1-\delta x_k+x_kx_m)}
\label{BAEblob}
\end{align}
for $1\leq m\leq M$. The associated energy~\eqref{energyu} is then simply
\begin{equation*}
E_b(\{\lambda_m\})=-\mu+\sum_{m=1}^M(\lambda_m-\delta) .
\end{equation*}
These equations were already derived in~\cite{Gier_2004} using coordinate Bethe ansatz.

A natural question is whether they ``completely'' describe the spectrum of $H_b$. To clarify what this means, we have to factor out the obvious redundancies of these equations. First, equations~\eqref{BAEu} as well as the corresponding eigenvalues~\eqref{eigenvalueBA} are invariant under permutations of the $v_k$, so we should consider solutions as unordered tuples $\{v_k\}_{1\leq k\leq M}$. This of course is just a~direct consequence of the commutation relation~\eqref{bcomrel} and of the definition of $\ket{\{v_m\}}$ in~\eqref{bansatz}. Second, note that if $\{v_k\}$ is a solution, then so are $\{v_1,\ldots,-v_l,\ldots,v_M\}$ and $\{v_1,\ldots,v_l+{\rm i}\pi r,\ldots,v_M\}$, $r\in\ZZ$, and with the same energy for any $1\leq l\leq M$. This means that we can look for non-zero solutions in the fundamental domain
\begin{equation}
\label{halfsp}
S_+ := \{ v_k\, |\, {\rm Re}~v_k\geq 0 , -\pi/2\leq{\rm Im}~v_k< \pi/2\}\backslash\{0\}
\end{equation}
for all $1\leq k\leq M$.\footnote{Solutions with $v_k=0$ for some $k$ have to be excluded too because they correspond to ``double roots'' of the BAE. For generic values of the parameters this situation does not occur.} Finally, note that if $\{v_1,\ldots,v_M\}$ is a solution then so is $\{v_1,\ldots,v_M,\infty\}$ and they both have the same energy. This is actually a consequence of the $\Uq$ symmetry. Indeed one can show that $\mathcal{B}(\infty)\propto\FF_{\Hb}$, either by direct computation or more easily using the construction of $R(u)$ from the universal $\RRR$-matrix of $\Uq$, as will be explained later on at the end of Section~\ref{alternative}. Since $t_b(u)$ commutes with $\Uq$, if $\ket{\{v_m\}}$ is an eigenvector then so is $\mathcal{B}(\infty)\ket{\{v_m\}}\propto\FF_{\Hb}\ket{\{v_m\}}$, and moreover it has the same energy $E_b$ because of (\ref{energyu}). Finite solutions $\{v_1,\ldots,v_M\}$ provide eigenstates, which we therefore expect to be $\Uq$ highest-weight vectors of weight $\qa^{\alpha-1+N-2M}$~\eqref{Kweight}.

For $\qa^\alpha\notin\pm\qa^\ZZ$, using the fusion rule~\eqref{fusrule} repeatedly, we know that the Hilbert space $\Hb$ decomposes into irreducible $\Uq$-modules as
\begin{equation*}
\Hb=\bigoplus_{M=0}^N{N\choose M}\VV_{\alpha+N-2M},
\end{equation*}
so there is exactly ${N\choose M}$ linearly independent highest-weight vectors in the $M$-magnon sector.\footnote{This can also be seen from the Schur--Weyl decomposition~\eqref{swb} and the dimensions of standard blob modules~\eqref{dimblob}.} Therefore, we conjecture that the system of $M$ equations $\eqref{BAEu}$ on an unordered set of $M$ complex numbers $\{v_k\}_{1\leq k\leq M}$, such that $v_k \in S_+$ for all $1\leq k\leq M$, has exactly ${N\choose M}$ distinct solutions, at least for generic values of $\alpha$ and $\mu$, and that the corresponding eigenvectors $\ket{\{v_m\}}$ are linearly independent and highest-weight for the $\Uq$ symmetry. Then an eigenbasis of $H_b$ is given by the vectors $\FF^k_{\Hb}\ket{\{v_m\}}$, $k\in\NN$. Establishing such statements is usually quite challenging and rigorous proofs are known only for very few integrable spin chains~\cite{Granet_2020a,Chernyak_2022,Granet_2020,MTV2, MTV}.

Let us also note that all the results above carry through mutatis mutandis to the root of unity cases without any obstacle.

\begin{Example}
When $M=1$, the BAE~\eqref{BAEblob} becomes
\begin{equation*}
\frac{x+x^{-1}-\delta+\mu((\delta-y)x-1)}{x+x^{-1}-\delta+\mu((\delta-y)x^{-1}-1)}=x^{2N} ,
\end{equation*}
which can be rewritten as
\begin{equation}
\label{waveqUq}
U_{N}(\lambda/2)-(\mu+\delta)U_{N-1}(\lambda/2)+(1+\mu(\delta-y))U_{N-2}(\lambda/2)=0 ,
\end{equation}
where $U_n$ is the $n$-th Chebyshev polynomial of the second kind. The corresponding eigenvalue of $H_b$ is
\begin{equation*}
E=-\mu+\lambda-\delta .
\end{equation*}
Equation~\eqref{waveqUq} has exactly ${N\choose 1}=N$ solutions as it should.

When $\delta=0$ ($\qa=i$), we recover the spectral equation from~\cite{previouspaper}. Note also that for $\delta=0$ the ``interaction term'' on the right-hand side of~\eqref{BAEblob} is always equal to unity\footnote{This simplification is of course not surprising as for $\delta=0$ the system reduces to free fermions.}, and so the BAE will just be~$M$ copies of the same equation~\eqref{waveqUq} \emph{for all} $M$. Denoting $(\lambda_i)_{1\leq i\leq N}$ the $N$ solutions of~\eqref{waveqUq} at $\delta=0$, any choice of $M$ pairwise distinct\footnote{This condition is needed to ensure that the BAE are not degenerate.} $\lambda_i$ will then be a solution of the BAE. Therefore, the eigenvalues of $H_b$ in the $M$-magnon sector are given by
\begin{equation*}
E_S=-\mu-\sum_{i\in S}\lambda_i
\end{equation*}
for all sets $S\subset \{1,\ldots, N\}$ of cardinality $|S|=M$, in accordance with the results of~\cite{previouspaper}. Since there are ${N\choose M}$ such sets $S$, this also means that the BAE~\eqref{BAEblob} have exactly ${N\choose M}$ solutions for this special value of $\delta$, as we conjectured.
\end{Example}

\subsection{The two-boundary system}
\label{twoBA}

We now turn to the two-boundary Hamiltonian $H_{2b}$. In principle, this case can also be treated using Sklyanin's boundary Bethe ansatz formalism. For this, in addition to the ``left'' $K$-matrix~\eqref{kmatrix} one also needs a ``right'' $K$-matrix satisfying an analogue of the bYBE~\eqref{BYBE} and implementing the desired integrable boundary conditions on the right. It is possible to find such a $K$-matrix by brute force but let us instead present an alternative and conceptually better approach. To this end, we first start by giving another construction of the left $K$-matrix~\eqref{kmatrix} and then extend this approach to the two-boundary case.

\subsubsection[An alternative construction of K(u)]{An alternative construction of $\boldsymbol{K(u)}$}
\label{alternative}

The basic idea is to put the boundary site carrying the $\VV_\alpha$ representation on an equal footing with the bulk sites, each of which carries a $\CC^2$ representation, by making apparent that their contribution to the transfer matrix~\eqref{tmatu} just comes from the same universal affine $R$-matrix, but evaluated in different representations of $\Uq$.\footnote{More precisely, in evaluation representations of the affine quantum group $U_\qa\widehat{\mathfrak{sl}}_2$ corresponding to different representations of $\Uq$.}

Computing such an evaluation in full generality is a hard task, complicated by the fact that we have to choose the correct gauge so that it is compatible with our conventions~\cite{Boos_2010}. However, if one of the factors of the affine $\RRR$-matrix is evaluated in the fundamental $\CC^2$-representation~-- as will always be the case in our construction~-- then there is a simpler procedure using the so-called ``baxterisation'' trick~\cite{Jones:1990hq}.

Concretely, recall that $\Uq$ admits a universal (non-affine) $R$-matrix given by~\cite{Drinfeld:1986in} (see also~\cite[Chapter~6.4]{qgroups})
\begin{equation}
\label{Rmatrix}
\RRR=\qa^{\frac{\HH\otimes\HH}{2}}\sum_{k\geq 0}\frac{\{1\}^{2k}}{\{k\}!}\qa^{k(k-1)/2}\EE^k\otimes \FF^k ,
\end{equation}
where
\begin{equation*}
\{n\}!:=\prod_{k=1}^n\{k\} .
\end{equation*}
Although strictly speaking $\RRR\notin\Uq\otimes\Uq$, it can be evaluated on the tensor product of any pair $(\XX, \YY)$ of representations of $\Uq$ as long as \emph{at least one} of them is finite-dimensional. We denote this evaluation by $\RRR_{\XX,\YY}$. One of the essential properties of $\RRR$ is that for any two such representations $\XX$ and $\YY$, the two operators
\begin{equation*}
P_{\XX,\YY}\circ\RRR_{\XX,\YY} \qquad \text{and} \qquad \RRR_{\YY,\XX}^{-1}\circ P_{\XX,\YY} ,
\end{equation*}
where
\begin{equation*}
\begin{aligned}
P_{\XX,\YY}\colon \ \XX\otimes \YY & \to \YY\otimes \XX,\\
 x\otimes y & \mapsto y\otimes x
\end{aligned}
\end{equation*}
is the operator permuting the two tensor factors, commute with the action of $\Uq$. In other words, $\RRR$ generates two (a priori different) $\Uq$-intertwiners between $\XX\otimes \YY$ and $\YY\otimes \XX$. These are precisely the building blocks we need to evaluate the affine $R$-matrix. Indeed, introducing for any representation $\XX$ of $\Uq$,
\begin{gather}
R_{\XX,\CC^2}(u) :=\dfrac{{\rm e}^{u+\frac{h}{2}}}{2}\RRR_{\XX,\CC^2}-\dfrac{{\rm e}^{-u-\frac{h}{2}}}{2}P_{\CC^2,\XX}\circ\RRR_{\CC^2,\XX}^{-1}\circ P_{\XX,\CC^2}, \nonumber\\
R_{\CC^2,\XX}(u) :=\dfrac{{\rm e}^{u+\frac{h}{2}}}{2}\RRR_{\CC^2,\XX}-\dfrac{{\rm e}^{-u-\frac{h}{2}}}{2}P_{\XX,\CC^2}\circ\RRR_{\XX,\CC^2}^{-1}\circ P_{\CC^2,\XX} ,
\label{Rgen}
\end{gather}
as well as
\begin{equation*}
\check{R}_{\XX,\CC^2}(u):=P_{\XX,\CC^2}\circ R_{\XX,\CC^2}(u) ,\qquad\check{R}_{\CC^2,\XX}(u):=P_{\CC^2,\XX}\circ R_{\CC^2,\XX}(u)
\end{equation*}
one checks that
\begin{itemize}\itemsep=0pt
\item[\emph{$(i)$}] $R_{\CC^2,\CC^2}(u)=R(u)$ from~\eqref{afrmat},
\item[\emph{$(ii)$}] $\check{R}_{\XX,\CC^2}(u)$ and $\check{R}_{\CC^2,\XX}(u)$ are $\Uq$-intertwiners,
\item[\emph{$(iii)$}] For any three representations $\XX_{1}$, $\XX_2$, $\XX_3$ of $\Uq$ with at least two of them isomorphic to~$\CC^2$ the generalisation of the YBE~\eqref{YBE}
\begin{equation}
\label{YBEX}
R_{\XX_1,\XX_2}(u-v)R_{\XX_1,\XX_3}(u)R_{\XX_2,\XX_3}(v)=R_{\XX_2,\XX_3}(v)R_{\XX_1,\XX_3}(u)R_{\XX_1,\XX_2}(u-v)
\end{equation}
is satisfied.
\end{itemize}
This means that $R_{\XX,\CC^2}(u)$ and $R_{\CC^2,\XX}(u)$ are precisely the evaluations of the universal affine $R$-matrix we are looking for. Note also that
\begin{equation}
\label{invgen}
R_{\CC^2,\XX}(u)R_{\XX,\CC^2}(-u)=\frac{1}{4}\left(\CCC_{\XX}-2\cosh(2u)\Id_{\XX}\right)\otimes\Id_{\CC^2},
\end{equation}
where $\CCC_{\XX}$ is the Casimir~\eqref{casimir} of $\XX$. In particular, for $\XX=\VV_\alpha$, by~\eqref{casimirva}
\begin{equation}
\label{invalpha}
R_{\CC^2,\VV_\alpha}(u)R_{\VV_\alpha,\CC^2}(-u)=\sinh(\frac{h\alpha}{2}+u)\sinh(\frac{h\alpha}{2}-u)\Id_{\XX}\otimes\Id_{\CC^2},
\end{equation}
so $R_{\CC^2,\VV_\alpha}(u)$ and $R_{\CC^2,\VV_\alpha}(u)$ are invertible for $u\neq\pm h\alpha/2$.

Now going back to the construction in~\eqref{trmat} and \eqref{trmatp}, we see that the simplest transfer matrix with an integrable $\Uq$-invariant boundary coupling to some $\Uq$-module $\XX$ one can construct is of the form
\begin{equation}
\label{trmatnew}
t_b(u)\propto\qtr_0{T_b(u-h/2)\hat{T}_b(u-h/2)}
\end{equation}
with
\begin{gather}
T_b(u) :=R_{0,N}(u)\cdots R_{0,1}(u)R_{0,\XX}(u-\zeta),\nonumber\\
\hat{T}_b(u) :=R_{\XX,0}(u+\zeta)R_{1,0}(u)\cdots R_{N,0}(u),
\label{trmatneww}
\end{gather}
where the index $i\in\{0,\ldots,N\}$ stands for the $i$-th $\CC^2$-site, and $\zeta\in\CC$ is some inhomogeneity parameter. Taking $\XX=\VV_\alpha$ and comparing~\eqref{trmat} with~\eqref{trmatnew}, we see that we should have
\begin{equation}
\label{kmatr}
K(u)\propto R_{0,\VV_\alpha}(u-\zeta)R_{\VV_\alpha,0}(u+\zeta) .
\end{equation}
From this form of the $K$-matrix one can easily show that it satisfies the bYBE~\eqref{BYBE} using the YBE~\eqref{YBEX} and~\eqref{idsol}. This is actually another instance of the general result~\cite[Proposition~2]{sklyanin1988boundary}, simply because the YBE~\eqref{YBEX} implies that $R_{0,\VV_\alpha}(u-\zeta)$ satisfies the RTT relation~\eqref{RTT}, $R_{\VV_\alpha,0}(u+\zeta)\propto R_{0,\VV_\alpha}(-u-\zeta)^{-1}$ by~\eqref{invalpha}, and $\Id_{\CC^2}$ is a solution of the bYBE~\eqref{BYBE} by~\eqref{idsol}.

It only remains to fix the normalisation factor in~\eqref{kmatr} and to express the coupling constant~$\mu$ in terms of the inhomogeneity $\zeta$. Computing~\eqref{kmatr} explicitly using~\eqref{Rmatrix} and \eqref{Rgen} and matching the result with~\eqref{kmatrix}, we obtain
\begin{equation}
\label{kmatbr1}
K(u)=\dfrac{R_{\CC^2,\VV_\alpha}(u-\zeta)R_{\VV_\alpha,\CC^2}(u+\zeta)}{\sinh(\frac{h\alpha}{2}-\zeta)\sinh(\frac{h\alpha}{2}+\zeta)}
\end{equation}
with
\begin{equation*}
\mu=\dfrac{\sinh(h)\sinh(h\alpha)}{\sinh(\zeta-\frac{h\alpha}{2})\sinh(\zeta+\frac{h\alpha}{2})} .
\end{equation*}

A fruitful consequence of this formalism is that we are now able to compute $\mathcal{B}(\infty)$ very easily. Indeed, define
\begin{equation}
\label{limdef}
\mathcal{T}(\infty)=
\begin{pmatrix}
\mathcal{A}(\infty) & \mathcal{B}(\infty)\\
\mathcal{C}(\infty) & \mathcal{D}(\infty)
\end{pmatrix}:=\lim_{u\to+\infty}{\rm e}^{-(2N+2)u}\mathcal{T}(u) .
\end{equation}
Then by~\eqref{Rgen} and~\eqref{trmatneww} this limit is finite and
\begin{equation}
\label{Rasymp}
\mathcal{T}(\infty)\propto\RRR_{0,N}\cdots\RRR_{0,1}\RRR_{0,\VV_\alpha}\RRR_{\VV_\alpha,0}\RRR_{1,0}\cdots\RRR_{N,0} .
\end{equation}
But for any representations $\XX$ and $\YY$ of $\Uq$
\begin{align}
\RRR_{\CC^2,\YY}\RRR_{\CC^2,\XX} & =
\begin{pmatrix}
\KK^{1/2}_\YY & \{1\}\KK^{1/2}_\YY\FF_\YY \\
0 & \KK^{-1/2}_\YY
\end{pmatrix}
\begin{pmatrix}
\KK^{1/2}_\XX & \{1\}\KK^{1/2}_\XX\FF_\XX \\
0 & \KK^{-1/2}_\XX
\end{pmatrix}\nonumber\\
& =\begin{pmatrix}
\KK^{1/2}_\XX\KK^{1/2}_\YY & \{1\}\KK^{1/2}_\XX\KK^{1/2}_\YY\left(\FF_\XX+\KK^{-1}_\XX\FF_\YY\right) \\
0 & \KK^{-1/2}_\XX\KK^{-1/2}_\YY
\end{pmatrix}\nonumber\\
& =
\begin{pmatrix}
\KK^{1/2}_{\XX\otimes\YY} & \{1\}\KK^{1/2}_{\XX\otimes\YY}\FF_{\XX\otimes\YY} \\
0 & \KK^{-1/2}_{\XX\otimes\YY}
\end{pmatrix}
=\RRR_{\CC^2,\XX\otimes\YY} ,
\label{rfus1}
\end{align}
where we used the coproduct formula~\eqref{coproduct}. Similarly,
\begin{equation}
\label{rfus2}
\RRR_{\XX,\CC^2}\RRR_{\YY,\CC^2}=\RRR_{\XX\otimes\YY,\CC^2} .
\end{equation}
Finally,
\begin{align}
\RRR_{\CC^2,\XX}\RRR_{\XX,\CC^2}&{}=
\begin{pmatrix}
\KK^{1/2}_\XX & \{1\}\KK^{1/2}_\XX\FF_\XX \\
0 & \KK^{-1/2}_\XX
\end{pmatrix}
\begin{pmatrix}
\KK^{1/2}_\XX & 0 \\
\{1\}\KK^{-1/2}_\XX\EE_\XX & \KK^{-1/2}_\XX
\end{pmatrix} \nonumber\\
&{}=
\begin{pmatrix}
\KK_\XX+\qam\{1\}^2\FF_\XX\EE_\XX & \qam\{1\}\FF_\XX\\
\{1\}\KK^{-1}\EE & \KK^{-1}
\end{pmatrix} ,
\label{RRrel}
\end{align}
where we used the defining relations of $\Uq$~\eqref{uqdefrel}. Therefore, using~\eqref{rfus1} and \eqref{rfus2} iteratively on~\eqref{Rasymp}, we have
\begin{equation*}
\mathcal{T}(\infty)\propto\RRR_{0,N}\cdots\RRR_{0,1}\RRR_{0,\VV_\alpha}\RRR_{\VV_\alpha,0}\RRR_{1,0}\cdots\RRR_{N,0}=\RRR_{0,\VV_\alpha\otimes(\CC^2)^{\otimes N}}\RRR_{\VV_\alpha\otimes(\CC^2)^{\otimes N},0}=\RRR_{0,\Hb}\RRR_{\Hb,0}
\end{equation*}
and so from~\eqref{RRrel} applied to $\XX=\Hb$, we obtain $\mathcal{B}(\infty)\propto\FF_{\Hb}$. The proportionality constant can easily be fixed but we will not need it. Note that the reasoning above applies to any integrable $\Uq$-invariant spin chain as long as we renormalise the corresponding monodromy $\mathcal{T}(u)$ by an appropriate power of ${\rm e}^{-u}$ in~\eqref{limdef} to make the $u\to+\infty$ limit finite.

The result that $\mathcal{B}(\infty)\propto\FF_{\Hb}$, which we have now established, was used in the arguments given towards the end of Section~\ref{oneBA}.

\subsubsection[Bethe ansatz for H\_\{2b\}]{Bethe ansatz for $\boldsymbol{H_{2b}}$}

From all the above, it is now clear how to proceed to construct the transfer matrix for the two-boundary system. We simply take
\begin{equation}
\label{t2b}
t_{2b}(u):=\dfrac{\qtr_0{T_{2b}(u-h/2)\hat{T}_{2b}(u-h/2)}}{\sinh(\frac{h\alpha_l}{2}-\zeta_l)\sinh(\frac{h\alpha_l}{2}+\zeta_l)\sinh(\frac{h\alpha_r}{2}-\zeta_r)\sinh(\frac{h\alpha_r}{2}+\zeta_r)}
\end{equation}
with
\begin{align}
&T_{2b}(u) :=R_{0,\VV_{\alpha_r}}(u-\zeta_r)R_{0,N}(u)\cdots R_{0,1}(u)R_{0,\VV_{\alpha_l}}(u-\zeta_l) ,\nonumber\\
&\hat{T}_{2b}(u) :=R_{\VV_{\alpha_l},0}(u+\zeta_l)R_{1,0}(u)\cdots R_{N,0}(u)R_{\VV_{\alpha_r},0}(u+\zeta_r) .\label{t2baux}
\end{align}
Using~\eqref{trmat}, \eqref{trmatp}, \eqref{invalpha}--\eqref{kmatbr1}, we have
\begin{equation*}
\begin{aligned}
\frac{\ddd}{\ddd u}\bigg\rvert_{u=\frac{h}{2}} t_{2b}(u) & = \frac{\ddd}{\ddd u}\bigg\rvert_{u=0} \frac{\qtr_0 R_{0,\VV_{\alpha_r}}(-\zeta_r) T(u)K(u)\hat{T}(u)R_{\VV_{\alpha_r},0}(\zeta_r)}{\sinh(\frac{h\alpha_r}{2}-\zeta_r)\sinh(\frac{h\alpha_r}{2}+\zeta_r)}\\
& \quad + \frac{\ddd}{\ddd u}\bigg\rvert_{u=0}\frac{\qtr_0 R_{0,\VV_{\alpha_r}}(u-\zeta_r)T(0)K(0)\hat{T}(0)R_{\VV_{\alpha_r},0}(u+\zeta_r)}{\sinh(\frac{h\alpha_r}{2}-\zeta_r)\sinh(\frac{h\alpha_r}{2}+\zeta_r)} \\
& =\frac{\ddd}{\ddd u}\bigg\rvert_{u=\frac{h}{2}} t_{b}(u)+4\sinh^{2N}(h)\coth(h)\left(-\mu_r b_r+\frac{\mu_r}{2}\right),
\end{aligned}
\end{equation*}
so (compare with~\eqref{Hbtm})
\begin{gather}
H_{2b}=-\mu_l b_l-\mu_r b_r-\sum_{i=1}^{N-1} e_i\nonumber\\
\hphantom{H_{2b}}{}=\frac{\tanh(h)}{4\sinh^{2N}(h)}\frac{\ddd}{\ddd u}\bigg\rvert_{u=\frac{h}{2}} t_{2b}(u)-N\cosh(h)
+\frac{1}{2\cosh(h)}-\frac{\mu_l+\mu_r}{2}\label{H2btm}
\end{gather}
with
\begin{equation}
\label{mulr}
\mu_{l/r}=\dfrac{\sinh(h)\sinh(h\alpha_{l/r})}{\sinh(\zeta_{l/r}-\frac{h\alpha_{l/r}}{2})\sinh(\zeta_{l/r}+\frac{h\alpha_{l/r}}{2})} .
\end{equation}

To find the BAE one does not need to redo all the computations of the previous section. Indeed, using the YBE~\eqref{YBEX} with appropriate choices of representations $\XX_{1}$, $\XX_2$, $\XX_3$ and spectral parameters $u$, $v$, we have
\begin{gather}
R_{0,\VV_{\alpha_r}}(u-\zeta_r)R_{0,i}(u)R_{\VV_{\alpha_r},i}(\zeta_r)=R_{\VV_{\alpha_r},i}(\zeta_r)R_{0,i}(u)R_{0,\VV_{\alpha_r}}(u-\zeta_r) ,\nonumber\\
R_{i,0}(u)R_{\VV_{\alpha_r},0}(u+\zeta_r)R_{\VV_{\alpha_r},i}(\zeta_r)=R_{\VV_{\alpha_r},i}(\zeta_r)R_{\VV_{\alpha_r},0}(u+\zeta_r)R_{i,0}(u) ,
\label{evYB}
\end{gather}
which we can use to bring the sites $\VV_{\alpha_l}$ and $\VV_{\alpha_r}$ together to the left by means of a similarity transformation. Concretely~\eqref{evYB} for $i=N$ implies that
\begin{equation*}
\begin{aligned}
T_{2b}(u)\hat{T}_{2b}(u)R_{\VV_{\alpha_r},N}(\zeta_r) & =R_{0,\VV_{\alpha_r}}(u-\zeta_r)R_{0,N}(u)\cdots R_{N,0}(u)R_{\VV_{\alpha_r},0}(u+\zeta_r)R_{\VV_{\alpha_r},N}(\zeta_r)\\
& =R_{0,\VV_{\alpha_r}}(u-\zeta_r)R_{0,N}(u)\cdots R_{\VV_{\alpha_r},N}(\zeta_r)R_{\VV_{\alpha_r},0}(u+\zeta_r)R_{N,0}(u)\\
& =R_{0,\VV_{\alpha_r}}(u-\zeta_r)R_{0,N}(u)R_{\VV_{\alpha_r},N}(\zeta_r)\cdots R_{\VV_{\alpha_r},0}(u+\zeta_r)R_{N,0}(u)\\
& =R_{\VV_{\alpha_r},N}(\zeta_r)R_{0,\VV_{\alpha_r}}(u-\zeta_r)R_{0,N}(u)\cdots R_{\VV_{\alpha_r},0}(u+\zeta_r)R_{N,0}(u),
\end{aligned}
\end{equation*}
so setting
\begin{equation*}
\mathcal{R}:=R_{\VV_{\alpha_r},N}(\zeta_r)R_{\VV_{\alpha_r},N-1}(\zeta_r)\cdots R_{\VV_{\alpha_r},2}(\zeta_r)R_{\VV_{\alpha_r},1}(\zeta_r),
\end{equation*}
we have\footnote{Note that by~\eqref{invalpha} $\mathcal{R}$ is invertible if and only if $\zeta_r\neq\pm h\alpha_r/2$, but this is just equivalent to requiring that $\mu_r$~\eqref{mulr} is finite.}
\begin{gather}
\mathcal{R}^{-1}T_{2b}(u)\hat{T}_{2b}(u)\mathcal{R} =T(u)R_{0,\VV_{\alpha_r}}(u-\zeta_r)R_{0,\VV_{\alpha_l}}(u-\zeta_l)R_{\VV_{\alpha_l},0}(u+\zeta_l)\nonumber\\
\hphantom{\mathcal{R}^{-1}T_{2b}(u)\hat{T}_{2b}(u)\mathcal{R} =}{}\times R_{\VV_{\alpha_r},0}(u+\zeta_r)\hat{T}(u)
\label{simtran}
\end{gather}
with $T(u)$, $\hat{T}(u)$ from~\eqref{trmatp}. Introducing the new $K$-matrix
\begin{equation}
\label{kmatr2b}
\tilde{K}(u)=\dfrac{R_{0,\VV_{\alpha_r}}(u-\zeta_r)R_{0,\VV_{\alpha_l}}(u-\zeta_l)R_{\VV_{\alpha_l},0}(u+\zeta_l)R_{\VV_{\alpha_r},0}(u+\zeta_r)}{\sinh\big(\frac{h\alpha_l}{2}-\zeta_l\big)\sinh\big(\frac{h\alpha_l}{2}+\zeta_l\big)\sinh\big(\frac{h\alpha_r}{2}-\zeta_r\big)\sinh\big(\frac{h\alpha_r}{2}+\zeta_r\big)}
\end{equation}
satisfying the bYBE~\eqref{BYBE} (again as a consequence of the YBE~\eqref{YBEX} and~\eqref{idsol} or~\cite[Proposition~2]{sklyanin1988boundary}), the corresponding monodromy
\begin{equation*}
\tilde{\mathcal{T}}(u):=T(u-h/2)\tilde{K}(u-h/2)\hat{T}(u-h/2) ,
\end{equation*}
and transfer matrix
\begin{equation*}
\tilde{t}_{2b}(u):=\qtr_0{\tilde{\mathcal{T}}(u)} ,
\end{equation*}
we have by~\eqref{simtran}
\begin{equation*}
t_{2b}(u)=\mathcal{R}\tilde{t}_{2b}(u)\mathcal{R}^{-1} ,
\end{equation*}
so $\tilde{t}_{2b}(u)$ has the same spectrum as $t_{2b}(u)$. But diagonalising $\tilde{t}_{2b}(u)$ is straightforward. Indeed, to implement ABA for the one-boundary system we only needed the commutation relations~\eqref{comrelA} and \eqref{comrelD} and the eigenvalues of $\mathcal{A}(u)$~\eqref{AonUp} and $\bar{\mathcal{D}}(u)$~\eqref{DonUp} when acting on the reference state~$\ket{\Uparrow}$. Replacing $K(u)$ by $\tilde{K}(u)$ does not change the commutation relations~\eqref{comrelA} and \eqref{comrelD} because they were solely derived from the bYBE~\eqref{BYBEexp} which the new monodromy $\tilde{\mathcal{T}}(u)$ equally satisfies as $\tilde{K}(u)$ is also a solution of the bYBE~\eqref{BYBE}. Therefore, we only need to compute the new eigenvalues of $\mathcal{A}(u)$ and $\bar{\mathcal{D}}(u)$ when acting on the new reference state\footnote{We use the same notations as for the one-boundary case for the entries of the new monodromy $\tilde{\mathcal{T}}(u)$ and the new reference state $\ket{0}\otimes\ket{0}\otimes\ket{\uparrow}^{\otimes N}$.}
\begin{equation*}
\ket{\Uparrow}:=\ket{0}\otimes\ket{0}\otimes\ket{\uparrow}^{\otimes N} .
\end{equation*}
This amounts to replacing the diagonal coefficients $a(u)$ and $d(u)$ of $K(u)$ in~\eqref{matform}--\eqref{hwaction} by the diagonal coefficients $\tilde{a}(u)$ and $\tilde{d}(u)$ of $\tilde{K}(u)$ (which are now operators acting on $\VV_{\alpha_l}\otimes\VV_{\alpha_r}$) in all the computations of Section~\ref{oneBA}.\footnote{The trick we just used is not essential to implement ABA in the two-boundary case and we could have equally worked with $t_{2b}(u)$ directly (see Appendix~\ref{genBAE}).} Doing so, we obtain, instead of~\eqref{AonUp} and \eqref{DonUp},
\begin{gather*}
\mathcal{A}(u)\ket{\Uparrow} = \sinh^{2N}(u+h/2)\Delta_l(u)\Delta_r(u)\ket{\Uparrow} , \\
\bar{\mathcal{D}}(u)\ket{\Uparrow} = \sinh^{2N}(u-h/2)\Delta_l(-u)\Delta_r(-u)\ket{\Uparrow} ,
\end{gather*}
while~\eqref{delta} is replaced by
\begin{align}
\Delta_{l/r}(u) & =1-\mu_{l/r}\frac{\sinh(u-h/2)\sinh(u+h(\alpha_{l/r}-1/2))}{\sinh(h)\sinh(h\alpha_{l/r})}\nonumber\\
& = \dfrac{\sinh(u+h\frac{\alpha_{l/r}-1}{2}-\zeta_{l/r})\sinh(u+h\frac{\alpha_{l/r}-1}{2}+\zeta_{l/r})}{\sinh(\frac{h\alpha_{l/r}}{2}-\zeta_{l/r})\sinh(\frac{h\alpha_{l/r}}{2}+\zeta_{l/r})} .\label{delta2}
\end{align}

Thus
\begin{equation*}
\ket{\{v_m\}}=\mathcal{B}(v_1)\cdots\mathcal{B}(v_M)\ket{\Uparrow}
\end{equation*}
is an eigenvector of $\tilde{t}_{2b}(u)$ -- or, equivalently, $\mathcal{R}\ket{\{v_m\}}$ is an eigenvector of $t_{2b}(u)$ -- with eigenvalue (compare with~\eqref{eigenvalueBA})
\begin{gather}
\Lambda_{2b}(\{v_m\};u)= \sinh^{2N}(u+h/2)\Delta_l(u)\Delta_r(u)\frac{\sinh(2u+h)}{\sinh(2u)}\nonumber\\
\hphantom{\Lambda_{2b}(\{v_m\};u)=+}{}\times\prod_{m=1}^M\frac{\sinh(u-v_m-h)\sinh(u+v_m-h)}{\sinh(u-v_m)\sinh(u+v_m)}\nonumber\\
\hphantom{\Lambda_{2b}(\{v_m\};u)=}{} + \sinh^{2N}(u-h/2)\Delta_l(-u)\Delta_r(-u)\frac{\sinh(2u-h)}{\sinh(2u)}\nonumber\\
\hphantom{\Lambda_{2b}(\{v_m\};u)=+}{}\times\prod_{m=1}^M\frac{\sinh(u-v_m+h)\sinh(u+v_m+h)}{\sinh(u-v_m)\sinh(u+v_m)}
\label{lambda2b}
\end{gather}
if and only if $\{v_m\}_{1\leq m\leq M}$ satisfy the Bethe ansatz equations (compare with~\eqref{BAEu})
\begin{gather}
\label{BAE2}
\frac{\Delta_l(v_m)\Delta_r(v_m)}{\Delta_l(-v_m)\Delta_r(-v_m)}\!\left(\frac{\sinh(v_m+h/2)}{\sinh(v_m-h/2)}\right)^{2N}\!\!=\!\prod_{\underset{k\neq m}{k=1}}^M\frac{\sinh(v_m-v_k+h)\sinh(v_m+v_k+h)}{\sinh(v_m-v_k-h)\sinh(v_m+v_k-h)}
\end{gather}
for all $1\leq m\leq M$. The corresponding eigenvalue of $H_{2b}$ is (compare with~\eqref{energyu})
\begin{equation*}
E_{2b}(\{v_m\})=-\mu_l-\mu_r+\sum_{m=1}^M\frac{\sinh^2(h)}{\sinh(v_m-h/2)\sinh(v_m+h/2)} .
\end{equation*}

Of course we could have guessed this result on physical grounds by interpreting $\Delta(u)/\Delta(-u)$ as the phase acquired by a quasi-particle of rapidity $u$ reflected at the boundary. For periodic integrable spin chains, this heuristic can actually be mathematically justified using the representation theory of the affine quantum group $U_\qa\widehat{\mathfrak{sl}}_2$ (the algebra defined by the RTT relation~\eqref{RTT}), establishing that the form of the eigenvalues of the transfer matrix and the BAE are completely fixed by the choice of a (trigonometric) Drinfeld polynomial, which uniquely specifies (up to isomorphism) an irreducible highest-weight representation of $U_\qa\widehat{\mathfrak{sl}}_2$ for the physical space~\cite{drinfeldpol}. This allows to implement ABA for any such choice of representation without almost any computation. To the best of our knowledge, a similar formalism has not been fully developed for open $\Uq$-invariant spin chains, especially for infinite-dimensional highest-weight representations, so for the sake of completeness we perform ABA for all open integrable $\Uq$-invariant highest-weight spin chains in Appendix~\ref{genBAE}. This also provides an alternative derivation of eigenvalues~\eqref{lambda2b} and BAE~\eqref{BAE2} which does not require $K$-matrices nor the similarity transformation~\eqref{simtran}.

Coming back to the two-boundary case, we have $\mathcal{B}(\infty)\propto\FF_{\Hbb}$,\footnote{As was mentioned at the end of Section~\ref{alternative}, this is true for any integrable $\Uq$-invariant spin chain. Note also that for ${\Hbb}$ we need to renormalise the monodromy $\mathcal{T}(u)$ by ${\rm e}^{-(2N+4)u}$ in~\eqref{limdef} to make the limit $u\to+\infty$ finite.} and we expect the finite (permutation invariant) solutions $\{v_k\}_{1\leq k\leq M}$ of the BAE~\eqref{BAE2} belonging to the fundamental domain $S_+$~\eqref{halfsp} to provide all the $\Uq$ highest-weight eigenstates of weight $\qa^{\alpha_l+\alpha_r-2+N-2M}$ of $H_{2b}$. Recalling the decomposition~\eqref{hbbdec}, valid for $\qa^{\alpha_{l/r}}, \qa^{\alpha_l+\alpha_r}\neq\pm \qa^\ZZ$,
\begin{equation*}
\Hbb=\bigoplus_{M\geq 0}\Hs_M\otimes\VV_{\alpha_l+\alpha_r-1+N-2M}
\end{equation*}
and the dimensions $d_M$~\eqref{multqgen}, there are $d_M$ such linearly independent vectors and so for generic values of the parameters we conjecture that the BAE~\eqref{BAE2} have $d_M$ such solutions. Note in particular that the number of magnons $M$ is not bounded as in the one-boundary case.

The BAE~\eqref{BAE2} are exactly the ones found in~\cite{cao2002exact,nep2003} for $\Hnd$ (\ref{hnd}) under the Nepomechie condition (\ref{nepcond}), which is not surprising: indeed, by Corollary~\ref{cor} (for $N$ even), $H_{2b}|_{\Hs_M}$ is an irreducible subblock of $\Hnd^{(M)}$ for $0\leq M\leq N-1$ and $H_{2b}|_{\Hs_M}=\Hnd^{(M)}$ for $M\geq N$. Note however that our derivation of the BAE is fully rigorous and relies on no additional assumptions. Also, we are now able to pinpoint \emph{the algebraic origin of the Nepomechie condition: it is just a direct consequence of the fusion rules~\eqref{fusrule} and more importantly~\eqref{vermafusrule}, yielding the $\Uq$-decomposition of $\Hbb$~\eqref{hbbdec} and restricting the generator $\YYY$~\eqref{Ygen} of the two-boundary TL algebra to take exactly the values $Y_M$~\eqref{ymlr} of the Nepomechie condition in its irreducible sectors $\Hs_M$.} Strictly speaking, this relation between $\Hnd$ and~$H_{2b}$ is only valid for generic values of $\qa$ and~$\alpha_{l/r}$, as our arguments are based on the Schur--Weyl duality from Theorem~\ref{thm}, proven under this assumption. However, by continuity of the (generalised) spectrum, and since the ``allowed" values of the parameters form a dense set, the \emph{spectral} equivalence between $\Hnd$ and (sectors of) $H_{2b}$ still holds for all $\qa,\alpha_{l/r}\in\CC$ even though their exact algebraic connection may be more involved in the non-generic non-semi-simple cases. An instance where ABA was applied to such a case can be found in~\cite{Gainutdinov_2016x}.

\subsubsection{Completeness}

Another interesting consequence of our formalism is the question of completeness of the BAE for $\Hnd$. For $M\geq N$, $H_{2b}|_{\Hs_M}=\Hnd^{(M)}$ so it is equivalent for both Hamiltonians. For $0\leq M\leq N-1$, however, $H_{2b}|_{\Hs_M}$ is only an irreducible subblock of $\Hnd^{(M)}$ by Corollary~\ref{cor}, but both Hamiltonians still have the same BAE. This means that for $0\leq M\leq N-1$, the BAE~\eqref{BAE2} cannot possibly provide all the eigenvalues of $\Hnd^{(M)}$, as is already quite clear for $M=0$. But thanks to Corollary~\ref{corr}, we know exactly which additional BAE we have to write to diagonalise the remaining block of $\Hnd^{(M)}$ for $0\leq M\leq N-1$: we simply need to replace $H_{2b}$ by $\overline{H}_{2b}$, which just amounts to the replacement (\ref{hamtr}), namely $\alpha_{l/r},\mu_{l/r}\to-\alpha_{l/r},-\mu_{l/r}$ up to an irrelevant additive constant, and by (\ref{ymlrtr}) the simultaneous replacement of the magnon number $M$ by the dual magnon number $\overline{M}:=N-M-1$. Thus (\ref{delta2}) is now replaced by
\begin{equation*}
\begin{aligned}
\bar{\Delta}_{l/r}(u) & =:1-\mu_{l/r}\frac{\sinh(u-h/2)\sinh(u-h(\alpha_{l/r}+1/2))}{\sinh(h)\sinh(h\alpha_{l/r})}\\
& = \dfrac{\sinh(u-h\frac{\alpha_{l/r}+1}{2}-\zeta_{l/r})\sinh(u-h\frac{\alpha_{l/r}+1}{2}+\zeta_{l/r})}{\sinh(\frac{h\alpha_{l/r}}{2}-\zeta_{l/r})\sinh(\frac{h\alpha_{l/r}}{2}+\zeta_{l/r})} ,
\end{aligned}
\end{equation*}
and we obtain the BAE
\begin{gather}
\label{dBAE}
\frac{\overline{\Delta}_l(v_m)\overline{\Delta}_r(v_m)}{\overline{\Delta}_l(-v_m)\overline{\Delta}_r(-v_m)}\!\left(\frac{\sinh(v_m+h/2)}{\sinh(v_m-h/2)}\right)^{2N}\!\!=\!\prod_{\underset{k\neq m}{k=1}}^{\overline{M}}\frac{\sinh(v_m-v_k+h)\sinh(v_m+v_k+h)}{\sinh(v_m-v_k-h)\sinh(v_m+v_k-h)}
\end{gather}
for $\overline{M}$ Bethe roots $\{v_m\}_{1\leq m\leq \overline{M}}$. Together with~\eqref{BAE2} we expect (\ref{dBAE}) to provide the complete spectrum of $\Hnd^{(M)}$ for $0\leq M\leq N-1$, as was previously conjectured in~\cite{Nepomechie_2003} (see also~\cite{Gier_2004}). Rigorously proving such a completeness statement is of course very hard and has only been achieved for a handful of integrable models~\cite{Chernyak_2022, MTV}. Still, our approach demonstrates that two sets of BAE are definitely needed and explains their algebraic origin. Note also that by Corollary~\ref{corr}\,$(iv)$,~\eqref{dBAE} are the BAE for $\Hnd^{(M)}$ with $M\leq -1$. Thus~\eqref{BAE2} and~\eqref{dBAE} taken together cover all the possible values of the parameters satisfying the Nepomechie constraint~\eqref{nepcond} for $M\in\ZZ$.

Finally, let us also mention that there exists a different set of BAE for the two-boundary which can be established using functional relations between various $Q$-functions~\cite{bookoff} or separation of variables~\cite{Kitanine_2014}. These equations are rather different from~\eqref{BAE2} and it is therefore quite surprising that they should yield the same spectrum. We discuss this question in more detail in Appendix~\ref{otherBAE}.

\begin{Remark}
If $\qa$ is a $2p$-th root of unity, the fusion rule~\eqref{vermafusrule} becomes
\begin{equation*}
\VV_{\alpha_l}\otimes\VV_{\alpha_r}\cong\bigoplus_{n=0}^{p-1} \VV_{\alpha_l+\alpha_r+p-1-2n},
\end{equation*}
so we have
\begin{equation*}
\Hbb=\bigoplus_{M=0}^{N+p-1}d_M'\VV_{\alpha_l+\alpha_r+p-1+N-2M}
\end{equation*}
with the magnon number $0\leq M\leq N+p-1$ now bounded and some different multiplicities~$d'_M$ whose explicit expression can be found in~\cite{previouspaper}. Thus we see that contrary to the one-boundary case where we expect the same number of solutions to the BAE~\eqref{BAEu} for generic and root of unity $\qa$, in the two-boundary case the BAE~\eqref{BAE2} should apparently behave quite differently in these two situations. This must also be related to the fact that the representation theory of $\twoBlob$ changes significantly at roots of unity and in particular the structure of the vacuum module $\WW_0$ -- which is still not fully known -- becomes much more complicated than for generic~$\qa$ as in Theorem~\ref{dgthm2} (see also \cite[Conjecture~2]{previouspaper}). This requires further study.
\end{Remark}

\section{Outlook}

In this paper, we have constructed a $\Uq$-invariant realisation $H_{2b}$ of the open XXZ Hamiltonian with non-diagonal boundary terms $\Hnd$ for all values of the parameters satisfying~\eqref{nepcond} by using the representation theory of the two-boundary Temperley--Lieb algebra $\twoBlob$. This enabled us to rigorously derive the BAE equations~\eqref{BAE2} for $\Hnd$ by ABA and to understand the algebraic origin of the Nepomechie condition~\eqref{nepcond} from the point of view of $\Uq$ fusion rules~\eqref{fusrule}--\eqref{vermafusrule}, restricting the possible values of the weight $Y$ of $\twoBlob$ to the discrete set $\{Y_M, M\in\ZZ\}$~\eqref{ymlr}.

Although the BAE we derived were previously known in the literature, a direct construction of the eigenstates by standard algebraic Bethe ansatz had never been performed until now and could be most useful in the computation of finer observables of the system, such as correlation functions and form factors, and in the study of some closely related models such as the asymmetric simple exclusion process (ASEP)~\cite{Cramp_2010, Simon_2009}. It is also worth mentioning that the algebraic Bethe ansatz formalism we presented generalises straightforwardly to open XXZ spin chains with additional inhomogeneity parameters at every site. Finally, our construction admits a well-defined $\qa\to 1$ limit, giving rise to non-compact boundary conditions for the open XXX spin chain.

It would be very interesting to construct an even more general $\Uq$-invariant spin chain which could reach arbitrary values of $Y$ and not just the discrete set subject to the Nepomechie condition~\eqref{nepcond}. The weights $y_{l/r}$ are entirely determined by the value of the Casimir $\CCC_{\VV_{\alpha_{l/r}}}$ and the possible values of $Y$ by the values $\CCC$ can take on the tensor product $\VV_{\alpha_l}\otimes\VV_{\alpha_r}$. Therefore, to find such a generalisation, one would need to construct two new one-parameter families of boundary $\Uq$-modules $\XX_{\alpha_{l/r}}$, with $\alpha_{l/r}$ parametrising $y_{l/r}$ through the Casimir $\CCC_{\XX_{\alpha_{l/r}}}$, and more crucially such that $\CCC$ can take any value on $\XX_{\alpha_l}\otimes\XX_{\alpha_r}$. This implies that the $\Uq$-decomposition of $\XX_{\alpha_l}\otimes\XX_{\alpha_r}$ should no longer be discrete as in~\eqref{vermafusrule} but continuous. For the $\qa\to 1$ XXX case, such fusion rules are known to arise in the context of principal series representations of $\mathsf{SL}(2,\CC)$~\cite{naimark}. For the general XXZ spin chain a natural guess would be to use their $\Uq$ $\qa$-deformed analogues. Such spin chains could shed light on the origin of the general BAE for $\Hnd$ (see Appendix~\ref{otherBAE}). Continuous fusion rules are also a central feature of Liouville CFT and the celebrated DOZZ formula for its 3-point function constants, which was recently proved~\cite{DOZZ}, including in the imaginary case~\cite{imDOZZ} where its relevance for lattice models has been established~\cite{Ikhlef_2016}. A well-defined lattice model with similar properties would certainly be of great help to further our understanding of the challenging questions still surrounding this theory. We will explore these ideas in future work.

It would be of course very desirable to extend our formalism to XXZ-type chains of spin-$1$ with non-diagonal integrable boundary conditions~\cite{IOZ96} and eventually to arbitrary spins~\cite{DN02,ML19}. This should reveal new lattice algebras generalising the two-boundary Temperley--Lieb algebra. One obvious guess would be here the fused Temperley--Lieb algebras, or fused Hecke algebras~\cite{CP21} for higher rank cases. Unfortunately, their boundary versions and the corresponding representation theory are poorly understood. A yet another interesting problem would be developing ABA in the non-semi-simple cases when the weights of the Verma modules take integer values, even at generic $\qa$, and the Hamiltonian is non-diagonalisable. This is similar to the problem studied in~\cite{Gainutdinov_2016x} but one should take limits in $\alpha_{l/r}$ variables instead of limits $\qa$ to a root of unity.

Finally, it is known that in the critical domain $|\qa|=1$, the boundary loop model defined by~$\mathbf{H}$~\eqref{Hbf} is conformally invariant in the large-$N$ scaling limit, with some explicit conjectures for its conformal spectrum based on the Coulomb gas approach~\cite{Dubail_2009,Jacobsen_2008} including for generalisations to anisotropic boundary conditions in the dilute $O(n)$ model~\cite{Dubail_2009O,Dubail_2010}. Using the explicit spin chain Hamiltonians $H_b$ and $H_{2b}$ and the corresponding BAE~\eqref{BAEu}--\eqref{BAE2} one can now hope to establish these results rigorously. This will be the subject of a forthcoming paper. Also, since from a physical perspective the two-boundary system can be seen as the fusion of two one-boundary systems, the representation theory of the discrete lattice algebras $\Blob$ and $\twoBlob$ may provide new insight into the fusion of Virasoro primary fields.

\appendix

\section{Proof of Theorem~\ref{thm}}
\label{proof}

The goal of this section is to prove the isomorphisms~\eqref{2bmodiso}. By Theorem~\ref{dgthm2}, this would imply that all the $\Hs_M$ appearing in the decomposition~\eqref{bimod2b} are irreducible representations of $\utwoBlob$ and therefore that $\Uq$ and $\utwoBlob$ are indeed mutual centralisers on $\Hbb$.

The main idea is to follow a more abstract approach to Schur--Weyl duality and rewrite the decomposition into irreducible $\Uq$-modules~\eqref{hbbdec} as
\begin{equation*}
\Hbb=\bigoplus_{M\geq 0}\Hom_{\Uq}\left(\VV_{\alpha_l+\alpha_r-1+N-2M},\Hbb\right)\otimes\VV_{\alpha_l+\alpha_r-1+N-2M},
\end{equation*}
where $\Hom_{\Uq}(\XX,\YY)$ denotes the space of $\Uq$-intertwiners between two $\Uq$-modules $\XX$ and~$\YY$. Of course $\Hom_{\Uq}\left(\VV_{\alpha_l+\alpha_r-1+N-2M},\Hbb\right)$ and $\Hs_M$ are isomorphic as vector spaces: a~$\Uq$-intertwiner $f\colon \VV_{\alpha_l+\alpha_r-1+N-2M}\to\Hbb$ is uniquely determined by the image of the highest-weight vector $\ket{0}$ of $\VV_{\alpha_l+\alpha_r-1+N-2M}$~\eqref{alpharepgen}, that is the choice of a highest-weight vector $f\ket{0}\in\Hbb$ of the same weight. Since the subspace of all such vectors is $\Hs_M$ by definition
\begin{equation}
\label{swiso}
\Hom_{\Uq}\left(\VV_{\alpha_l+\alpha_r-1+N-2M},\Hbb\right)\cong\Hom_\CC\left(\CC\ket{0},\Hs_M\right)\cong\Hs_M .
\end{equation}
Actually, this isomorphism is even an isomorphism of $\twoBlobm$-modules. Indeed, the $\Uq$-invariant action of $\twoBlobm$ on $\Hbb$ induces an action on the target space of intertwiners $f\in\Hom_{\Uq}\left(\VV_{\alpha_l+\alpha_r-1+N-2M},\Hbb\right)$ which is equivalent to the action $\twoBlobm$ on $\Hs_M$ via the isomorphism~\eqref{swiso}.

The advantage of this rewriting is that we can construct morphisms\footnote{We shall use the term ``morphism" instead of ``homomorphism" to lighten the exposition.} between standard $\twoBlobm$-modules and $\Hom_{\Uq}\left(\VV_{\alpha_l+\alpha_r-1+N-2M},\Hbb\right)$ in a much more canonical way. The idea is to use the diagrammatical calculus for $\Uq$-intertwiners to map well-chosen $\utwoBlob$-modules to some bigger spaces of $\Uq$-intertwiners in a way compatible with the lattice algebra action and then to appropriately specialise these maps to the spaces $\Hom_{\Uq}(\VV_{\alpha_l+\alpha_r-1+N-2M},\allowbreak\Hbb)$.

Let us first recall a few basic facts about this diagrammatical formalism and explain how it can be used to rederive Schur--Weyl duality for the simpler case of the TL algebra~\eqref{swtl}. To any TL diagram, that is a planar configuration of non-intersecting strings between two sets of~$N$ points, one can assign a $\Uq$-intertwiner from $\Hs:=(\CC^2)^{\otimes N}$ to itself, that is an element of $\End_{\Uq}(\Hs)$, by mapping the elementary generators of TL diagrams $e_i$, $1\leq i\leq N-1$, to the corresponding $\Uq$-intertwiners~\eqref{eiexp}. Since their composition rules are the same as the defining relations of the TL algebra~\eqref{TLrel}, this provides a $\Uq$-invariant representation $\rho_{\tl}\colon \TL\to\End_{\Uq}(\Hs)$.

More generally, let us denote $\WW_{\leq j}$ the vector space spanned by all planar configurations of non-intersecting strings between a set of $2j\leq N$ points (at the bottom) and a set of $N$ points (at the top). Although $\WW_{\leq j}$ is not an algebra, (except for $j=N/2$ where we recover the TL algebra), $\TL$ naturally acts on $\WW_{\leq j}$ by stacking TL diagrams on top of elements of $\WW_{\leq j}$. Moreover, using the elementary building blocks\footnote{There is a more abstract categorical construction of these morphisms in terms of evaluation and coevaluation maps which we will not present here (see~\cite{previouspaper} for more details).}
\begin{equation*}
\begin{tikzpicture}
\draw (0.3,0) node {$C_{i}=$};
\draw (1,0.4) [upup=1];
\draw (1,0.7) node {$i$};
\draw (2,0.7) node {$i+1$};
\draw (8,0) node {$=\qa^{1/2}\ket{\uparrow\downarrow}-\qa^{-1/2}\ket{\downarrow\uparrow}\in\Hom_{\Uq}\big(\CC,\CC^2\otimes\CC^2\big) ,\qquad 1\leq i\leq N-1 ,$};
\draw (0.3,-1) node {$\bar{C}_{i}=$};
\draw (1,-1.3) [dndn=1];
\draw (1,-1.5) node {$i$};
\draw (2,-1.5) node {$i+1$};
\draw (8,-1) node {$=\qa^{1/2}\bra{\uparrow\downarrow}-\qa^{-1/2}\bra{\downarrow\uparrow}\in\Hom_{\Uq}\big(\CC^2\otimes\CC^2,\CC\big) ,\qquad 1\leq i\leq N-1 ,$};
\end{tikzpicture}
\end{equation*}
any diagram of $\WW_{\leq j}$ can be mapped to an element of $\Hom_{\Uq}\big((\CC^2)^{\otimes 2j},\Hs\big)$. As these diagrams suggest, $C_{i}\bar{C}_{i}=e_i$, $\bar{C}_{i}C_{i}=\qa+\qam=\delta$ and more generally one can check that $C_{i}$, $\bar{C}_{i}$, $1\leq i\leq N-1$ satisfy all the natural diagrammatical rules inherited from the TL algebra. This means that the mapping
\begin{equation}
\label{Psijtl}
\Psi_j^{\tl}\colon \ \WW_{\leq j}\to\Hom_{\Uq}\big((\CC^2)^{\otimes 2j},\Hs\big)
\end{equation}
is a morphism of $\TL$-modules, where $\TL$ acts on the target space of intertwiners $f\in\Hom_{\Uq}\big((\CC^2)^{\otimes 2j},\Hs\big)$ via the representation map $\rho_\tl$.\footnote{One can also construct the morphisms $\Psi_j^{\tl}$ by first embedding $\WW_{\leq j}$ into $\TL$ by adding $N/2-j$ ``spectator" caps $\cap$ at the bottom of all diagrams, then mapping it to $\End_{\Uq}(\Hs)$ via $\rho_{\tl}$ and finally removing the spectator caps by precomposing with $\delta^{j-N/2}\Id_{(\CC^2)^{\otimes 2j}}\otimes\bigotimes_{i=j}^{N-1} C_{2i+1}$. This method will be used for the two-boundary case.}

Evaluating all $f\in\Hom_{\Uq}\big((\CC^2)^{\otimes 2j},\Hs\big)$ at $\ket{\uparrow}^{\otimes 2j}$, we obtain a morphism of $\TL$-modules
\begin{equation}
\label{psijtl}
\psi_j^{\tl}\colon \ \WW_{\leq j}\to\Hs^{\tl}_j ,\qquad \ell\mapsto\Psi_j^{\tl}(\ell)\ket{\uparrow}^{\otimes 2j},
\end{equation}
where $\Hs^{\tl}_j\subset\Hs$ is the subspace of highest-weight vectors of weight $\qa^{2j}$. Indeed, since $\Psi_j^{\tl}(\ell)\in\Hom_{\Uq}\big((\CC^2)^{\otimes 2j},\Hs\big)$
\begin{gather}
\EE_{\Hs}\psi_j^{\tl}(\ell) =\EE_{\Hs}\Psi_j^{\tl}(\ell)\ket{\uparrow}^{\otimes 2j}=\Psi_j^{\tl}(\ell)\EE_{(\CC^2)^{\otimes 2j}}\ket{\uparrow}^{\otimes 2j}=0,\nonumber\\
\KK_{\Hs}\psi_j^{\tl}(\ell) =\KK_{\Hs}\Psi_j^{\tl}(\ell)\ket{\uparrow}^{\otimes 2j}=\Psi_j^{\tl}(\ell)\KK_{(\CC^2)^{\otimes 2j}}\ket{\uparrow}^{\otimes 2j}=\qa^{2j}\psi_j^{\tl}(\ell)\label{tlhw}
\end{gather}
for all $\ell\in\WW_{\leq j}$ so $\Im\psi_j^{\tl}\subset\Hs^{\tl}_j$. Equivalently, $\psi_j^{\tl}$ can be seen as a map from $\WW_{\leq j}$ to $\Hom_{\Uq}\left(\CC^{2j+1},\Hs\right)\cong\Hs^{\tl}_j$ obtained by precomposing $\Psi_j^{\tl}(\ell)\in\Hom_{\Uq}\big((\CC^2)^{\otimes 2j},\Hs\big)$ by the unique (up to normalisation) $\Uq$-intertwiner from the spin-$j$ representation $\CC^{2j+1}$ of $\Uq$ to~$(\CC^2)^{\otimes 2j}$.

Finally, $\WW_{\leq j}$ contains a stable $\TL$-subspace $\WW_{<j}$ spanned by all diagrams of $\WW_{\leq j}$ containing strictly less than $2j$ through lines (recall that the action of the TL algebra can only decrease the number of through lines). By definition, the quotient $\WW_{\leq j}/\WW_{<j}$ is isomorphic, as a $\TL$-module, to the standard module $\WW_j$. For all $\ell\in\WW_{<j}$, $\psi_j^{\tl}(\ell)$ contains one or more caps $\cap$ linking bottom points and moreover at least one of these caps has to connect two neighbouring sites. This means that $\Psi_j^{\tl}(\ell)$ is of the form $A\otimes\bar{C}_i$ for some $1\leq i\leq 2j-1$ and some $A\in\Hom_{\Uq}\big((\CC^2)^{\otimes 2(j-1)},\Hs\big)$. Since $\bar{C}_i\ket{\uparrow\uparrow}=0$, $\psi_j^{\tl}(\ell)=0$, and so $\ell\in\Ker \psi_j^{\tl}$. Therefore, $\WW_{<j}\subseteq\Ker \psi_j^{\tl}$ which implies that $\psi_j^{\tl}$ induces a morphism of $\TL$-modules
\begin{equation}
\label{psijtltilde}
\tilde{\psi}_{j}^{\tl}\colon \ \WW_{\leq j}/\WW_{<j}\cong\WW_j\to\Hs^{\tl}_j .
\end{equation}
It is clearly non-zero as
\begin{equation*}
\begin{tikzpicture}[scale=0.5]
\draw (3,0.4) [upup=1];
\draw (5,0) node{$\ldots$};
\draw (6,0.4) [upup=1];
\draw (0,0.4) [hdnup=0];
\draw (1,0) node{$\ldots$};
\draw (2,0.4) [hdnup=0];
\draw (12.5,0) node {$\mapsto~\ket{\uparrow}^{\otimes 2j}\otimes\bigotimes_{i=j}^{N-1} C_{2i+1}\neq 0$};
\end{tikzpicture}
\end{equation*}
and since $\WW_j$ is irreducible $\tilde{\psi}_{j}^{\tl}$ must be injective. But by~\eqref{tlmoddim}, $\dim\WW_j=\dim\Hs^{\tl}_j$ so $\Hs^{\tl}_j\cong\WW_j$. This proves Schur--Weyl duality~\eqref{swtl} between the actions of $\TL$ and $\Uq$ on $\Hs:=(\CC^2)^{\otimes N}$.

We would now like to extend this formalism to the actions of $\utwoBlob$ and $\Uq$ on $\Hbb:=\VV_{\alpha_l}\otimes(\CC^2)^{\otimes N}\otimes\VV_{\alpha_r}$, in particular to construct (universal) two-boundary analogues of the morphisms $\Psi_j^{\tl}$~\eqref{Psijtl}, $\psi_j^{\tl}$~\eqref{psijtl} and $\tilde{\psi}_{j}^{\tl}$~\eqref{psijtltilde}.

Let us define $\WW_{\leq j}^{2b}$, the vector space spanned by all two-boundary TL diagrams from $2j\leq N$ points (at the bottom) to $N$ points (at the top), that is TL diagrams of $\WW_{\leq j}$ decorated by left/right blobs/anti-blobs in all admissible ways. $\twoBlob$ naturally acts on $\WW_{\leq j}^{2b}$ by stacking two-boundary TL diagrams on top of elements of $\WW_{\leq j}^{2b}$. Since we will be working with the universal two-boundary TL algebra $\utwoBlob$, we need to slightly extend this definition by promoting $\WW_{\leq j}^{2b}$ to a free $\CC[\YYY]$-module $\WW_{\leq j}^{2b}[\YYY]$ of rank $\dim\WW_{\leq j}^{2b}$, generated by the basis elements of $\WW_{\leq j}^{2b}$. Then $\WW_{\leq j}^{2b}[\YYY]$ admits an action of $\utwoBlob$, where the additional $\YYY$ generator simply acts by multiplication.

We would 
like to define a morphism of $\utwoBlob$-modules $\Psi_j$ from $\WW_{\leq j}^{2b}[\YYY]$ to $\Hom_{\Uq}\big(\VV_{\alpha_l}\otimes(\CC^2)^{\otimes 2j}\otimes\VV_{\alpha_r},\Hbb\big)$. For $j=N/2$ this is straightforward: $\WW_{\leq N/2}^{2b}[\YYY]$ and $\utwoBlob$ are isomorphic as (left) $\utwoBlob$-modules and so we can just take the representation map
\begin{equation*}
\rho_{\Hbb}\colon \ \utwoBlob\to\End_{\Uq}(\Hbb)
\end{equation*}
constructed in~\cite{previouspaper}. To build $\Psi_j$ for $0\leq j\leq N/2-1$, let us take advantage of this simpler case by embedding $\WW_{\leq j}^{2b}[\YYY]$ into $\utwoBlob$.

Define $\II_j$ the subspace of all two-boundary TL diagrams $\ell$ of the form $\ell'\otimes \cap^{\otimes (N/2-j)}$ where $\ell'\in\WW_{\leq j}^{2b}$ is a two-boundary TL diagram from $2j$ points labelled $\{1,\ldots,2j\}$ (at the bottom) to $N$ points (at the top) and $\cap^{\otimes (N/2-j)}$ are $N/2-j$ caps linking the remaining pairs of points $(2j+1, 2j+2)$, $(2j+3, 2j+4),\dots , (N-1,N)$ at the bottom. For example, for $N=4$ and $j=1$
\begin{equation*}
\begin{tikzpicture}[scale=0.75]
\draw (-0.9,0) node{$\ell'=$};
\draw (0,0.4) [hdnup=0];
\draw (0,0) node[scale=1.5]{$\circ$};
\draw (1,0.4) [upup=1];
\draw (3,0.4) [hdnup=0];
\draw (3,0) node[scale=0.75]{$\blacksquare$};
\draw (4,0) node{$\mapsto$};
\draw (5,-1) [dnup=0];
\draw (5,0) node[scale=1.5]{$\circ$};
\draw (6,1) [upup=1];
\draw (6,-1) [dnup=2];
\draw (7,0) node[scale=0.75]{$\blacksquare$};
\draw (7,-1) [dndn=1];
\draw (10,0) node{$=\ell=\ell'\otimes\cap$\,.};
\draw (0,-0.75) node{$1$};
\draw (3,-0.75) node{$2$};
\draw (5,-1.5) node{$1$};
\draw (6,-1.5) node{$2$};
\draw (7,-1.5) node{$3$};
\draw (8,-1.5) node{$4$};
\end{tikzpicture}
\end{equation*}
Clearly
\begin{equation*}
\{0\}\subsetneq\II_0\subsetneq\II_1\subsetneq\cdots\subsetneq\II_{N/2}:=\twoBlob
\end{equation*}
and since the left action of $\twoBlob$ on $\II_j$ preserves the $\cap^{\otimes (N/2-j)}$ part, $\II_j$ is a left $\twoBlob$ ideal. We can extend this construction to the universal two-boundary TL algebra $\utwoBlob$ by promoting $\II_j$ to a free $\CC[\YYY]$-module $\II_j[\YYY]$ of rank $\dim\II_j$ generated by the basis diagrams of $\II_j$. Then
\begin{equation*}
\{0\}\subsetneq\II_0[\YYY]\subsetneq\II_1[\YYY]\subsetneq\cdots\subsetneq\II_{N/2}[\YYY]:=\utwoBlob
\end{equation*}
is an increasing sequence of ideals of $\utwoBlob$. Obviously, $\WW_{\leq j}^{2b}[\YYY]$ and $\II_j[\YYY]$ are isomorphic as (left) $\utwoBlob$-modules via the map $\ell'\mapsto\ell:=\ell'\otimes\cap^{\otimes (N/2-j)}$ so we can (and will) identify them.

Now the restriction map
\begin{equation*}
\rho_{\Hbb}|_{\II_j[\YYY]} \colon \ \II_j[\YYY]\to\End_{\Uq}(\Hbb)
\end{equation*}
is a $\utwoBlob$-module morphism (because $\rho_{\Hbb}$ is a representation of $\utwoBlob$ on $\Hbb$) and its image consists of elements of $\End_{\Uq}(\Hbb)$ of the form $A{\otimes \bar{C}_{2j+1}\otimes\cdots\otimes\bar{C}_{N-1}}$ with $A\in\Hom_{\Uq}\big(\VV_{\alpha_l}\otimes(\CC^2)^{\otimes 2j}\otimes\VV_{\alpha_r},\Hbb\big)$. Let us define the map
\begin{align*}
\pi_j \colon \ &\Im \big(\rho_{\Hbb}|_{\II_j[\YYY]}\big)\to \Hom_{\Uq}\big(\VV_{\alpha_l}\otimes(\CC^2)^{\otimes 2j}\otimes\VV_{\alpha_r},\Hbb\big) ,\\
& A\otimes \bar{C}_{2j+1}\otimes\cdots\otimes\bar{C}_{N-1}\mapsto A .
\end{align*}
Concretely, using the fact that $\bar{C}_iC_i=\delta\neq 0$,\footnote{Recall that $\delta=0$ corresponds to $\qa=i$, whereas we have assumed that $\qa$ is not a root of unity.}
\begin{equation*}
\pi_j(X)=\delta^{j-N/2}X\circ\bigl(\Id_{\VV_{\alpha_l}\otimes(\CC^2)^{\otimes 2j}}\otimes C_{2j+1}\otimes\cdots\otimes C_{N-1}\otimes\Id_{\VV_{\alpha_r}}\bigr)
\end{equation*}
for all $X\in\Im \big(\rho_{\Hbb}|_{\II_j[\YYY]}\big)$, from which it is clear that
\begin{itemize}\itemsep=0pt
\item $\pi_j$ is well-defined,
\item $\pi_j(X)$ is a $\Uq$-intertwiner as the composition of two $\Uq$-intertwiners,
\item $\pi_j$ is a $\utwoBlob$-module morphism as $\utwoBlob$ acts only on the target space of the $\Uq$-intertwiner~$X$.
\end{itemize}
Thus, for all $0\leq j\leq N/2$, we have constructed a $\utwoBlob$-module morphism
\begin{equation}
\label{Psij2b}
\Psi_j:=\pi_j\circ\rho_{\Hbb}|_{\II_j[\YYY]} \colon \ \WW_{\leq j}^{2b}[\YYY]\cong\II_j[\YYY]\to \Hom_{\Uq}\big(\VV_{\alpha_l}\otimes(\CC^2)^{\otimes 2j}\otimes\VV_{\alpha_r},\Hbb\big)
\end{equation}
as we wanted.

This abstractly-defined map $\Psi_j$ has actually a very natural diagrammatical interpretation. Indeed, two-boundary TL diagrams of $\WW_{\leq j}^{2b}$ are just decorated TL diagrams of $\WW_{\leq j}$ so constructing $\Psi_j$ amounts to implementing these decorations in terms of $\Uq$-intertwiners (and mapping~$\YYY$ to~\eqref{Ygen}). Concretely, one can introduce the diagrams
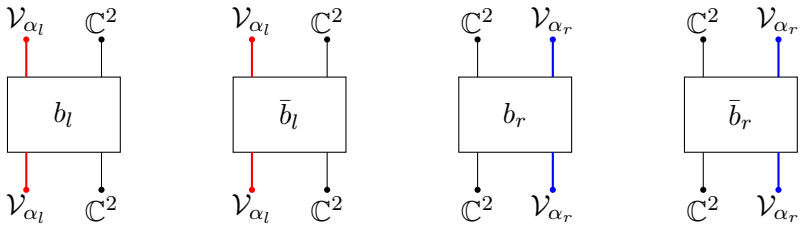
\begin{equation*}
\begin{tikzpicture}
\draw (0.25,0) rectangle (1.75,-1);
\draw[red, thick] (0.5,0) -- (0.5, 0.5);
\draw[red, thick] (0.5,-1) -- (0.5,-1.5);
\draw (1.5,0) -- (1.5, 0.5);
\draw (1.5,-1) -- (1.5,-1.5);
\draw (1.5, 0.5) node[scale=2]{.};
\draw (1.5,-1.5) node[scale=2]{.};
\draw[red] (0.5, 0.5) node[scale=2]{.};
\draw[red] (0.5,-1.5) node[scale=2]{.};
\draw (1, -0.5) node{$b_l$};
\draw (1.5, 0.75) node{$\CC^2$};
\draw (1.5,-1.75) node{$\CC^2$};
\draw (0.5, 0.75) node{$\VV_{\alpha_l}$};
\draw (0.5,-1.75) node{$\VV_{\alpha_l}$};

\draw (3.25,0) rectangle (4.75,-1);
\draw[red, thick] (3.5,0) -- (3.5, 0.5);
\draw[red, thick] (3.5,-1) -- (3.5,-1.5);
\draw (4.5,0) -- (4.5, 0.5);
\draw (4.5,-1) -- (4.5,-1.5);
\draw (4.5, 0.5) node[scale=2]{.};
\draw (4.5,-1.5) node[scale=2]{.};
\draw[red] (3.5, 0.5) node[scale=2]{.};
\draw[red] (3.5,-1.5) node[scale=2]{.};
\draw (4, -0.5) node{$\bar{b}_l$};
\draw (4.5, 0.75) node{$\CC^2$};
\draw (4.5,-1.75) node{$\CC^2$};
\draw (3.5, 0.75) node{$\VV_{\alpha_l}$};
\draw (3.5,-1.75) node{$\VV_{\alpha_l}$};

\draw (6.25,0) rectangle (7.75,-1);
\draw (6.5,0) -- (6.5, 0.5);
\draw (6.5,-1) -- (6.5,-1.5);
\draw[blue, thick] (7.5,0) -- (7.5, 0.5);
\draw[blue, thick] (7.5,-1) -- (7.5,-1.5);
\draw[blue] (7.5, 0.5) node[scale=2]{.};
\draw[blue] (7.5,-1.5) node[scale=2]{.};
\draw (6.5, 0.5) node[scale=2]{.};
\draw (6.5,-1.5) node[scale=2]{.};
\draw (7, -0.5) node{$b_r$};
\draw (6.5, 0.75) node{$\CC^2$};
\draw (6.5,-1.75) node{$\CC^2$};
\draw (7.5, 0.75) node{$\VV_{\alpha_r}$};
\draw (7.5,-1.75) node{$\VV_{\alpha_r}$};

\draw (9.25,0) rectangle (10.75,-1);
\draw (9.5,0) -- (9.5, 0.5);
\draw (9.5,-1) -- (9.5,-1.5);
\draw[blue, thick] (10.5,0) -- (10.5, 0.5);
\draw[blue, thick] (10.5,-1) -- (10.5,-1.5);
\draw[blue] (10.5, 0.5) node[scale=2]{.};
\draw[blue] (10.5,-1.5) node[scale=2]{.};
\draw (9.5, 0.5) node[scale=2]{.};
\draw (9.5,-1.5) node[scale=2]{.};
\draw (10, -0.5) node{$\bar{b}_r$};
\draw (9.5, 0.75) node{$\CC^2$};
\draw (9.5,-1.75) node{$\CC^2$};
\draw (10.5, 0.75) node{$\VV_{\alpha_r}$};
\draw (10.5,-1.75) node{$\VV_{\alpha_r}$};
\end{tikzpicture}
\end{equation*}
and decorate the TL of strings a $\Uq$-intertwiner belonging to $\Hom_{\Uq}\big((\CC^2)^{\otimes 2j},\Hs\big)$ by deforming them until they are in contact with the left (red line) or right (blue line) boundary and then inserting the diagrams $b_{l/r}$, $\bar{b}_{l/r}$ above to obtain a two-boundary $\Uq$-intertwiner belonging to $\Hom_{\Uq}\big(\VV_{\alpha_l}\otimes(\CC^2)^{\otimes 2j}\otimes\VV_{\alpha_r},\Hbb\big)$. Of course such a deformation is only possible if one never intersects any other string while doing so. But this is consistent with the decoration rules for two-boundary diagrams because a string can acquire a left/right blob/anti-blob only when touching the left/right boundary.

Using the above procedure, we can uniquely map diagrams of $\WW_{\leq j}^{2b}$ to $\Hom_{\Uq}\big(\VV_{\alpha_l}\otimes(\CC^2)^{\otimes 2j}\otimes\VV_{\alpha_r},\Hbb\big)$. For example, for $N=4$ and $j=0$,
\begin{equation*}
\begin{tikzpicture}
\draw (1,1.5) [upup=1];
\draw (3,1.5) [upup=1];
\draw (1.5,0.75) node[scale=2]{$\circ$};
\draw (3.25,0.8) node[scale=2]{$\bullet$};
\draw (3.75,0.8) node{$\blacksquare$};

\draw (4.6,1) node{$\mapsto$};

\draw (5.25,0) rectangle (6.75,-1);
\draw (5.25,3) rectangle (6.75,2);
\draw (6.5,2) [upup=1];
\draw[red, thick] (5.5,3) -- (5.5, 3.5);
\draw[red] (5.5, 3.5) node[scale=2]{.};
\draw (6.5,3) -- (6.5, 3.5);
\draw (6.5, 3.5) node[scale=2]{.};
\draw (7.5,2) -- (7.5, 3.5);
\draw (7.5, 3.5) node[scale=2]{.};
\draw (8.5,0) -- (8.5, 3.5);
\draw (8.5, 3.5) node[scale=2]{.};
\draw[red, thick] (5.5,0) -- (5.5, 2);
\draw[red, thick] (5.5,-1) -- (5.5,-1.5);
\draw (6.5,-1) [upup=3];
\draw (6.5,0) [dndn=1];
\draw (7.5,0) [upup=1];
\draw[red] (5.5,-1.5) node[scale=2]{.};
\draw (6, -0.5) node{$b_l$};
\draw (6, 2.5) node{$\bar{b}_l$};

\draw (9.25,0) rectangle (10.75,-1);
\draw (10, -0.5) node{$b_r$};
\draw[blue, thick] (10.5,-1) -- (10.5,-1.5);
\draw[blue] (10.5,-1.5) node[scale=2]{.};
\draw[blue, thick] (10.5,0) -- (10.5,3.5);
\draw (9.5,0) -- (9.5,3.5);
\draw (9.5,3.5) node[scale=2]{.};
\draw[blue] (10.5,3.5) node[scale=2]{.};

\draw (10.5,-1.75) node{$\VV_{\alpha_r}$};
\draw (5.5,-1.75) node{$\VV_{\alpha_l}$};
\draw (10.5, 3.75) node{$\VV_{\alpha_r}$};
\draw (5.5, 3.75) node{$\VV_{\alpha_l}$};
\draw (6.5, 3.75) node{$\CC^2$};
\draw (7.5,3.75) node{$\CC^2$};
\draw (8.5, 3.75) node{$\CC^2$};
\draw (9.5,3.75) node{$\CC^2$};

\draw (13.15,1) node{$ =\bar{b}_l (C_1\bar{C}_1)(b_lC_2\bar{C}_2b_r)(C_1C_3) .$};

\end{tikzpicture}
\end{equation*}
For all $\ell\in\WW_{\leq j}^{2b}$, it is technically possible to write an explicit expression of the corresponding $\Uq$-intertwiner $\Psi_j(\ell)\in\Hom_{\Uq}\big(\VV_{\alpha_l}\otimes(\CC^2)^{\otimes 2j}\otimes\VV_{\alpha_r},\Hbb\big)$ in terms of the $C_i$, $\bar{C}_i$, ${1\leq i\leq N-1}$, $b_{l/r}$ and $\bar{b}_{l/r}$, but the shortcut we took by adding the ``spectator" part $\cap^{\otimes N/2-j}$ to embed them into $\utwoBlob$ and mapping them to intertwiners using the representation map~$\rho_{\Hbb}$ achieves the same goal faster while also providing a direct proof that $\Psi_j$ commutes with the action of $\utwoBlob$.

Coming back to our proof, let us first consider the morphism of $\utwoBlob$-modules
\begin{equation*}
\Psi_0 \colon \ \WW_{\leq 0}^{2b}[\YYY]\cong\II_0[\YYY]\to\Hom_{\Uq}(\VV_{\alpha_l}\otimes\VV_{\alpha_r},\Hbb) .
\end{equation*}
Note that, by definition, $\WW_{\leq 0}^{2b}[\YYY]=\WW_0[\YYY]$, the universal vacuum module of $\utwoBlob$.

Recalling the fusion rule~\eqref{vermafusrule}, rewritten in a convenient way,
\begin{equation}
\label{vermafusruleM}
\VV_{\alpha_l}\otimes\VV_{\alpha_r}\cong\bigoplus_{M\geq N/2} \VV_{\alpha_l+\alpha_r-1+N-2M},
\end{equation}
we have maps $\varphi_M\in\Hom_{\Uq}\left(\VV_{\alpha_l+\alpha_r-1+N-2M},\VV_{\alpha_l}\otimes\VV_{\alpha_r}\right)$, $M\geq N/2$, which are unique up to normalisation. They induce maps
\begin{equation*}
\Psi_0^{(M)} \colon \ \WW_0[\YYY]\to\Hom_{\Uq}(\VV_{\alpha_l+\alpha_r-1+N-2M},\Hbb) , \qquad \ell\mapsto\Psi_0(\ell)\circ\varphi_M,
\end{equation*}
which are also morphisms of $\utwoBlob$-modules. Moreover, for any $f\!\in\!\Hom_{\Uq}\!(\VV_{\alpha_l+\alpha_r-1+N-2M},\allowbreak\Hbb)$, $M\geq 0$,
\begin{equation}
\label{Ycom}
\YYY_{\Hbb}f(-)=f(\YYY_{\VV_{\alpha_l+\alpha_r-1+N-2M}}-)=Y_{M}f(-) .
\end{equation}
In other words, $\YYY$ acts as the scalar $Y_M$ on $\Hom_{\Uq}\left(\VV_{\alpha_l+\alpha_r-1+N-2M},\Hbb\right)$ which means that ${(\YYY-Y_M)\WW_0[\YYY]\subseteq\Ker \Psi_0^{(M)}}$ and so, for all $M\geq N/2$, $\Psi_0^{(M)}$ induces a morphism of $\twoBlobm$-modules between $\WW_0[\YYY]/(\YYY-Y_M)\WW_0[\YYY]\cong\WW_0$ and $\Hom_{\Uq}\left(\VV_{\alpha_l+\alpha_r-1+N-2M},\Hbb\right)\cong \Hs_M$, which we still denote $\Psi_0^{(M)}$.

Note that, concretely, for all $\ell\in\WW_0$, $\Psi_0^{(M)}(\ell)\in\Hs_M$ is simply the evaluation of the $\Uq$-intertwiner $\Psi_0(\ell)\in\Hom_{\Uq}\left(\VV_{\alpha_l}\otimes\VV_{\alpha_r},\Hbb\right)$ at the highest-weight vector $\ket{w_M}$ of the $\VV_{\alpha_l+\alpha_r-1+N-2M}$ summand of $\VV_{\alpha_l}\otimes\VV_{\alpha_r}$ in~\eqref{vermafusruleM}. This evaluation $\Psi_0^{(M)}(\ell)=\Psi_0(\ell)\ket{w_M}$ is indeed an element of $\Hs_{M}$ -- the subspace of highest-weight vectors of weight $\qa^{\alpha_l+\alpha_r-2+N-2M}$ -- simply because for all $\ell\in\WW_0[\YYY]$,
\begin{gather}
\EE_{\Hbb}\!\Psi_0^{(M)}\!(\ell) = \EE_{\Hbb}\Psi_0(\ell)\!\ket{w_M}= \Psi_0(\ell)\EE_{\VV_{\alpha_l}\otimes\VV_{\alpha_r}}\!\ket{w_M}=0, \nonumber\\
\KK_{\Hbb}\!\Psi_0^{(M)}\!(\ell) = \KK_{\Hbb}\Psi_0(\ell)\!\ket{w_M}= \Psi_0(\ell)\!\KK_{\VV_{\alpha_l}\otimes\VV_{\alpha_r}}\!\ket{w_M}=\qa^{\alpha_l+\alpha_r-2+N-2M}\Psi_0(\ell)\!\ket{w_M}\! .
\label{HWcomp}
\end{gather}

Clearly $\Psi_0^{(M)}\neq 0$ as, for example,
\begin{equation*}
\begin{tikzpicture}[scale=0.5]
\draw (1,0.4) [upup=1];
\draw (3,0) node {$\ldots$};
\draw (4,0.4) [upup=1];
\draw (9,0) node {$\mapsto~\bigotimes_{i=1}^{N/2}C_{2i-1}\neq 0 .$};
\end{tikzpicture}
\end{equation*}
By Theorem~\ref{dgthm2}\,$(iii)$, the $\twoBlobm$-module $\WW_0$ is irreducible for all $M\geq N$, so $\Psi_0^{(M)}$ must be injective. Since $\dim\WW_0=\dim\Hs_M=2^N$ by~\eqref{dim2blob} and~\eqref{multqgen}, $\Psi_0^{(M)}$ is an isomorphism and so $\Hs_M\cong\WW_0$ for all $M\geq N$.

Now for $N/2\leq M\leq N-1$, $\Ker \Psi_0^{(M)}$ is a $\twoBlobm$-submodule of $\WW_0$ and so by Theorem~\ref{dgthm2}\,$(ii)$ it must be equal to either $\{0\}$, $\WW_{M+1-N/2}^{\bar{b}\bar{b}}$ or $\WW_0$. It cannot be equal to $\WW_0$ because $\Psi_0^{(M)}\neq 0$. It cannot be equal to $\{0\}$ because $d_M=\dim \Hs_M<\dim\WW_0=2^N$ by~\eqref{multqgen}. Therefore, $\Ker \Psi_0^{(M)}=\WW_{M+1-N/2}^{\bar{b}\bar{b}}$ and so $\Psi_0^{(M)}$ induces a non-zero injective $\twoBlobm$-module morphism between $\WW_0/\WW_{M+1-N/2}^{\bar{b}\bar{b}}$ and $\Hs_M$. Since by~\eqref{dim2blob} and~\eqref{multqgen} $\dim\WW_0/\WW_{M+1-N/2}^{\bar{b}\bar{b}}=\dim \Hs_M=d_M$ we have $\Hs_M \cong\WW_0/\WW_{M+1-N/2}^{\bar{b}\bar{b}}$ for all $N/2\leq M\leq N-1$.

It remains to deal with the cases $0\leq M\leq N/2-1$. Here we can no longer use $\Psi_0$ because the tensor product $\VV_{\alpha_r}\otimes\VV_{\alpha_r}$~\eqref{vermafusruleM} does not contain summands $\VV_{\alpha_l+\alpha_r-1+N-2M}$ for $0\leq M\leq N/2-1$. The idea is then to consider the above-constructed $\utwoBlob$-module morphisms~\eqref{Psij2b}
\begin{equation*}
\Psi_j\colon \ \II_j[\YYY]\to\Hom_{\Uq}\big(\VV_{\alpha_l}\otimes(\CC^2)^{\otimes 2j}\otimes\VV_{\alpha_r},\Hbb\big)
\end{equation*}
now with $1\leq j\leq N/2$. Evaluating the elements of $\Hom_{\Uq}\big(\VV_{\alpha_l}\otimes(\CC^2)^{\otimes 2j}\otimes\VV_{\alpha_r},\Hbb\big)$ at $\ket{0}\otimes\ket{\uparrow}^{\otimes 2j}\otimes\ket{0}$ or, equivalently, pre-composing them with the only (up to normalisation) $\Uq$-intertwiner
\begin{equation*}
\chi_j\colon \ \VV_{\alpha_l+\alpha_r-1+2j}\to\VV_{\alpha_l}\otimes(\CC^2)^{\otimes 2j}\otimes\VV_{\alpha_r} ,
\end{equation*}
we obtain, for each $1\leq j\leq N/2$, a morphism of $\utwoBlob$-modules\footnote{Using the same reasoning as in~\eqref{tlhw} and~\eqref{HWcomp}, we easily see that $\EE_{\Hbb}\Psi_{j}(\ell)\ket{0}\otimes\ket{\uparrow}^{\otimes 2j}\otimes\ket{0}=0$ and $\KK_{\Hbb}\Psi_{j}(\ell)\ket{0}\otimes\ket{\uparrow}^{\otimes 2j}\otimes\ket{0}=\qa^{\alpha_l+\alpha_r-2+2j}\Psi_{j}(\ell)\ket{0}\otimes\ket{\uparrow}^{\otimes 2j}\otimes\ket{0}$ for all $\ell\in\II_j[\YYY]$ so indeed $\Im\psi_{j}\subseteq\Hs_{N/2-j}$ for all $1\leq j\leq N/2$.}
\begin{gather*}
\psi_{j}\colon \ \II_j[\YYY] \to\Hom_{\Uq}\left(\VV_{\alpha_l+\alpha_r-1+2j},\Hb\right)\cong\Hs_{N/2-j},\\
\hphantom{\psi_{j}\colon}{} \ \ell \mapsto\Psi_{j}(\ell)\circ\chi_j\cong\Psi_{j}(\ell)\ket{0}\otimes\ket{\uparrow}^{\otimes 2j}\otimes\ket{0} .
\end{gather*}
By~\eqref{Ycom}, $\YYY$ acts as the scalar $Y_M$ on $\Hom_{\Uq}\left(\VV_{\alpha_l+\alpha_r-1+N-2M},\Hbb\right)\cong\Hs_{M}$ and so for all $1\leq j\leq N/2$, $\psi_{j}$ induces a morphism of $\twoBlobm$-modules, $M=N/2-j$, between $\II_j[\YYY]/(\YYY-Y_M)\II_j[\YYY]\cong\II_j$ and $\Hom_{\Uq}\left(\VV_{\alpha_l+\alpha_r-1+2j},\Hbb\right)$ which we still denote $\psi_{j}$.

Now consider the subspace $\UU_j\subset\II_j$ spanned by all two-boundary TL diagrams $\ell$ of the form $\ell'\otimes\cap^{\otimes N/2-j}$ where $\ell'\in\WW_{\leq j}^{2b}$ is a two-boundary TL diagram from $2j$ to $N$ points such that it has either
\begin{itemize}\itemsep=0pt
\item[$(i)$] strictly less than $2j$ through lines,
\item[$(ii)$] or exactly $2j$ through lines with the leftmost or rightmost through line of $\ell'$ carrying a~left/right anti-blob.
\end{itemize}
The left action of $\twoBlobm$ on $\II_j$ can only decrease the number of through lines and can never change the left/right blob/anti-blob decoration of the leftmost/rightmost through line of a diagram with $2j$ through lines,\footnote{It is actually this property that makes the standard modules $\WW_j^{bb}$, $\WW_j^{b\bar{b}}$, $\WW_j^{\bar{b}b}$, $\WW_j^{\bar{b}\bar{b}}$ well-defined.} which implies that $\UU_j$ is a stable $\twoBlobm$ subspace. Moreover, by definition of standard modules, the quotient $\II_j/\UU_j$ -- which is exactly the space of two-boundary TL diagrams with $2j$ through lines with leftmost and rightmost through lines both carrying left and right blobs -- is isomorphic to $\WW_j^{bb}$ as a $\twoBlobm$-module.

Let us show the following.

\begin{Lemma}
For all $1\leq j\leq N/2$, $\UU_j\subseteq\Ker\psi_{j}$.
\end{Lemma}

\begin{proof}
First consider $\ell\in\UU_j$ of the form $(i)$. This means that $\ell=\ell'\otimes\cap^{\otimes N/2-j}$ with ${\ell'\in\WW_{\leq j}^{2b}}$ containing at least one cap at the bottom. Moreover, at least one of these caps connects neighbouring sites: if that was not the case, it would be impossible to fill the bottom $2j$ points of~$\ell'$ with non-intersecting caps and through lines. If $j=1$, $\ell'$ contains no through lines and the bottom of~$\ell'$ has a single cap linking points $1$ and $2$ which can carry any left/right blob/anti-blob configuration. Therefore, $\Psi_1(\ell)$ is of the form $A\bar{C}_1b_lb_r$, or $A\bar{C}_1b_l\bar{b}_r$, or $A\bar{C}_1\bar{b}_lb_r$, or $A\bar{C}_1\bar{b}_l\bar{b}_r$ for some $A\in\Hom_{\Uq}(\VV_{\alpha_l}\otimes\VV_{\alpha_r},\Hbb)$ and so
\begin{equation*}
\psi_1(\ell)=\Psi_1(\ell)\ket{0}\otimes\ket{\uparrow\uparrow}\otimes\ket{0}=0
\end{equation*}
because $\bar{C}_1b_lb_r\ket{0}\otimes\ket{\uparrow\uparrow}\otimes\ket{0}=\bar{C}_1\ket{0}\otimes\ket{\uparrow\uparrow}\otimes\ket{0}=0$ and $\bar{b}_l\ket{0}\otimes\ket{\uparrow}=\bar{b}_r\ket{\uparrow}\otimes\ket{0}=0$.

Now take $j\geq 2$ and first suppose $\ell'$ contains through lines. We have the following cases:
\begin{itemize}\itemsep=0pt
\item Either the bottom points $1$ and $2j$ of $\ell'$ are both occupied by through lines. Then the caps in between cannot carry blobs/anti-blobs and so $\ell'$ contains an undecorated cap linking points $i$ and $i+1$ for some $2\leq i \leq 2j-2$.
\item Or the bottom point $1$ is connected to some other bottom point $2\leq k\leq 2j-2$ by a~cap (for~$k$ even). If $k\neq 2$, then the points strictly between $1$ and $k$ can only contain undecorated caps, at least one of them linking neighbouring points $i$ and $i+1$ for some $2\leq i \leq k-2$. If $k=2$, $\ell'$ contains a cap linking points $1$ and $2$ which can only carry a~left blob/anti-blob, because it is separated from the right boundary by a through line.
\item Or the bottom point $2j$ is connected to some other bottom point $3\leq k\leq 2j-1$ by a~cap (for $k$ odd). If $k\neq 2j-1$, then the points strictly between $k$ and $2j$ can only contain undecorated caps, at least one of them linking neighbouring points $i$ and $i+1$ for some $k+1\leq i \leq 2j-2$. If $k=2j-1$, $\ell'$ contains a cap linking points $2j-1$ and $2j$ which can only carry a~right blob/anti-blob, because it is separated from the left boundary by a~through line.
\end{itemize}
On the other hand if $\ell'$ contains no through lines and $j\geq 2$ then:
\begin{itemize}\itemsep=0pt
\item Either the bottom of $\ell'$ contains some ``long" cap linking points $1\leq k< k'\leq 2j$ with $k+3\leq k'$ (for $k$ and $k'$ of opposite parities) so there is an undecorated cap linking points~$i$ and $i+1$ for some $k+1\leq i\leq k'-1$.
\item Or the bottom of $\ell'$ contains only nearest-neighbour caps. Then either the leftmost cap (linking points $1$ and $2$) is undecorated or carries only a left blob/antiblob, or the rightmost cap (linking points $2j-1$ and $2j$) is undecorated or carries only a right blob/antiblob. Indeed, the leftmost and rightmost caps cannot both carry left and right blobs/anti-blobs as they would both need to touch the two boundaries of the system, which is impossible without them crossing.
\end{itemize}
To summarise, if $j\geq 2$, the bottom of $\ell'$ contains
\begin{itemize}\itemsep=0pt
\item either an undecorated cap linking neighbouring points $i$ and $i+1$ for some $1\leq i\leq 2j$,
\item or a cap linking points $1$ and $2$ and carrying only a left blob/anti-blob,
\item or a cap linking points $2j-1$ and $2j$ and carrying only a right blob/anti-blob.
\end{itemize}
This means that $\Psi_j(\ell)$ is of the form $A\bar{C}_i$, for some $1\leq i\leq 2j$, or $A\bar{C}_1b_l$, or $A\bar{C}_1\bar{b}_l$, or $A\bar{C}_{2j-1}b_r$, or $A\bar{C}_{2j-1}\bar{b}_r$ for some $A\in\Hom_{\Uq}\big(\VV_{\alpha_l}\otimes(\CC^2)^{\otimes 2(j-1)}\otimes\VV_{\alpha_r},\Hbb\big)$. Therefore,
\begin{equation*}
\psi_j(\ell)=\Psi_j(\ell)\ket{0}\otimes\ket{\uparrow}^{\otimes 2j}\otimes\ket{0}=0 .
\end{equation*}

Finally, if $\ell$ is of the form $(ii)$, $\Psi_j(\ell)$ is of the form $A\bar{b}_l$ or $A\bar{b}_r$ with some $A\in\Hom_{\Uq}\big(\VV_{\alpha_l}\otimes(\CC^2)^{\otimes 2j}\otimes\VV_{\alpha_r},\Hbb\big)$ and since $\bar{b}_l\ket{0}\otimes\ket{\uparrow}=\bar{b}_r\ket{\uparrow}\otimes\ket{0}=0$, $\psi_j(\ell)=0$.
\end{proof}

Now $\UU_j\subseteq\Ker \psi_{j}$ implies that $\psi_{j}$ induces a morphism of $\twoBlobm$-modules
\begin{equation*}
\tilde{\psi}_j \colon \ \II_j/\UU_j\cong\WW_j^{bb}\to\Hs_M
\end{equation*}
for all $1\leq j\leq N/2$, with $M=N/2-j$. Clearly, $\tilde{\psi}_{j}$ is non-zero as, for example,
\begin{equation*}
\begin{tikzpicture}[scale=0.5]
\draw (0,0.4) [hdnup=0];
\draw (0,0) node{$\bullet$};
\draw (8,0.4) [upup=1];
\draw (7,0) node{$\ldots$};
\draw (5,0.4) [upup=1];
\draw (1,0.4) [hdnup=0];
\draw (2,0) node{$\ldots$};
\draw (3,0.4) [hdnup=0];
\draw (4,0.4) [hdnup=0];
\draw (4,0) node[scale=0.5]{$\blacksquare$};
\draw (17,0) node {$\mapsto~\ket{0}\otimes\ket{\uparrow}^{\otimes 2j}\otimes\left(\bigotimes_{i=j}^{N-1}C_{2i+1}\right)\otimes\ket{0}\neq 0 .$};
\end{tikzpicture}
\end{equation*}
By Theorem~\ref{dgthm2}, the $\twoBlobm$-module $\WW_{N/2-M}^{bb}$ is irreducible for all $1\leq M\leq N/2-1$ so~$\tilde{\psi}_{j}$ is injective. Since $\dim\WW_{N/2-M}^{bb}=\dim\Hs_M$ by~\eqref{dim2blob} and~\eqref{multqgen}, we have $\WW_{N/2-M}^{bb}\cong\Hs_M$ for all $0\leq M\leq N/2-1$ which completes the proof.

\section[ABA for a general U\_q sl\_2-invariant highest-weight spin chain]{ABA for a general $\boldsymbol{\Uq}$-invariant highest-weight spin chain}
\label{genBAE}

Consider a tensor product $\XX:=\bigotimes_{i=1}^n\XX_i$ of irreducible highest-weight $\Uq$-modules of weight $\qa^{\alpha_i-1}$ (that is, $\XX_i$ is a spin-$\frac{\alpha_i-1}{2}$ representation if $\alpha_i\in\NN^*$ and a Verma module otherwise) and define the monodromy
\begin{equation*}
\mathcal{T}(u):=T(u-h/2)\hat{T}(u-h/2)=
\begin{pmatrix}
\mathcal{A}(u) & \mathcal{B}(u)\\
\mathcal{C}(u) & \mathcal{D}(u)
\end{pmatrix}
\end{equation*}
with
\begin{gather*}
T(u-h/2) :=
\begin{pmatrix}
A(u) & B(u)\\
C(u) & D(u)
\end{pmatrix}
=R_{0,\XX_n}(u-h/2-\zeta_n)\cdots R_{0,\XX_1}(u-h/2-\zeta_1) ,\\
\hat{T}(u-h/2) :=
\begin{pmatrix}
\hat{A}(u) & \hat{B}(u)\\
\hat{C}(u) & \hat{D}(u)
\end{pmatrix}
=R_{\XX_1,0}(u-h/2+\zeta_1)\cdots R_{\XX_n,0}(u-h/2+\zeta_n) ,
\end{gather*}
for some inhomogeneity parameters $\zeta_i\in\CC$, $1\leq i\leq n$, and the transfer matrix
\begin{equation*}
t(u) := \qtr_0 \mathcal{T}(u)={\rm e}^h\mathcal{A}(u)+{\rm e}^{-h}\mathcal{D}(u) .
\end{equation*}

By~\cite[Proposition 2]{sklyanin1988boundary}, $\mathcal{T}(u)$ is a solution of the bYBE~\eqref{BYBEexp}: using the YBE~\eqref{YBEX} one checks that $T(u)$ satisfies the RTT relation~\eqref{RTT}, $\hat{T}(u)\propto T(-u)^{-1}$ because of~\eqref{invgen}, and $\Id_{\CC^2}$ is a~solution of the bYBE~\eqref{BYBE} by~\eqref{idsol}, so the product $\mathcal{T}(u+h/2):=T(u)\Id_{\CC_2}\hat{T}(u)$ is also one.\footnote{This statement can also be proved by induction on $n$ using only the YBE~\eqref{YBEX} and~\eqref{idsol}.} Therefore, the commutation relations~\eqref{comrelA} and \eqref{comrelD} remain the same.

It remains to compute the eigenvalues of $\mathcal{A}(u)$ and $\bar{\mathcal{D}}(u)$ when acting on the reference state%
\begin{equation*}
\ket{\mathbf{0}}:=\bigotimes_{i=1}^n\ket{0_i},
\end{equation*}
where $\ket{0_i}$ is the highest-weight vector of $\XX_i$. To do so, first write
\begin{align*}
R_{0,\XX_i}(u-h/2-\zeta_i) & =\dfrac{{\rm e}^{u-\zeta_i}}{2}\RRR_{\CC^2,\XX_i}-\dfrac{{\rm e}^{-u+\zeta_i}}{2}P_{\XX_i,\CC^2}\circ\RRR_{\XX_i,\CC^2}^{-1}\circ P_{\CC^2,\XX_i}\\
& = \dfrac{{\rm e}^{u-\zeta_i}}{2}
\begin{pmatrix}
\KK^{1/2}_i & \{1\}\KK^{1/2}_i\FF_i \\
0 & \KK^{-1/2}_i
\end{pmatrix}
-\dfrac{{\rm e}^{-u+\zeta_i}}{2}
\begin{pmatrix}
\KK^{-1/2}_i & 0\\
-\{1\}\EE_i\KK^{-1/2}_i & \KK^{1/2}_i
\end{pmatrix}\\
& =\begin{pmatrix}
\sinh(u-\zeta_i+h\HH_i/2) & \sinh(h)\exp(u-\zeta_i+h\HH_i/2)\FF_i\\
\sinh(h)\EE_i\exp(-u+\zeta_i-h\HH_i/2) & \sinh(u-\zeta_i-h\HH_i/2)
\end{pmatrix} \\
& :=\begin{pmatrix}
\mathsf{a}_i & \mathsf{b}_i\\
\mathsf{c}_i & \mathsf{d}_i
\end{pmatrix},
\end{align*}
where $\qa^{\HH_i}={\rm e}^{h\HH_i}:=\KK_i$. Similarly,
\begin{align*}
R_{\XX_i,0}(u-h/2+\zeta_i)& =\dfrac{{\rm e}^{u+\zeta_i}}{2}\RRR_{\XX_i,\CC^2}-\dfrac{{\rm e}^{-u-\zeta_i}}{2}P_{\XX_i,\CC^2}\circ\RRR_{\CC^2,\XX_i}^{-1}\circ P_{\XX_i,\CC^2}\\
& =\begin{pmatrix}
\sinh(u+\zeta_i+h\HH_i/2) & \sinh(h)\FF_i\exp(-u-\zeta_i+h\HH_i/2)\\
\sinh(h)\exp(u+\zeta_i-h\HH_i/2)\EE_i & \sinh(u+\zeta_i-h\HH_i/2)
\end{pmatrix}\\
&:=\begin{pmatrix}
\hat{\mathsf{a}}_i & \hat{\mathsf{b}}_i\\
\hat{\mathsf{c}}_i & \hat{\mathsf{d}}_i
\end{pmatrix} .
\end{align*}
Since $C(u)\ket{\mathbf{0}}=\hat{C}(u)\ket{\mathbf{0}}=0$, we have
\begin{gather}
\mathcal{A}(u)\ket{\mathbf{0}} = A(u)\hat{A}(u)\ket{\mathbf{0}} ,\nonumber\\
\mathcal{D}(u)\ket{\mathbf{0}} = \big(C(u)\hat{B}(u)+D(u)\hat{D}(u)\big)\ket{\mathbf{0}} .
\label{ADgen}
\end{gather}
Introducing a basis $\{\ket{\uparrow},\ket{\downarrow}\}$ of the auxiliary space, we have, similarly to the one-boundary computation in Section~\ref{oneBA},
\begin{gather}
A(u)\ket{\mathbf{0}} = \bra{\uparrow}T(u)\ket{\uparrow}\otimes\ket{\mathbf{0}} = \prod_{i=1}^n \mathsf{a}_i\ket{\mathbf{0}} =\prod_{i=1}^n\sinh(u-\zeta_{i}+h\frac{\alpha_{i}-1}{2})\ket{\mathbf{0}} ,\nonumber\\
\hat{A}(u)\ket{\mathbf{0}} = \bra{\uparrow}\hat{T}(u)\ket{\uparrow}\otimes\ket{\mathbf{0}} = \prod_{i=1}^n \hat{\mathsf{a}}_i\ket{\mathbf{0}} =\prod_{i=1}^n\sinh(u+\zeta_{i}+h\frac{\alpha_{i}-1}{2})\ket{\mathbf{0}} ,\nonumber\\
D(u)\ket{\mathbf{0}} = \bra{\downarrow}T(u)\ket{\downarrow}\otimes\ket{\mathbf{0}} = \prod_{i=1}^n \mathsf{d}_i\ket{\mathbf{0}} =\prod_{i=1}^n\sinh(u-\zeta_{i}-h\frac{\alpha_{i}-1}{2})\ket{\mathbf{0}} ,\nonumber\\
\hat{D}(u)\ket{\mathbf{0}} = \bra{\downarrow}\hat{T}(u)\ket{\downarrow}\otimes\ket{\mathbf{0}} = \prod_{i=1}^n \hat{\mathsf{d}}_i\ket{\mathbf{0}} =\prod_{i=1}^n\sinh(u+\zeta_{i}-h\frac{\alpha_{i}-1}{2})\ket{\mathbf{0}} .
\label{adgen}
\end{gather}
and
\begin{align}
C(u)\hat{B}(u)\ket{\mathbf{0}} & = \bra{\downarrow}T(u)\ket{\uparrow}\bra{\uparrow}\hat{T}(u)\ket{\downarrow}\otimes\ket{\mathbf{0}}\nonumber\\
& =\bra{\downarrow}T(u)\ket{\uparrow}\otimes\sum_{k=1}^{n}\left(\prod_{i=1}^{k-1}\hat{\mathsf{a}}_i\right)\hat{\mathsf{b}}_k\left(\prod_{i=k+1}^{n}\hat{\mathsf{d}}_i\right)\ket{\mathbf{0}}\nonumber\\
& =\sum_{k=1}^{n}\left(\prod_{i=k+1}^{n}\mathsf{d}_i\right)\mathsf{c}_k\left(\prod_{i=1}^{k-1}\mathsf{a}_i\right)\left(\prod_{i=1}^{k-1}\hat{\mathsf{a}}_i\right)\hat{\mathsf{b}}_k\left(\prod_{i=k+1}^{n}\hat{\mathsf{d}}_i\right)\ket{\mathbf{0}}\nonumber\\
& =\sum_{k=1}^{n}\left(\prod_{i=1}^{k-1}\mathsf{a}_i\hat{\mathsf{a}}_i\right)\left(\prod_{i=k+1}^{n}\mathsf{d}_i\hat{\mathsf{d}}_i\right)\mathsf{c}_k\hat{\mathsf{b}}_k\ket{\mathbf{0}} .
\label{cbgen}
\end{align}
Note also that
\begin{align}
\mathsf{c}_k\hat{\mathsf{b}}_k & =\sinh^2(h){\rm e}^{-2u}\EE_k\KK_k^{-1/2}\FF_k\KK_k^{1/2}\nonumber\\
& =\sinh^2(h){\rm e}^{-2u+h}\EE_k\FF_k\nonumber\\
& =\frac{{\rm e}^{-2u+h}}{2}\left(\cosh(h\alpha_k)-\cosh(h\HH_k-h)\right)\nonumber\\
& ={\rm e}^{-2u+h}\sinh(\frac{h\alpha_k+h\HH_k-h}{2})\sinh(\frac{h\alpha_k-h\HH_k+h}{2}),
\label{cbk}
\end{align}
where we used~\eqref{casimir}--\eqref{casimirva} to write
\begin{equation*}
\CCC_{\XX_k}=4\sinh^2(h)\EE_k\FF_k+2\cosh(h\HH_k-h)=2\cosh(h\alpha_k) .
\end{equation*}

Define
\begin{gather*}
\mathsf{D}_n:={\rm e}^{-2u}\sinh(h)\left(\prod_{i=1}^n\mathsf{d}_i\hat{\mathsf{d}}_i-\prod_{i=1}^n\mathsf{a}_i\hat{\mathsf{a}}_i\right)+{\rm e}^{-h}\sinh(2u)\sum_{k=1}^{n}\left(\prod_{i=1}^{k-1}\mathsf{a}_i\hat{\mathsf{a}}_i\right)\left(\prod_{i=k+1}^{n}\mathsf{d}_i\hat{\mathsf{d}}_i\right)\mathsf{c}_k\hat{\mathsf{b}}_k .
\end{gather*}
Let us show by induction on $n\geq 1$ that $\mathsf{D}_n\ket{\mathbf{0}}=0$. For $n=1$, using~\eqref{cbk}, we can check by direct computation that
\begin{equation}
\label{D1}
\mathsf{D}_1\ket{0_1}=\big({\rm e}^{-2u}\sinh(h)\big(\mathsf{d}_1\hat{\mathsf{d}}_1-\mathsf{a}_1\hat{\mathsf{a}}_1\big)+{\rm e}^{-h}\sinh(2u)\mathsf{c}_1\hat{\mathsf{b}}_1\big)\ket{0_1}=0 .
\end{equation}
Now assume that $\mathsf{D}_{n-1}\ket{\mathbf{0}}=0$. We have
\begin{equation*}
\mathsf{a}_n\hat{\mathsf{a}}_n\mathsf{D}_{n-1}=\mathsf{D}_{n}-\big({\rm e}^{-2u}\sinh(h)\big(\mathsf{d}_n\hat{\mathsf{d}}_n-\mathsf{a}_n\hat{\mathsf{a}}_n\big)+{\rm e}^{-h}\sinh(2u)\mathsf{c}_n\hat{\mathsf{b}}_n\big)\prod_{i=1}^{n-1}\mathsf{d}_i\hat{\mathsf{d}}_i,
\end{equation*}
so by~\eqref{D1} and the induction hypothesis $\mathsf{D}_{n}\ket{\mathbf{0}}=0$. By~\eqref{dbar}, \eqref{ADgen}--\eqref{cbgen},
\begin{equation*}
\mathsf{D}_{n}\ket{\mathbf{0}}=\sinh(2u-h)\left(\bar{\mathcal{D}}(u)-\prod_{i=1}^{n}\mathsf{d}_i\hat{\mathsf{d}}_i\right)\ket{\mathbf{0}}=0,
\end{equation*}
so finally
\begin{gather}
\mathcal{A}(u)\ket{\mathbf{0}} =\prod_{i=1}^{n}\mathsf{a}_i\hat{\mathsf{a}}_i\ket{\mathbf{0}}=\prod_{i=1}^n\sinh(u-\zeta_{i}+h\frac{\alpha_{i}-1}{2})\sinh(u+\zeta_{i}+h\frac{\alpha_{i}-1}{2})\ket{\mathbf{0}} ,\nonumber\\
\bar{\mathcal{D}}(u)\ket{\mathbf{0}} =\prod_{i=1}^{n}\mathsf{d}_i\hat{\mathsf{d}}_i\ket{\mathbf{0}}=\prod_{i=1}^n\sinh(u-\zeta_{i}-h\frac{\alpha_{i}-1}{2})\sinh(u+\zeta_{i}-h\frac{\alpha_{i}-1}{2})\ket{\mathbf{0}} .
\label{eigADgen}
\end{gather}

Therefore, using the commutation relations~\eqref{comrelA} and~\eqref{comrelD} and performing exactly the same computations~\eqref{auxA} and~\eqref{auxD} as in the one-boundary case but with the new eigenvalues of $\mathcal{A}(u)$ and $\bar{\mathcal{D}}(u)$~\eqref{eigADgen}, we find that
\begin{equation*}
\ket{\{v_m\}}=\mathcal{B}(v_1)\cdots\mathcal{B}(v_M)\ket{\mathbf{0}}
\end{equation*}
is an eigenvector of the transfer matrix $t(u)$ with eigenvalue
\begin{equation*}
\begin{aligned}
\Lambda(\{v_m\};u)= & \frac{\sinh(2u+h)}{\sinh(2u)}\left(\prod_{i=1}^n\Delta_{\alpha_i,\zeta_i}(u)\right)\prod_{m=1}^M\frac{\sinh(u-v_m-h)\sinh(u+v_m-h)}{\sinh(u-v_m)\sinh(u+v_m)}\\
& + \frac{\sinh(2u-h)}{\sinh(2u)}\left(\prod_{i=1}^n\Delta_{\alpha_i,\zeta_i}(-u)\right)\prod_{m=1}^M\frac{\sinh(u-v_m+h)\sinh(u+v_m+h)}{\sinh(u-v_m)\sinh(u+v_m)},
\end{aligned}
\end{equation*}
where
\begin{equation*}
\Delta_{\alpha_i,\zeta_i}(u):=\sinh(u+h\frac{\alpha_{i}-1}{2}-\zeta_{i})\sinh(u+h\frac{\alpha_{i}-1}{2}+\zeta_{i})
\end{equation*}
if and only if $\{v_m\}_{1\leq m\leq M}$ satisfy the Bethe ansatz equations
\begin{equation}
\label{BAEgen}
\prod_{i=1}^n\frac{\Delta_{\alpha_i,\zeta_i}(v_m)}{\Delta_{\alpha_i,\zeta_i}(-v_m)}=\prod_{\underset{k\neq m}{k=1}}^M\frac{\sinh(v_m-v_k+h)\sinh(v_m+v_k+h)}{\sinh(v_m-v_k-h)\sinh(v_m+v_k-h)}
\end{equation}
for all $1\leq m\leq M$. As before $\mathcal{B}(\infty):=\lim_{u\to+\infty}{\rm e}^{-2nu}\mathcal{B}(u)\propto\FF_\XX$ and so we expect the finite (permutation invariant) solutions $\{v_k\}_{1\leq k\leq M}$ of the BAE~\eqref{BAEgen} belonging to the fundamental domain $S_+$ to provide all the $\Uq$ highest-weight eigenstates of weight $\qa^{\left(\sum_{i=1}^n\alpha_i\right)-n-2M}$ of $t(u)$.

Note that for $\XX_i=\CC^2$, that is $\alpha_i=2$, $\Delta_{2,0}(u)=\sinh(u-h/2)^2$, and for $\XX_i=\VV_{\alpha_{l/r}}$, $\Delta_{\alpha_{l/r},\zeta_{l/r}}(u)\propto\Delta_{l/r}(u)$~\eqref{delta2}. Thus for $n=N+2$, $\XX_1=\VV_{\alpha_l}$, $\zeta_1=\zeta_l$, $\XX_{N+2}=\VV_{\alpha_r}$, $\zeta_{N+2}=\zeta_r$ and $\XX_i=\CC^2$, $\zeta_i=0$ for all $2\leq i\leq N+1$, we recover~\eqref{lambda2b} and \eqref{BAE2} (up to normalisation of~$t_{2b}(u)$) as we should.

\section{A different set of BAE for the two-boundary system}
\label{otherBAE}

Following~\cite{bookoff}, it is possible to find the general form of the eigenvalues of $t_{2b}(u)$ solely from the knowledge of some functional relations it satisfies and their analytic properties. The end result is that for any eigenvalue $\Lambda_{2b}(\{v_m\};u)$ of $t_{2b}(u)$ one can write a TQ relation of the form
\begin{align*}
\Lambda_{2b}(\{v_m\};u)= & \sinh^{2N}(u+h/2)\Delta_l(u)\Delta_r(u)\dfrac{\sinh(2u+h)}{\sinh(2u)}\dfrac{Q(u-h)}{Q(u)} \\ & +\sinh^{2N}(u-h/2)\Delta_l(-u)\Delta_r(-u)\dfrac{\sinh(2u-h)}{\sinh(2u)}\dfrac{Q(u+h)}{Q(u)}\\
& +c\dfrac{\sinh^{2N}(u+h/2)\sinh^{2N}(u-h/2)\sinh(2u+h)\sinh(2u-h)}{Q(u)},
\end{align*}
where
\begin{equation*}
Q(u)=\prod_{m=1}^N\sinh(u-v_m)\sinh(u+v_m)
\end{equation*}
for some Bethe roots $v_m$, $1\leq m\leq N$ and $c$ some constant. The Bethe ansatz equations are obtained by imposing that $\Lambda_{2b}(\{v_m\};u)$ has no poles at all the $v_m$, that is
\begin{gather}
 \sinh^{2N}(v_m+h/2)\Delta_l(v_m)\Delta_r(v_m)\prod_{\underset{k\neq m}{k=1}}^N\sinh(v_m-v_k-h)\sinh(v_m+v_k-h)\nonumber\\
\qquad{}-\sinh^{2N}(v_m-h/2)\Delta_l(-v_m)\Delta_r(-v_m)\prod_{\underset{k\neq m}{k=1}}^N\sinh(v_m-v_k+h)\sinh(v_m+v_k+h)\nonumber\\
\quad\qquad{}=c\sinh^{2N}(v_m+h/2)\sinh^{2N}(v_m-h/2)
\label{2BAEc}
\end{gather}
for all $1\leq m\leq N$. Note that if $c\neq 0$, the trigonometric polynomial $Q(u)$ has to be of degree~$2N$ for the BAE~\eqref{2BAEc} to admit at least one solution.

The right-hand side of~\eqref{2BAEc} is known as the ``inhomogeneous term" and this equation provides the BAE for $\Hnd$~\eqref{hnd} for arbitrary values of $h$, $\delta_{r/l}$, $\kappa_{r/l}$ and $\Theta$, whether they satisfy the Nepomechie condition or not, with $c$ an explicit function of these parameters. Actually, the Nepomechie cases~\eqref{nepcond} with $0\leq M\leq N$ correspond to $c=0$. Indeed, if $c=0$, then a subset of the $v_m$ can be sent to $\infty$ and we obtain
\begin{align*}
\Lambda_{2b}(\{v_m\};u)={} & \sinh^{2N}(u+h/2)\Delta_l(u)\Delta_r(u)\dfrac{\sinh(2u+h)}{\sinh(2u)}\dfrac{Q(u-h)}{Q(u)} \\ & {}+\sinh^{2N}(u-h/2)\Delta_l(-u)\Delta_r(-u)\dfrac{\sinh(2u-h)}{\sinh(2u)}\dfrac{Q(u+h)}{Q(u)}
\end{align*}
with
\begin{equation*}
Q(u)=\prod_{m=1}^M\sinh(u-v_m)\sinh(u+v_m)
\end{equation*}
for some magnon number $0\leq M\leq N$ as in~\eqref{lambda2b} and the BAE~\eqref{2BAEc} then reduce to the form~\eqref{BAE2}
\begin{equation*}
\dfrac{\Delta_l(v_m)\Delta_r(v_m)}{\Delta_l(-v_m)\Delta_r(-v_m)}\left(\dfrac{\sinh(v_m+h/2)}{\sinh(v_m-h/2)}\right)^{2N}=\prod_{\underset{k\neq m}{k=1}}^M\frac{\sinh(v_m-v_k+h)\sinh(v_m+v_k+h)}{\sinh(v_m-v_k-h)\sinh(v_m+v_k-h)}
\end{equation*}
for all $1\leq m\leq M$. Thus the BAE from~\cite{bookoff} are consistent with~\eqref{BAE2} as long as $0\leq M\leq N$. However, we have seen that in the two-boundary case the magnon number $M$ can take any (positive) integer value. The question is then what happens for $M>N$.

For this we need to come back to the general $c\neq 0$ case. It turns out that the constant $c$ can be entirely fixed from the knowledge of the large $u$ asymptotic. Indeed, if $c\neq 0$, then
\begin{equation}
\label{lamasymp}
\Lambda_{2b}(\{v_m\};u) \underset{u\to\infty}{\sim} 4^{-N-2}{\rm e}^{2(N+2)u} \left(\dfrac{2\mu_l\mu_r\cosh(h\alpha_l+h\alpha_r-(N+1)h)}{\sinh(h)^2\sinh(h\alpha_l)\sinh(h\alpha_r)}+4c\right) .
\end{equation}
On the other hand, we can compute the large $u$ limit of $t_{2b}(u)$ directly. We have~\eqref{Rgen}
\begin{equation*}
R_{\XX,\CC^2}(u) \underset{u\to\infty}{\sim} \dfrac{{\rm e}^{u+\frac{h}{2}}}{2}\RRR_{\XX,\CC^2},\qquad R_{\CC^2,\XX}(u) \underset{u\to\infty}{\sim} \dfrac{{\rm e}^{u+\frac{h}{2}}}{2}\RRR_{\CC^2,\XX}
\end{equation*}
for any $\Uq$ module $\XX$ and so, using~\eqref{t2b} and~\eqref{t2baux} and then~\eqref{rfus1}--\eqref{RRrel}
\begin{align*}
t_{2b}(u)\underset{u\to\infty}{\sim}{} & \dfrac{4^{-N-2}{\rm e}^{2(N+2)u}\mu_l\mu_r\qtr_0\RRR_{0,\VV_{\alpha_r}}\RRR_{0,N}\cdots \RRR_{0,1}\RRR_{0,\VV_{\alpha_l}}\RRR_{\VV_{\alpha_l},0}\RRR_{1,0}\cdots \RRR_{N,0}\RRR_{\VV_{\alpha_r},0}}{\sinh(h)^2\sinh(h\alpha_l)\sinh(h\alpha_r)}\\
={}&\dfrac{4^{-N-2}{\rm e}^{2(N+2)u}\mu_l\mu_r\qtr_0\RRR_{0,\Hbb}\RRR_{\Hbb,0}}{\sinh(h)^2\sinh(h\alpha_l)\sinh(h\alpha_r)}\\
={}&\dfrac{4^{-N-2}{\rm e}^{2(N+2)u}\mu_l\mu_r\left(\qa\left(\KK_{\Hbb}+\qam\{1\}^2\FF_{\Hbb}\EE_{\Hbb}\right)+\qam\KK_{\Hbb}^{-1}\right)}{\sinh(h)^2\sinh(h\alpha_l)\sinh(h\alpha_r)}\\
={}&\dfrac{4^{-N-2}{\rm e}^{2(N+2)u}\mu_l\mu_r\CCC_{\Hbb}}{\sinh(h)^2\sinh(h\alpha_l)\sinh(h\alpha_r)},
\end{align*}
where $\CCC_{\Hbb}$ is the Casimir element~\eqref{casimir} evaluated in the representation ${\Hbb}$. Recalling that $\CCC$ is constant on any irreducible $\Uq$ module, in particular~\eqref{casimirva}
\begin{equation*}
\CCC_{\VV_\alpha}=\qa^\alpha+\qa^{-\alpha}=2\cosh(h\alpha)
\end{equation*}
and the $\Uq$ decomposition~\eqref{hbbdec}
\begin{equation*}
\Hbb=\bigoplus_{M\geq 0}\Hs_M\otimes\VV_{\alpha_l+\alpha_r-1+N-2M}
\end{equation*}
it is then clear that
\begin{equation}
\label{tasymp}
t_{2b}(u)|_{\Hs_M}\underset{u\to\infty}{\sim} 4^{-N-2}{\rm e}^{2(N+2)u}\dfrac{2\mu_l\mu_r\cosh(h\alpha_l+h\alpha_r+h(N-2M-1))}{\sinh(h)^2\sinh(h\alpha_l)\sinh(h\alpha_r)} .
\end{equation}
Now comparing~\eqref{lamasymp} and~\eqref{tasymp}, we finally arrive at
\begin{equation}
\label{cvalue}
c=\dfrac{\mu_l\mu_r\sinh(h(N-M))\sinh(h\alpha_l+h\alpha_r-h(M+1))}{\sinh(h)^2\sinh(h\alpha_l)\sinh(h\alpha_r)} .
\end{equation}

We are thus led to conjecture that for $M\geq N$, the BAE~\eqref{BAE2} on $M$ Bethe roots are equivalent to the BAE~\eqref{2BAEc} with only $N$ Bethe roots and $c$ as in~\eqref{cvalue}. Note that the case $M=N$ is trivial since $c$ vanishes precisely for $M=N$.

\subsection*{Acknowledgements}

We are thankful to Samuel Belliard, Rouven Frassek and Karol Kozlowski for useful discussions. We thank the anonymous referees for their constructive comments. This work was supported by the French Agence Nationale de la Recherche (ANR) under grant ANR-21-CE40-0003 (project CONFICA). The work of A.M.G.\ was supported by the CNRS, and partially by the ANR grant JCJC ANR-18-CE40-0001 and the RSF Grant No.~20-61-46005. A.M.G.\ is also grateful to IPHT Saclay and LPENS in Paris for their kind hospitality during the visits in 2022.

\pdfbookmark[1]{References}{ref}
\LastPageEnding

\end{document}